\providecommand{\pgfsyspdfmark}[3]{}
\documentclass[aps, prl, reprint, superscriptaddress, nofootinbib]{revtex4-2}

\usepackage[utf8]{inputenc}
\usepackage[english]{babel}
\usepackage{float, comment, physics}
\usepackage{times}
\usepackage{graphicx,epsfig} 
\usepackage[caption=false]{subfig}
\graphicspath{{images/}}
\usepackage{color}
\usepackage[dvipsnames]{xcolor}
\usepackage{amsmath,bbm,amssymb, amsthm, mathtools}
\usepackage{dsfont} 
\usepackage{stmaryrd} 
\usepackage{tikz}
\definecolor{linkcolor}{rgb}{0,0,0.6}	
\usepackage[colorlinks=true, pdfstartview=FitV, linkcolor= linkcolor, citecolor= linkcolor, urlcolor= linkcolor, hyperindex=true, hyperfigures=false]{hyperref}
\hypersetup{colorlinks,linkcolor={MidnightBlue},citecolor={MidnightBlue},urlcolor={MidnightBlue}}
\usepackage{numprint}
\addto\captionsenglish{}

\newcommand\C[1]{\tikz[remember picture]{\node(#1)[inner sep=0pt]{\sffamily#1};}}

\newcommand{\newbrace}[2]{
	\begin{tikzpicture}[baseline=-0.5ex]
		\draw[->] (0,0) -- node[midway,fill=white] {#1}  (1.,1.);
		\draw[->] (0,0)-- node[midway,fill=white] {#2}  (1.,-1.);
	\end{tikzpicture}
}

\newcommand{\newbracebig}[2]{
	\begin{tikzpicture}[baseline=-0.5ex]
		\draw[->] (0,0) -- node[midway,fill=white] {#1}  (1.4,2.7);
		\draw[->] (0,0)-- node[midway,fill=white] {#2}  (1.4,-2.7);
	\end{tikzpicture}
}

\newenvironment{casesnew}[2]
{\;\newbrace{#1}{#2}\;\begin{array}{@{}l@{}}}{\end{array}}

\newenvironment{casesnewbig}[2]
{\;\newbracebig{#1}{#2}\;\begin{array}{@{}l@{}}}{\end{array}}

\newcommand{\ii}{\mathrm{i}}

\newcommand*{\toccontents}{\@starttoc{toc}}

\begin{document}

\title{Run-and-tumble exact work statistics in a lazy quantum measurement engine: \\ stochastic information processing }

\author{L\'{e}a Bresque}
\thanks{Lead and corresponding author} 
\email[]{lea.bresque@protonmail.com}
\affiliation{ICTP -- The Abdus Salam International Centre for Theoretical Physics, Strada Costiera 11, 34151 Trieste, Italy}

\author{Debraj Das}
\affiliation{ICTP -- The Abdus Salam International Centre for Theoretical Physics, Strada Costiera 11, 34151 Trieste, Italy}

\author{\'Edgar Rold\'an}
\thanks{Corresponding author}
\email[]{edgar@ictp.it}
\affiliation{ICTP -- The Abdus Salam International Centre for Theoretical Physics, Strada Costiera 11, 34151 Trieste, Italy}

\date{\today}

\begin{abstract} 

We introduce a single-qubit quantum measurement engine fuelled by backaction energy input. To reduce energetic costs associated with information processing, the measurement outcomes are only used with a prescribed {\em laziness} probability in the feedback step. As a result, we show that the work extracted over consecutive cycles is a second-order Markov process, 
analogous to a  run-and-tumble process with transient anomalous diffusion. We derive exact analytical expressions for the work finite-time moments and first-passage-time statistics. Furthermore, we find the optimal laziness probability  maximizing the mean power extracted per cycle. 
\end{abstract}


\maketitle

\textit{Introduction}. Information engines use measurements outcomes to manipulate energy flows in ways seemingly incompatible with the classical laws of thermodynamics. 
As exemplified by Maxwell's demon thought experiment, 
processing information allows extracting work from a single heat reservoir or moving a particle against a potential, at no apparent work cost. More generally, having information on a system allows decreasing its entropy. Although it was later understood that information processing comes with an energy cost~\cite{leff2002maxwell}, thereby solving the apparent discrepancies, the energetic implications of manipulating information have become pivotal in a plethora  of cutting-edge  research avenues~\cite{parrondoSzilardEngineRevisited2001,berutExperimentalVerificationLandauer2012,parrondoThermodynamicsInformation2015,strasberg2017quantum,pekola2019thermodynamics,bechhoefer2021control,garcia2024optimal}.   

Quantum measurement engines (QME) not only use the information obtained by measuring, but can also leverage on a purely quantum resource: backaction, i.e., their disturbance on the state of quantum systems. QMEs can be fuelled solely from backaction, i.e., without feedback control, when a thermalization step resets the working substance in each cycle~\cite{dingMeasurementdrivenSingleTemperature2018,yiSingletemperatureQuantumEngine2017,buffoniQuantumMeasurementCooling2019}. 
They can also run without connecting the working substance to a bath, using backaction 
and information processing~\cite{bresqueTwoQubitEngineFueled2021,jussiauManybodyQuantumVacuum2023}. 
A simple example is that of a qubit measured once per cycle, with possible outcomes ``$+$" and ``$-$", respectively associated with the post measurement states $\ket{+}$ and~$\ket{-}$~\cite{elouardExtractingWorkQuantum2017}. This QME uses backaction as a source of energy and the measurement outcome information to deterministically obtain a qubit state most favourable for extracting this work in a field battery. Information is therefore used to rectify for the stochastic backaction, which most significantly changes the qubit state with probability $p_j$. 

If the measurement outcomes are not (or not always) used, the states of the qubit at the end of successive cycles are not deterministic anymore and even become correlated. Therefore, it is only on average and in the steady state that the working substance starts and end each cycle from the same state. At the stochastic level, we call trajectory a realization of a discrete number of consecutive runs. All trajectories start from the same state but will reach it again after a different number of cycles. To characterize the statistics of these numbers, we obtain the probability of the stopping times given the stopping condition of reaching back the qubit initial state together with having extracted a specific amount of work.

In this Letter, we introduce a lazy quantum measurement engine (LQME) where $p_l \in [0,1]$ designates the laziness probability, i.e., the probability not to use the measurement outcome. When $p_l=0$, the protocol, adapted from~\cite{elouardExtractingWorkQuantum2017}, always implement a feedback protocol according to the outcomes. When $p_l=1$, the feedback assumes the outcome is "+" no matter its actual value, i.e., the engine runs in absence of information. We solve the fixed-time and first-passage time statistics of the work extracted by mapping it to a Run-and-Tumble (RnT) process, in the most general $0\leq p_l\leq 1$ case. Moreover, we derive and optimize the steady-state power extracted according to $p_l$.

\begin{figure}
\centering
\includegraphics[width=0.98\columnwidth]{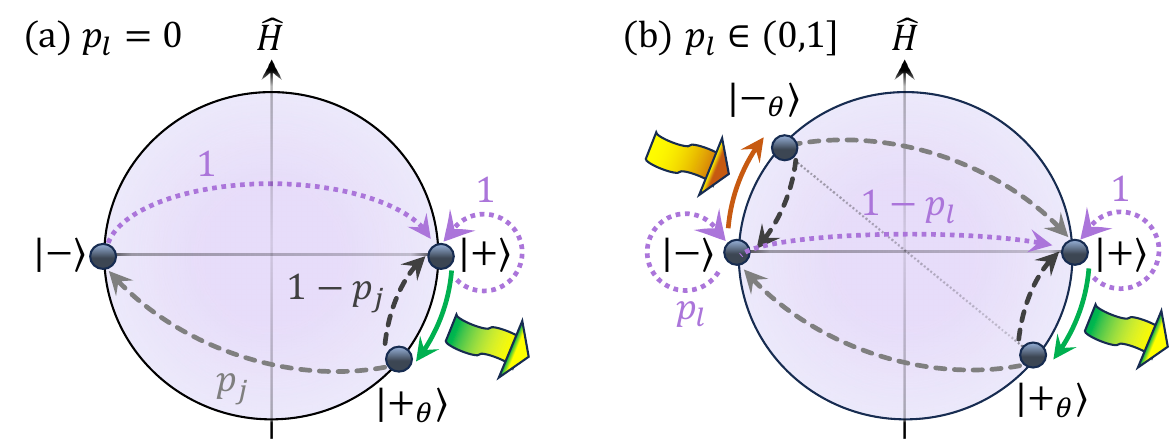}
\vspace{-8pt}
\caption{Illustration of a single-qubit LQME's (Lazy Quantum Measurement Engine) working principle on a Bloch sphere. (a) When the measurement outcomes are systematically used, $p_l=0$, a cycle consists of (i) a measurement in the $\{\ket{+},\ket{-}\}$ basis (dashed line), followed by (ii) a feedback step (dotted line), and (iii) a work extraction step (solid line). Lastly, (iv) the memory containing the measurement outcome is erased (not shown). (b) Allowing for a {\em lazy} feedback step, i.e., not using the outcome information with probability $p_l>0$, work can be either extracted from (outgoing green arrow) or exerted to (incoming orange arrow) the qubit, the latter occurring when step (iii) is applied from $\ket{-}$ (red line).}
\label{fig:working principle}
\end{figure}

\textit{Working principle ($p_l=0$).} We consider a single qubit measured once per cycle. Key to its working principle, the initial state of the qubit $\ket{+_\theta} = \cos(\theta/2)\ket{+}- \sin(\theta/2)\ket{-}$ and the measurement basis $\{\ket{+},\ket{-}\}$ are such that, no matter the measurement outcome, the average qubit's energy is always greater after the measurement. The bare qubit Hamiltonian being $\hat{H}_0 =\hbar\omega_0 \sigma_z$, this translates into $\Tr \big[ \hat{H}_0\ket{+_\theta}\bra{+_\theta} \big] < \Tr \big[ \hat{H}_0\ket{\pm}\bra{\pm} \big]$. Therefore, as illustrated in Fig~\ref{fig:working principle}.(a), 
the measurement can act as a purely quantum 
energy source
~\cite{brandnerCoherenceenhancedEfficiencyFeedbackdriven2015,elouardRoleQuantumMeasurement2017}. Following this measurement step (i), which bring the qubit state $\ket{+_\theta}$ to $\ket{-}$ with probability
\begin{equation}
    p_j = \sin^2(\theta/2),
\end{equation}
and to $\ket{+}$ with probability $1-p_j = \cos^2(\theta/2)$, a feedback loop (ii) is implemented based on the measurement outcome. Its purpose simply is to bring the $\ket{-}$ state to the $\ket{+}$ state at no energy cost. Then a work extraction step (iii) occurs, in which  the qubit is connected to a classical electromagnetic field 
that it aims to amplify for a finite duration $\tau$ before being disconnected. Lastly, in step~(iv) the memory needed for the feedback step is erased.  
Due to the feedback step, this four-step cycle deterministically starts and ends from the same state $\ket{+_\theta}$, and the energy provided by the measurement's  backaction is entirely transmitted to the field to amplify.
This protocol is inspired by former work~\cite{elouardExtractingWorkQuantum2017} although the step order has been modified for technical reasons discussed in the Supplementary Material (SM)~\ref{SM-sec:Protocol_step_order}. 

\textit{Running the engine blindly ($p_l=1$). }
When $\tau$ 
is small enough 
$\theta<\pi/2$, i.e., $p_j<1/2$. 
It means that, blindly applying the feedback associated with the outcome $+$, i.e., applying no feedback, it most often as good 
as adapting the feedback to the measurement outcome, at least for the first few cycles. The measurement outcomes are thus stored in a memory without need.  
This is a waste of resources, as processing information cost work, typically in the erasure~\cite{sagawaMinimalEnergyCost2009}.

One could therefore take the risk to lazily run the engine without looking at the measurement outcomes, always assuming that the post measurement state is $\ket{+}$.  
As illustrated in Fig~\ref{fig:working principle}.(b), this absence of feedback implies that, in the unlucky case in which the post measurement state is $\ket{-}$, the qubit takes energy from the field, instead of giving it. If this occurs early on in the repeating cycles, and since we assume $p_j<1/2$, it is likely that the engine will keep loosing energy for several cycles. Denoting $\langle .\rangle$ the average over many realizations,  
the average work increment extracted at cycle $t\geq 1$ reads 
$\langle \delta W_t \rangle= W_\theta(1-2p_j)^t $, 
where 
\begin{equation}
W_\theta=(\hbar\omega_0/2) \sin(\theta),
\end{equation}
is the quanta of work extracted starting from $\ket{+}$ or lost from $\ket{-}$, and $p_j$ is the probability to change state, i.e., jump, from $\ket{+_\theta}$ to $\ket{-}$ (or from $\ket{-_\theta}$ to $\ket{+}$).  As the number of cycles per trajectory, $t$, increases, $\langle \delta W_t \rangle$ tends to zero because the measurement outcomes become equally likely. The total average work extracted up to the end of cycle $t$ takes the form of a geometric sum,
\begin{equation}
    \langle W_t \rangle = \sum_{i=1}^t  \langle \delta W_i \rangle =  W_\theta \left[\frac{1-(1-2p_j)^{t+1}}{2p_j} -1 \right];
\end{equation}
it is a positive finite quantity that converges to $W_{\infty}= W_{\theta}/2p_j$ at large time $t$ due to the bias towards energy extraction from the initial state preparation (only the initial state of the first cycle). A LQME can thus run without information processing via the so-called open loop protocol, yet  feedback is still needed to ensure a non-zero long-time work extraction.

We now wonder how many cycles should we let the engine run blindly so that most trajectories reached, at least once, the work $W_\infty$?
The first  passage time (FPT) distribution, precisely 
characterizes the distribution of the first time each trajectory reached a prescribed threshold value $W_{T}$. Finding such distribution is  an arduous challenge, even for continuous Markovian dynamics~\cite{Mamede_2024}.
In our model, although the stochastic work increments are Markovian ($    \delta W_t = -\delta W_{t-1}$ with probability  $p_j$, and  $\delta W_t =    \delta W_{t-1}$ with probability  $1-p_j$), the work $W_t$ is a 2nd order Markov process: 
\begin{align}
    W_t 
    &= W_{t-1} + \begin{cases}\delta W_{t-1} \text{ with probability } 1-p_j \\ 
    -\delta W_{t-1} \text{ with probability } p_j \\ \end{cases}  \\
    & = W_{t-1} + \begin{cases}(W_{t-1} - W_{t-2})\text{ with probability } 1-p_j \\ 
    -(W_{t-1} - W_{t-2}) \text{ with probability } p_j. \end{cases} \nonumber
\end{align}
Notably, as shown below, taking $W_t$ together with the state obtained at the end of the cycle $s_t$ is analogous to a Run-and-Tumble (RnT) process---a paradigmatic model in active matter~\cite{bergColiMotion2004,tailleurStatisticalMechanicsInteracting2008,angelaniFirstpassageTimeRuntumble2014,shreshthaThermodynamicUncertaintyRuntumble2019,tucciFirstpassageTimeRuntumble2022}---in discrete space and time. Here, the work $W_t$ plays the role of the position, and $s_t$ that of the velocity, and $p_j$ is the tumble probability. Indeed, omitting the ket symbols for concision, when $s_t= +_\theta$, resp. $s_t=-_\theta$, the work is more likely to increase, resp. decrease and 
$(W_t, s_t)$ is Markovian. Since the work always varies in
quanta of $W_\theta$, i.e., $W_t= n_t W_\theta$, it is more convenient to consider $(n_t,s_t)$ which becomes, after one cycle: 
\begin{equation}
    \label{Eq:(n,s)} \hspace{-6pt}
    \begin{cases}
    (n_{t} + 1,+_\theta) 
    \text{ with probability} \begin{cases} 1-p_j &\hspace{-5pt}\text{if }s_{t} = +_\theta \\  p_j 
    &\hspace{-5pt}\text{if }s_{t}=-_\theta \end{cases} \\
    \hspace{-1pt} (n_{t} -1, -_\theta) 
    \text{ with probability} \begin{cases} p_j &\hspace{-5pt}\text{if }s_{t}=+_\theta \\  1-p_j 
    &\hspace{-5pt}\text{if }s_{t} = -_\theta.\end{cases} \end{cases}
\end{equation}

\begin{figure}[t]
\centering
\includegraphics[width=0.95\columnwidth]{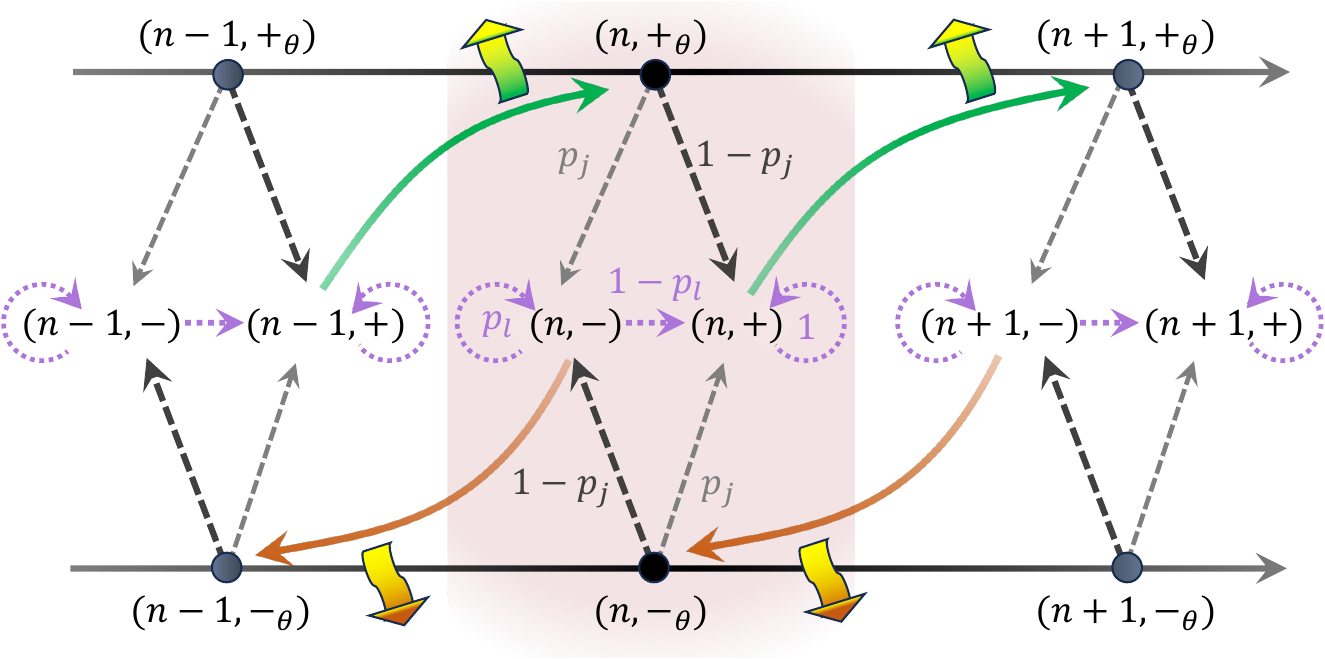}
\caption{Work extracted and qubit state dynamics mapped to a RnT process. A cycle always connects a state $\ket{\pm_\theta,n}$ to $\ket{\pm_\theta,n\pm 1}$. The measurement backaction (dashed grey) first leads to $\ket{\pm,n}$ states, the feedback step (dotted purple), occurring with probability $(1-p_l)$, adds a bias towards $\ket{+,n}$ from which work is extracted (solid green). Work can also be lost when the work extraction step is applied from $\ket{-,n}$. Outgoing yellow arrows pointing up, resp. down, symbolizes an increase, resp., decrease of the field's energy. 
The probabilities given in the central cell (light pink background) are the same for all cells.} 
\label{fig:RunAndTumble}
\end{figure}

\textit{Stochastic information processing $p_l\in ( 0,1 )$.} When the engine runs blindly, the evolution of the work extracted is not biased towards positive values and is a martingale. Therefore, the FPT distribution does not have a finite mean when $t\rightarrow\infty$.
We hence consider a more general situation in which there is a probability $1-p_l$ to use the measurement outcome information at each cycle.  When $p_l=0$, Fig.~\ref{fig:working principle}.(a), information is used deterministically at each cycle whereas for $p_l=1$, Fig.~\ref{fig:working principle}.(b), the measurement outcomes are lazily ignored. 

Stochastically using information in the feedback loop, simply amounts to changing the probabilities in Eq.~\eqref{Eq:(n,s)} in favour of the state $\ket{+_\theta}$ such that the probabilities to obtain $(n_t,s_t)$ now read:
\begin{align}
    &\hspace{-5pt} P_{t+1}(n, \hspace{-1pt}+_\theta) = (1-p_jp_l ) P_t(n-1,\hspace{-1pt}+_\theta) \nonumber\\
    &\hspace{55pt} +  (1-p_r p_l) P_t(n-1,\hspace{-1pt}-_\theta) , 
    \label{eq:Master+_feedback} 
    \\
    &\hspace{-5pt} P_{t+1}(n,\hspace{-1pt}-_\theta) =   p_r p_l P_t(n+1,\hspace{-1pt}-_\theta) 
      + p_j p_lP_t(n+1,\hspace{-1pt}+_\theta), 
    \label{eq:Master_feedback}
\end{align}
where $p_r=1-p_j$. 
Hence, the work still follows a run-and-tumble evolution, but with a different tumbling probability starting from $\ket{+_\theta}$ and ending the cycle in $\ket{-_\theta}$: $p_jp_l$ than tumbling from $\ket{-_\theta}$ to $\ket{+_\theta}$: $1-p_l+p_jp_l$, see Fig.~\ref{fig:RunAndTumble}.

We solve the Master equations~\eqref{eq:Master+_feedback} and~\eqref{eq:Master_feedback}   via $\mathcal{Z}$-transforms in space and time as follows.  First, we transform the discrete time $t\to u$ as $\widetilde{P}_{u}(n,\pm_\theta) \equiv \sum_{t=0}^{\infty} u^t P_{t} (n,\pm_\theta)$, and the discrete work increments $n\to z$ as $\widehat{\widetilde{P}}_{u}(z,\pm_\theta) \equiv \sum_{n=-\infty}^{\infty} z^n \widetilde{P}_{u} (n,\pm_\theta)$. Then, we solve Eq.~\eqref{eq:Master+_feedback} and \eqref{eq:Master_feedback} in the~$(z,u)$ space, which yields $\widehat{\widetilde{P}}_{u}(z,\pm_\theta)$, which  inverting back in $n$~\cite{dasDynamicsLatticeRandom2023a} gives the  generating functions of the propagators (see SM~\ref{SM-Sec:Results_RnTMapping}): 
\begin{align}
    \label{eq:P++tilde_u}
    \widetilde{P}_{u}(n,\hspace{-1pt}+_\theta|n_0,\hspace{-1pt}+_\theta) &= \frac{z_u^{n+1-n_0} (u (1-p_j)  p_l z_u -  z^+_u z^-_u)}{u(z^+_u z^-_u)^{n+1-n_0}(1 - p_j p_l)  (z_u^2 - z^+_u z^-_u) }, \\
    \label{eq:P+-tilde_u}
    \widetilde{P}_{u}(n,\hspace{-1pt}+_\theta|n_0,\hspace{-1pt}-_\theta) &= \frac{ -  (1\hspace{-1pt}-\hspace{-1pt}(1\hspace{-1pt}-\hspace{-1pt}p_j) p_l) (z^+_u z^-_u)^{n_0-n+1}}{(1\hspace{-1pt} - \hspace{-1pt}p_j p_l)  (z_u^2 \hspace{-1pt}- \hspace{-1pt}z^+_u z^-_u) z_u^{n_0-n} }. \hspace{-5pt}
\end{align}
These probabilities are given conditionally on starting with the state $\ket{+_\theta}$, resp. $\ket{-_\theta}$ and $n_0 W_\theta$ of work,  with 
\begin{align}
   z^{\pm}_u = &\frac{1+ (1 - 2p_j) p_l u^2 }{2 u (1 - p_j p_l)} \nonumber \\
   &\pm \sqrt{\left[\frac{1+ (1 - 2p_j) p_l u^2 }{2 u (1 - p_j p_l)}\right]^2- \frac{(1-p_j) p_l }{1-p_jp_l} },\\
   z_u =& z^+_u\Theta(|z^-_u|-|z^+_u|) +  z^-_u\Theta(|z^+_u|-|z^-_u|),
\end{align}
where $\Theta$ denotes the Heaviside function. 
\begin{figure}[htb]
\centering
\includegraphics[width=0.95\columnwidth]{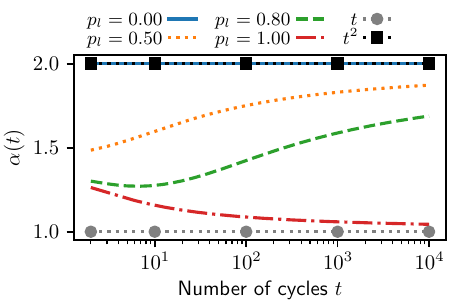}
\vspace{-3mm}
\caption{Anomalous diffusion exponent  $\alpha(t)=\ln\langle(n_t-n_0)^2\rangle/\ln(t)$ as a function of  the number of cycles, $t$, for $p_j=0.4$ and $n_0=0$ (analytical values). When $p_l<1$, the mean square displacement becomes ballistic at large times, whereas for $p_l=1$ (without  information processing), it converges to a diffusive one. At intermediate times, the work follows an anomalous, superdiffusive motion.}
\label{fig_MSD}
\end{figure}
From these solutions,  
we obtain the mean squared displacement $\expval{(n_t-n_0)^2}$ of the work extracted with respect to time. Strikingly, upon changing the laziness probability $p_l$, the work scaling can go from ballistic, $\expval{n_t} = t +n_0$ for perfect feedback $p_l=0$, to diffusive $\expval{(n_t-n_0)^2} \underset{t\rightarrow\infty}{\rightarrow} (t+1)(1-p_j)/p_j$ without feedback $p_l=1$, to intermediate superdiffusion~\ref{SM-Sec:averageMSD} for $p_l\in(0,1)$ where: 
\begin{align}
    \hspace{-3pt}\label{eq:avgn_feedback}
    \expval{n_t} \hspace{-1pt}
    &=  n_0 \hspace{-1pt}- \hspace{-1pt}1 \hspace{-1pt}+\hspace{-1pt}\frac{ (1-p_l) (t+1)}{ 1-p_l + 2p_j p_l } \hspace{-1pt}+ \hspace{-1pt}\frac{ 2p_j p_l [1\hspace{-1pt}-\hspace{-1pt} (1-2 p_j)^{t+1} p_l^{t+1}  ]}{(1\hspace{-1pt}-\hspace{-1pt}p_l + 2 p_jp_l)^2 }.
\end{align}
To quantify the diffusivity, we use the anomalous diffusion exponent $\alpha(t)$ such that $\expval{(n_t-n_0)^2} = t^{\alpha(t)}$ where $\alpha=2$ in the ballistic case, $\alpha=1$ in the diffusive case, and $1<\alpha<2$  for the superdiffusive cases~\cite{metzlerRandomWalksGuide2000,metzlerAnomalousDiffusionModels2014}, see Fig.~\ref{fig_MSD}. 
\begin{figure}[hb]
  \centering
  \includegraphics[width=0.95\linewidth]{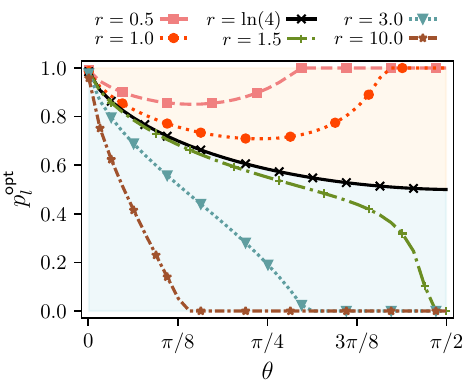}
  \vspace{-13pt}
  \caption{Optimal laziness probability $p_l$  maximizing the average steady-state power extracted $\langle\dot{W}\rangle$ as a function of  $\theta$, for different ratio values $r=\hbar\omega/k_{\rm B}T$. Symbols are analytical results obtained from Eq.~\eqref{eq:power} and lines a guide to the eye. 
  The shaded areas highlight the separatrix line $r_c=\ln(4)$ between   the classical-dominated (top orange region)  and quantum-dominated (bottom blue region)  parameter regions.
  }
  \label{fig:pl_opt}
\end{figure}

\textit{Steady-state power.} As shown in the SM,  the net energetic (Landauer) cost per cycle associated with  information processing  reads $\langle W_\text{info} \rangle = (1-p_l)k_B T S(p_j)$, where $S(p_j)=-p_j\ln(p_j)-(1-p_j)\ln(1-p_j)$ is the binary Shannon entropy with probability $p_j$ (in nats), 
and $T$ is the temperature of the bath  used for the memory erasure (see SM~\ref{SM-sec:Energetics}). The bath erases information by thermalizing the memory no matter its initial state, and work is then necessary to transform this thermal state into a known, zero entropy state, ``0'' typically. Notice that, when $p_l=1$, information is not needed and hence its associated erasure cost vanishes. In general, the net amount of work extracted per cycle, after $t$ cycles, reads $W_\theta(\langle  n_t\rangle-\langle n_{t-1}\rangle ) - \langle W_\text{info} \rangle $. Approximating the cycle duration to $\tau= \theta/\Omega$, where $\Omega$ is the coupling strength between the qubit and the field to amplify, and taking $t\rightarrow\infty$, we obtain the average steady-state power extracted (see SM~\ref{SM-sec:Energetics}): 
\begin{equation}
   \hspace{-8pt} \langle\dot{W}\rangle_{\rm ss} = \frac{\Omega (1-p_l)}{\theta}
    \left[\frac{(\hbar\omega/2)\sin\theta}{ 1-p_l\cos\theta}  - k_{\rm B}TS\left(\hspace{-2pt} \sin^2\frac{\theta}{2}\hspace{-2pt} \right)\right]\hspace{-3pt} .\label{eq:power}
\end{equation}
A tug-of-war between the first term, quantum work extracted, and the second term, classical information processing cost, quantified by the ratio $r=\hbar\omega/k_B T$, is at stake.
Figure~\ref{fig:pl_opt} shows that the optimal $p_l$ to maximize the power can be a non-monotonic function of the angle $\theta$ and transitions from convex to concave at a critical value $r_c=\ln(4)$.
In the classical limit, $r\ll 1$,
the erasure cost is higher than the energy gained from the measurement at each cycle, and therefore reducing information processing is beneficial. 
When $r\gg 1$ information plays a key role, except near the Zeno limit of very frequent measurements of deterministic outcomes, $\theta\rightarrow 0$. 

\begin{figure}
\centering 
\includegraphics[width=0.9\columnwidth]{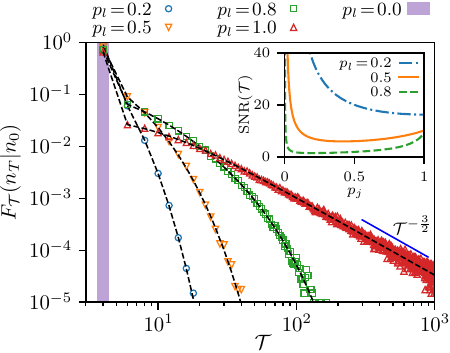}
\vspace{-10pt}
\caption{First-passage time probability ${F}_{\mathcal{T}}(n_T|n_0)$ for different values of laziness probability $p_l$ with $p_j=0.1$, $n_T=4$, and $n_0=0$. The points are obtained from $4\times10^5$ numerical simulations, while the dashed lines are the analytical solutions obtained by numerically inverting Eq.~\eqref{eq:FPT_generating_feedback}~\cite{abateNumericalInversionProbability1992}. For $p_l=0$, the purple solid bar denotes  $\delta_{{\mathcal{T}}, n_T-n_0}$. For $p_l=1$, ${F}_{\mathcal{T}}(n_T|n_0)$ decays as ${\mathcal{T}^{-3/2}}$ as illustrated by the blue solid line. Inset: signal-to-noise ratio associated with the first-passage time as a function of $p_j$ and $p_l$ obtained from Eq.~\eqref{eq:feedback-snr}, with  $n_T=10$ and $n_0=1$.}
\label{fig:FPT}
\end{figure}

\textit{First-passage time probability distribution.} 
To ensure the resetting of the qubit to its initial state $\ket{+_\theta}$ and a given work extracted, we introduce the stopping condition to reach the work threshold $n_T W_\theta$ starting from $n_0 W_\theta$. Instead of having stochastic final state and work extracted, it is the trajectory length, i.e., the first-passage time  $\mathcal{T} \equiv \text{inf}\{t\geq0|W_t=n_TW_0\}$ which become stochastic.   
We denote its generating function by 
$\widetilde{F}_u(n_T|n_0) = \sum_{t=0}^{\infty} u^t F_t(n_T|n_0) $, where $F_t(n_T|n_0)$ is the first-passage time density. Since we always start from the state $\ket{+_\theta}$, we have~\cite{balakrishnanRenewalEquationPersistent1988}
\begin{align}
    \widetilde{F}_u(n_T|n_0) &=\frac{\widetilde{P}_{u}(n',+_\theta| n_0,+_\theta)}{ \widetilde{P}_{u}(n',+_\theta| n_T,+_\theta) } , \quad n_0 < n_T < n',
\end{align}
which takes a form reminiscent of that of Markovian processes, although $n_t$ follows a second-order Markovian process. From Eq.~\eqref{eq:P++tilde_u}, when $n_0 < n_T < n'$, the exact analytical expression for the FPT probability generating function reads
\begin{align}
\label{eq:FPT_generating_feedback}
    \widetilde{F}_u(n_T|n_0) = \begin{cases}
    (z^+_u)^{n_0-n_T} &\text{ if } |z^+_u|>|z^-_u| \\
    (z^-_u)^{n_0-n_T} &\text{ if }|z^-_u|>|z^+_u|.
    \end{cases}
\end{align}
Note that the first-passage probability is independent of $n'$, as one expects~\cite{balakrishnanRenewalEquationPersistent1988}, and is also normalized, i.e., $\lim_{u\to 1} \widetilde{F}_u(n_T|n_0)
=1$, as we show in~\ref{SM-sec:FPS}.
In the purely ballistic case, where $p_l=0$ and $z_u= 1/u$, the generating function can be transformed back in time, giving $F_\mathcal{T}(n_T|n_0)= \delta_{\mathcal{T},n_T-n_0}$. As plotted on Fig.~\ref{fig:FPT}, in this case, all trajectories reach the threshold at the fastest stopping time: $n_T-n_0$. 

The mean first-passage time (MFPT) to reach $n_T$ starting from $n_0$, can be retrieved from the generating function as $\expval{\mathcal{T}} \equiv \lim_{u\to 1} \partial_u \widetilde{F}_u(n_T|n_0)$, and is given by 
\begin{align}
   \label{eq:MFPT_feedback}
   \expval{\mathcal{T}} = \frac{ (n_T-n_0) (1-p_l+2 p_jp_l) }{1-p_l} .
\end{align} 
Equation~\eqref{eq:MFPT_feedback} shows that the MFPT diverges when $p_l=1$, since some trajectories can take an infinite time to reach the stopping condition, i.e. never reach it.  
Following analogous steps, we derive the signal-to-noise ratio ${\rm{SNR}}(\mathcal{T})\equiv  \expval{\mathcal{T}}^2/(  \expval{\mathcal{T}^2} - \expval{\mathcal{T}}^2 )$ associated with the FPT, 
\begin{align}
\label{eq:feedback-snr}
    {\rm{SNR}}(\mathcal{T})= \frac{ (n_T-n_0) (1-p_l)   [1-p_l+2 p_j p_l]^2  }{ 4 p_j p_l (1-p_l+p_j p_l) [1+p_l-2 p_j p_l]   },
\end{align}
which varies non-monotonically as a function of $p_j$ for a given $p_l$.
As shown in the inset of Fig.~\ref{fig:FPT}, when the jump probability $p_j\rightarrow 0$ or when $p_l=0$, 
all cycles lead to work extraction and hence the SNR diverges. Similarly, when $p_j=1$, if $p_l=1$, all trajectories would also be the same, with switches from $\ket{+}$ to~$\ket{-}$ in every cycle.

\textit{Conclusion.} Quantum measurement engines can run on the sole quantum resource of measurement backaction at intermediate time scales. However, the total absence of feedback, leads to zero average power in the long run. Information is therefore necessary, but being lazy in its usage can be rewarding, for instance regarding the average steady-state power extracted, and even more so towards the classical limit, $\hbar\omega/k_B T\gg1$. However, when targeting a specific amount of total work extracted, more information processing leads to a lower spread of the first-passage time distribution and hence a higher predictability of the work statistic of the engine. 
Nonetheless, using all available information is not always beneficial for quantum measurement engines. 
However, the exact quantification of stochastic information processing consequences is a highly non-trivial problem which lead us to derive exact anomalous diffusion exponents and first-passage-time statistics for discrete run-and-tumble motion, for which most available results are in the continuum limit~\cite{angelaniFirstpassageTimeRuntumble2014,malakarSteadyStateRelaxation2018,moriUniversalPropertiesRuntumble2020,bruyneSurvivalProbabilityRuntumble2021,garcia-millanRuntumbleMotionHarmonic2021,tucciFirstpassageTimeRuntumble2022,gueneauRuntumbleParticleOnedimensional2024}---and similarly for work first-passage statistics~\cite{Mamede_2024}. We expect our work will inspire further fundamental connections between quantum mechanics and active matter---an exciting open avenue yet  still on its infancy~\cite{maes2022diffraction,krekels2024zig,kewmingFirstPassageTimes2024a}.

\textit{Acknowledgement} We thank Rosario Fazio for fruitful discussions. 
L. B. and \'E. R.  acknowledge funding from the PNRR MUR project PE0000023-NQSTI.

\let\oldaddcontentsline\addcontentsline
\renewcommand{\addcontentsline}[3]{}

\let\addcontentsline\oldaddcontentsline

\clearpage

\onecolumngrid

\widetext
\begin{center}
\textbf{\large ``Supplementary Material to \\``Run-and-tumble exact work statistics in a lazy quantum measurement engine: \\ stochastic information processing''}
\end{center}

\setcounter{equation}{0}
\setcounter{figure}{0}
\setcounter{table}{0}
\setcounter{page}{1}
\makeatletter
\renewcommand{\thesection}{S\arabic{section}}
\renewcommand{\theequation}{S\arabic{equation}}
\renewcommand{\thefigure}{S\arabic{figure}}
\renewcommand{\bibnumfmt}[1]{[S#1]}
\renewcommand{\citenumfont}[1]{S#1}

\setcounter{secnumdepth}{2}
\renewcommand{\thesection}{\Roman{section}} 
\
\tableofcontents

\pagebreak

\section*{Organization, Notations and $\mathcal{Z}$-transforms} 
This Supplementary Material (SM) is organized as follows: first we provide a table of notations and a set of useful $\mathcal{Z}$-transform relations, then, Sec.~\ref{SM-sec:Q_unit_evol}, we apply Schr\"odinger equation to the qubit in the work extraction step for readers unfamiliar with such quantum unitary evolutions.
The following sections contain the derivations of our main results. 
In Sec.~\ref{SM-Sec:Stocha_evo}, we derive the work statistics, in Sec.~\ref{SM-Sec:Results_RnTMapping} the mapping to the corresponding Run-and-Tumble (RnT) process and its solution. It covers all possible laziness values $p_l$, specifically dealing with the limits $p_l=0$ and $p_l=1$.
From this solution, the moments of the extracted work are derived in Sec.~\ref{SM-Sec:averageMSD} and the First-Passage (FP) statistics explicitly derived, Sec.~\ref{SM-sec:FPS}. Sec.~\ref{SM-sec:Energetics}, regards the derivation of our energetic section's results. 
The last two sections put our results in context, comparing them with preliminary derivations~\cite{elouardExtractingWorkQuantum2017_SM} in Sec.~\ref{SM-sec:Protocol_step_order} and generalizing them in Sec.~\ref{SM-sec:Generalization}. 
\\
\\

\label{sec:Notations}
As a help to the reader, we provide here a short list of notations and definition.
\begin{table}[!htbp]
\centering
\begin{tabular}{ |p{2cm}|p{6cm}||p{2cm}|p{6cm}|  }
 \hline
 \hline
 Symbol & Meaning or definition &Symbol & Meaning or definition\\
 \hline
 \hline
 $(n,\pm_\theta)$    &  Work $n W_\theta$ and qubit state $\ket{\pm_\theta}$  &$P_{t} (n,\pm_\theta)$ & Probability of the work to be $n W_\theta$ and qubit state $\ket{\pm_\theta}$ after the $t$-th cycle\\
  $\widetilde{P}_{u}(n,\pm_\theta)$    & $\sum_{t=0}^{\infty} u^t P_{t} (n,\pm_\theta)$    &$W_\theta$ & $\dfrac{\hbar \omega \sin\theta}{2}$\\
 $\widehat{\widetilde{P}}_{u}(z,\pm_\theta)$    & $\sum_{n=-\infty}^{\infty} z^n \widetilde{P}_{u} (n,\pm_\theta)$  &  $p_j$  & Probabilities of the qubit state to jump from $\ket{+_\theta}$ to $\ket{-}$ or from $\ket{-_\theta}$ to $\ket{+}$. We also define $p_r=1-p_j$. \\
 $n_0$ & Work of $W_0 = n_0 W_\theta$ at $t=0$ & 
 $p_l$ & Probability of being "lazy", i.e., to act as if the output state was $\ket{+}$, and hence apply no feedback, without actually looking and storing the measurement outcome. \\
 $n_T$ & (Cumulative) work of $n_T W_\theta$ as a stopping condition  & $s_t$ & Qubit in the state $\ket{s_t}\in \{\ket{+_\theta}, \ket{-_\theta}\}$ at the end of 
cycle $t$\\
 $z^{\pm}_u$ & $\frac{1+ u^2 p_l (p_r - p_j)  }{2 u (1- p_j p_l)} \pm \sqrt{\left[\frac{1+ u^2 p_l (p_r - p_j)  }{2 u (1- p_j p_l)}\right]^2- \chi }$  & $z_u$ & $\begin{cases}
    z^+_u &\text{ if } |z^-_u|>1>|z^+_u|  ,\\
    z^-_u &\text{ if } |z^+_u|>1>|z^-_u|,
    \end{cases}$  \\
    $\chi$ & $ z^+_u z^-_u  = \frac{(1-p_j) p_l }{1-p_j p_l} \leq 1$ & $F_t(n)$ & First passage time probability to reach the work $n_t W_\theta$ at time $t$ \\
 \hline
\end{tabular}
\end{table}
\\
\\

To compute the generating functions of the propagator for the work, 
we make explicit use of $\mathcal{Z}$-transforms.
In doing so, we recall readers that we transform the discrete work increment $n\to z$ of a $n$-dependent function $g(n)$ via the bilateral relation~\cite{gradshteyn2014table}
\begin{align}
\widehat{g}(z) \equiv {\mathcal{Z}}\{g\}(n) = \sum_{n=-\infty}^{\infty} z^n g(n),
\end{align}
where $z \in {\mathbb{C}}$ with $|z|<1$~\cite{abateNumericalInversionProbability1992_SM}. Similarly, we transform the discrete time $t\to u$, which we refer to as the $\mathcal{U}$-transform of a time-dependent function $f(t)$, via the unilateral relation~\cite{gradshteyn2014table}
\begin{align}
\widetilde{f}(u) 
 \equiv {\cal{U}} \{f\} (t) = \sum_{t=0}^{\infty} u^t f(t),
\end{align}
where $u \in {\mathbb{C}}$ with $|u|<1$~\cite{abateNumericalInversionProbability1992_SM}. The function $g(n)$ (or, $f(t)$) may be retrieved from $\widehat{g}(n)$ (or, $\widetilde{f}(u)$) using the inverse $\mathcal{Z}$-transform (or, inverse $\mathcal{U}$-transform) denoted by $g(n) = {\cal{Z}}^{-1} \{\widehat{g} \,\} (z)$ (or, $f(t) = {\cal{U}}^{-1} \{\widetilde{f} \,\} (u)$).
The inverse $\mathcal{Z}$-transform operation involves computing of a Cauchy contour integral as~\cite{gradshteyn2014table,abateNumericalInversionProbability1992_SM}
\begin{align}
    g(n) = {\cal{Z}}^{-1} \{\widehat{g} \,\} (z) = \frac{1}{2\pi \ii} \oint_{C} \dd{z} \frac{\widehat{g}(z)}{z^{n+1}} , 
    \label{eq:inv_Z_primo}
\end{align}
where $\ii=\sqrt{-1}$ is the imaginary unit, and $C$ is a circular closed contour about the origin $z=0$ with $|z|<1$. A similar relation as Eq.~\eqref{eq:inv_Z_primo} also holds for the inverse $\mathcal{U}$-transform. Here, we list a few examples of  $\mathcal{U}$-transform pairs that would appear handy for the calculations in the following sections:
\begin{alignat}{4}
& {\cal{U}} \{1\} = \frac{1}{1-u}, \quad 
&&  {\cal{U}} \{t+1\} = \frac{1}{(1-u)^2} , \quad 
&&  {\cal{U}} \qty{ \frac{1}{2} t (t+1)  } =  \frac{u}{(1-u)^3}  , \quad 
&&  {\cal{U}} \qty{ \frac{1}{2} t (t - 1)  } =  \frac{u^2}{(1-u)^3}  \nonumber \\[-1ex]
\label{eq:U_trans_rels} \\[-1ex]
& {\cal{U}} \qty{ a^t  } = \frac{1}{1-a u}  , \quad 
&&  {\cal{U}} \qty{ a^t (t+1) } =\frac{1}{(1-a u)^2}  , \quad 
&&  {\cal{U}} \qty{ \frac{1}{2} (t+1) (t+2) }  = \frac{1}{(1-u)^3}  . 
&& \nonumber 
\end{alignat}
The inverse ${\mathcal{U}}$-transform denoted by $\mathcal{U}^{-1}$ when applied to the right-hand side of the above relations, i.e., the $\widetilde{f}(u)$ functions, yields the corresponding time-dependent $f(t)$ functions given within the braces on the left-hand side. 
Note that the $\mathcal{Z}$-transform and the $\mathcal{U}$-transform are slightly different because of different summation limits of $n$ and $t$, respectively. 

\section{Work extraction evolution}
\label{SM-sec:Q_unit_evol}
We here provide all details about the qubit state's evolution during the work extraction step. We aim at guiding a reader unfamiliar with derivations in quantum mechanics, in understanding the unitary evolution from the state $\ket{+}$ to $\ket{+_\theta}$ during this important first step. 
The total Hamiltonian can be divided into a bare, constant, qubit Hamiltonian $H_0$ and a time-dependent term $H_I(t)$ coming from the interaction of the qubit with the classical electromagnetic field that we aim to amplify, such that: 
\begin{align}
H(t) = H_0 + H_I(t) = \hbar \omega_0 \sigma_z + i\frac{\hbar \Omega}{2}(\sigma_- e^{i\omega_0t}- \sigma_+ e^{-i\omega_0t})
\end{align}
with the lowering operator $\sigma_- = \ket{0}\bra{1}$ and raising one $\sigma_+ = \ket{1}\bra{0}$ are such that $\sigma_+\ket{0}=\ket{1}$ and  $\sigma_-\ket{1}=\ket{0}$. The Pauli matrice $\sigma_z = \ket{1}\bra{1}- \ket{0}\bra{0}$ sets the excited state to be $\ket{1}$ and to have $\hbar\omega_0$ of bare energy, whereas the ground state $\ket{0}$ has $-\hbar\omega_0$. When the initial state is $\ket{+} = \frac{\ket{1}+\ket{0}}{\sqrt{2}}$, evolving under $H$ for the time duration $\tau$, according to Schr\"odinger equation $i\hbar \frac{d\ket{\psi}}{dt}(t) = H \ket{\psi}(t)$, will lead to: 
\begin{equation}
\ket{\psi}(t) = e^{-\frac{i}{\hbar}\int_0^{\tau}H(t) dt} \ket{+}. 
\end{equation}
In the rotating frame with $H_0= \hbar \omega_0 \sigma_z$, the state $\ket{\psi}(t)$ becomes: 
\begin{equation}
    \label{eq:phi}
    \ket{\phi}(t) = U_0^\dagger(t) \ket{\psi}(t) \quad \text{with} \quad U_0(t)=e^{-i\omega_0\sigma_z}.
\end{equation}
Given that $\frac{d}{dt}\ket{\phi}(t) = \frac{d}{dt}\left(U_0^\dagger(t)\right) \ket{\psi}(t) + U_0^\dagger(t) \frac{d}{dt}\left(\ket{\psi}(t)\right)$, and that $\frac{d}{dt}\left(U_0^\dagger(t)\right) = \frac{i}{\hbar}U_0^\dagger(t) H_0$, 
\begin{equation}
\ket{\phi}(t) = e^{-\frac{i}{\hbar}\int_0^{\tau}U_0^\dagger(t) H_I(t) U_0(t) dt} \ket{+},  
\end{equation}
using the fact that  $[U_0^\dagger(t),H_0]=0$.
The interaction part of the Hamiltonian, becomes, in the rotating frame: 
\begin{align}
    \label{eq:rotU0}
    U_0^\dagger(t) H_I(t) U_0(t) &= i\frac{\hbar \Omega}{2} e^{i\omega_0\sigma_z t} (\sigma_- e^{i\omega_0t}- \sigma_+ e^{-i\omega_0t}) e^{-i\omega_0\sigma_z t}\nonumber \\
    &= i\frac{\hbar \Omega}{2} \sum_{k=0} \frac{(i\omega_0\sigma_z t)^k}{k!} \left(\ket{0}\bra{1} e^{i\omega_0t}- \ket{1}\bra{0}e^{-i\omega_0t}\right) \sum_{k=0} \frac{(-i\omega_0\sigma_z t)^k}{k!} \nonumber \\
    &= i\frac{\hbar \Omega}{2} \left(\sum_{k=0} \frac{(i\omega_0 t)^k}{k!} \ket{1}\bra{1}^k  \right)\left(\ket{0}\bra{1} e^{i\omega_0t}- \ket{1}\bra{0}e^{-i\omega_0t}\right) \left( \sum_{k=0} \frac{(-i\omega_0 t)^k}{k!} \ket{1}\bra{1}^k \right) \nonumber \\
     &= i\frac{\hbar \Omega}{2}  \left(\ket{0}\bra{1} e^{i\omega_0t}- \sum_{k=0} \frac{(i\omega_0 t)^k}{k!} 
    \ket{1}\bra{0}e^{-i\omega_0t}\right) \left( \sum_{k=0} \frac{(-i\omega_0 \tau)^k}{k!} \ket{1}\bra{1}^k \right) \nonumber \\
    &= i\frac{\hbar \Omega}{2}  \left(\ket{0}\bra{1} e^{i\omega_0t} \sum_{k=0} \frac{(-i\omega_0 \tau)^k}{k!} -e^{i\omega_0t}
    \ket{1}\bra{0}e^{-i\omega_0t}\right)  
    = i\frac{\hbar \Omega}{2}(\sigma_- - \sigma_+).
\end{align}
Using Eqs.~\eqref{eq:rotU0} and \eqref{eq:phi}, we get
\begin{align}
    \ket{+_\theta} = \ket{\phi}(\tau) &= e^{\frac{\Omega\tau}{2}(\sigma_- - \sigma_+)} \ket{+} 
    = \sum_{k=0} \frac{1}{k!} \left(\frac{\Omega\tau}{2}\right)^k (\sigma_- - \sigma_+)^k \ket{+} \nonumber \\
    &= \sum_{k=2p} \frac{1}{k!} \left(\frac{\Omega\tau}{2}\right)^k (-1)^p (\ket{0}\bra{0} + \ket{1}\bra{1}) \ket{+} + \sum_{k=2p+1} \frac{1}{k!} \left(\frac{\Omega\tau}{2}\right)^k (-1)^{p-1} (\ket{0}\bra{1} - \ket{1}\bra{0}) \ket{+} \nonumber \\
     &= \sum_{k=2p} \frac{(-1)^p}{(2p)!} \left(\frac{\Omega\tau}{2}\right)^{2p}   \ket{+} + \sum_{k=2p+1} \frac{(-1)^{p-1}}{k!} \left(\frac{\Omega\tau}{2}\right)^k  \frac{\ket{0} - \ket{1}}{\sqrt{2}} \nonumber \\
     &= \cos(\tau \Omega/2) \ket{+} - \sin(\tau \Omega/2) \ket{-} 
     = \cos(\theta/2) \ket{+} - \sin(\theta/2) \ket{-}
\end{align}
where $\ket{-}= \frac{\ket{1} - \ket{0}}{\sqrt{2}}$ and $\theta = \Omega\tau$.

Note that it is possible to extract work from $\ket{+}$, given that this state carries maximal coherences in the energy eigenbasis, by using a thermal bath to reversibly remove them~\cite{kammerlander2016coherence}. However, this implies connecting the qubit working substance to a thermal bath, which we want to avoid, and such total coherence removal would result in the mixed state $\frac{\ket{+}\bra{+}+\ket{-}\bra{-}}{2}$ which would then need to be purified again. 


\section{Stochastic evolution}
\label{SM-Sec:Stocha_evo}
For didactic purposes, we first study the case $p_l=0$, Sec.~\ref{subsec:noInfo}, where the measurement outcome information is deterministically used for feedback at each cycle. We reveal the non-Markovian evolution of the work, which also hold in the general case, $p_l\in[0,1]$, that we generalize to in Sec.~\ref{Subsec:info}. 

\subsection{Without information processing}
\label{subsec:noInfo}
The probability of the qubit state to remain in the same state between two cycles reads $p_r= \cos^2(\theta/2)$ whilst the one to jump is $p_j= \sin^2(\theta/2)$. 
The work increment, extracted during the $t$-th cycle, $\delta W_t$, is: 
\begin{align}
	\delta W_1  &=                 \begin{cases}
		W_\theta  & \text{with probability} \quad p_r \\
		-W_\theta    & \text{with probability} \quad p_j
	\end{cases}\nonumber \\
	\delta W_2 &= 	\begin{cases}
		\delta W_1  & \text{with probability} \quad p_r \\
		-\delta W_1    & \text{with probability} \quad p_j
	\end{cases} \nonumber \\
	\delta W_3 &= 	\begin{cases}
	\delta W_2  & \text{with probability} \quad p_r \\
	- \delta  W_2    & \text{with probability} \quad p_j 
\end{cases} \nonumber \\
    \delta W_t &= 	\begin{cases}
		\delta W_{t-1}  & \text{with probability} \quad p_r \\
		-\delta W_{t-1}    & \text{with probability} \quad p_j
	\end{cases}
\end{align}
where the work increment unit reads $W_\theta=( \hbar \omega \sin\theta )/2$. 
The work extracted up to the end of cycle $t\geq 1$, $W_t$, is:
\begin{equation}
	W_t  =  \sum_{k=1}^{t}  \delta W_k.
 \label{eq:stoch_work_pl1}
\end{equation}
\\
Note that while the stochastic work increments, $\delta W_t$, follow a \textbf{Markovian process}: 
\begin{equation}
	P(\delta W_t|\delta W_0,...\delta W_{t-1})= P(\delta W_t|\delta W_{t-1}),
\end{equation}
the (cumulative) work is such that: 
\begin{align}
	P(W_t |W_0 ,...W_{t-1} )&= P(W_t |W_{t-1} ,\delta W_t),\nonumber \\
	&= P(W_t |W_{t-1} ,W_{t-2} )
\end{align} 
where, to obtain the last equality, we used the fact that $\delta W_t$ only depends on $ \delta W_{t-1}$ and that
\begin{equation}
	\delta W_{t-1} = W_{t-1} -W_{t-2}.
\end{equation}
Therefore, the work follows a \textbf{second-order Markovian process}.
\\
Interestingly, the work extracted after $t+1$ cycles  can also be written as: 
\begin{align}
	W_{t+1}  &=    W_{t}  + \delta {W}_{t+1} \quad \text{where} \nonumber \\
	\delta {W}_{t+1}  &= 
	\begin{cases}
		\delta W_{t}  & \text{with probability} \quad 1-p_j \\
		-\delta W_{t}    & \text{with probability} \quad p_j.
	\end{cases}
\end{align}

\begin{figure}[!htbp] 
\subfloat[$p_l=1$ ; $p_j=0.2$\label{trajs_Wext_pl1}]{\includegraphics[width=0.48\columnwidth]{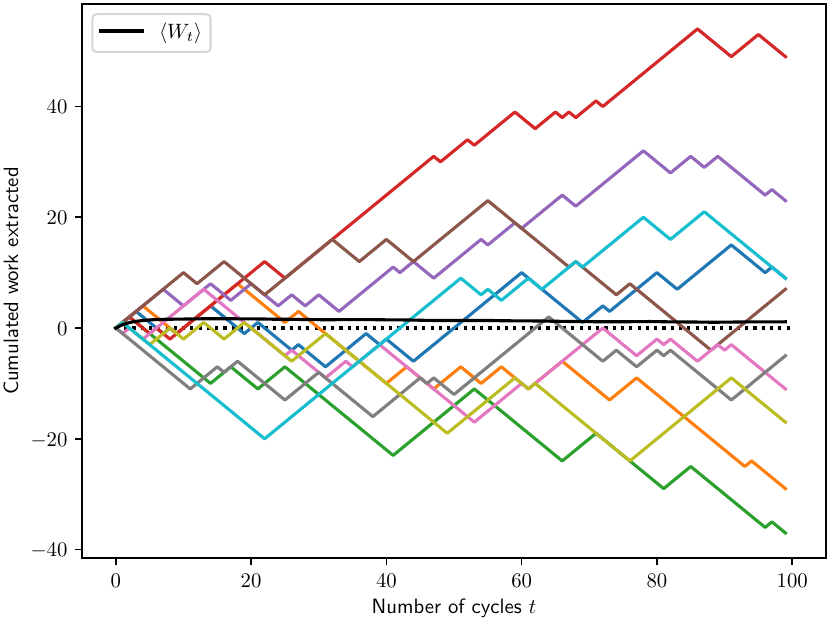}
}
\hspace*{\fill}
\subfloat[$p_l=0.7$ ; $p_j=0.2$\label{trajs_Wext_pl09}]{\includegraphics[width=0.48\columnwidth]{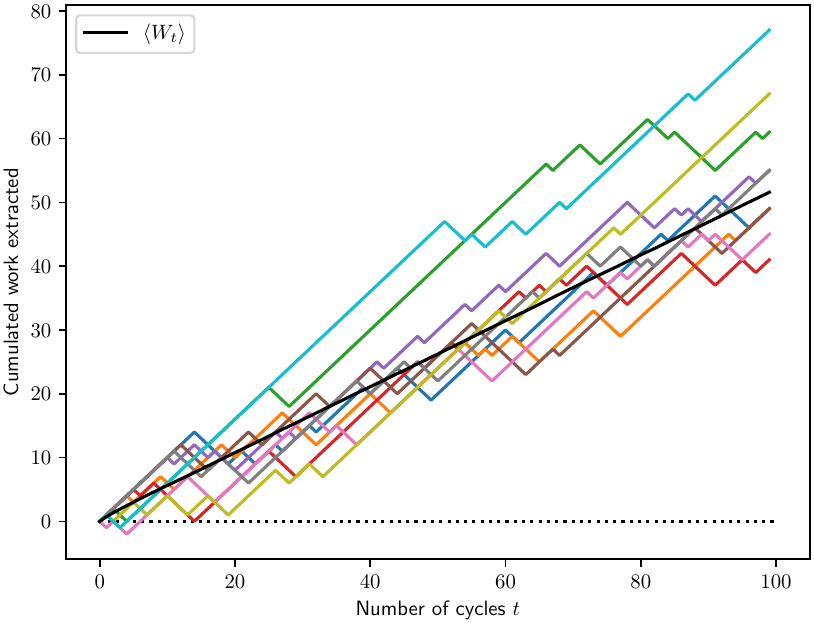}
}
		\caption{Example evolution of the work extracted, $W_t$ with respect to the number of cycles $t$, (a) without information processing, $p_l=1$ or (b) with little information processing, $p_l=0.7$. The average $\langle W_t\rangle$, solid black, is computed over 5000 simulations, i.e., trajectories.}
	\label{Example_trajectories}
\end{figure}

\subsection{With information processing}
\label{Subsec:info}
Adding a probability $(1-p_l)$, to flip the qubit state to $\ket{+}$ if it was measured in $\ket{-}$, during the feedback step, just after the measurement, the stochastic work increment reads: 
\begin{align}
	\delta W_1 &= 	\begin{cases}
		W_\theta  & \text{with probability} \quad p_r + p_j (1-p_l) \\
		-W_\theta    & \text{with probability} \quad p_jp_l
	\end{cases} \nonumber \\
	\delta W_2 &= 	\begin{cases}
	\delta W_1  & \text{with probability} \quad p_r + p_j (1-p_l) \quad \text{if} \quad \delta W_1>0 \\
	-\delta W_1    & \text{with probability} \quad p_jp_l \quad \text{if} \quad \delta W_1>0 \\
        \delta W_1  & \text{with probability} \quad p_rp_l \quad \text{if} \quad \delta W_1<0 \\
        -\delta W_1    & \text{with probability} \quad p_j+ p_rp_l \quad \text{if} \quad \delta W_1<0 \\
\end{cases} \nonumber \\
        &=\begin{cases}
	W_\theta  & \text{with probability} \quad (1- p_j p_l)^2  + (p_j+ p_rp_l)p_jp_l\\
	-W_\theta    & \text{with probability} \quad p_jp_l(1- p_j p_l) + p_r^2p_l^2 \\
\end{cases} \nonumber \\
	\delta W_t &= 	\begin{cases}
		\delta W_{t-1}  & \text{with probability} \quad p_r + p_j (1-p_l) \quad \text{if} \quad \delta W_{t-1}>0 \\
	-\delta W_{t-1}    & \text{with probability} \quad p_jp_l \quad \text{if} \quad \delta W_{t-1}>0 \\
        \delta W_{t-1}  & \text{with probability} \quad p_rp_l \quad \text{if} \quad \delta W_{t-1}<0 \\
        -\delta W_{t-1}    & \text{with probability} \quad p_j+ p_rp_l \quad \text{if} \quad \delta W_{t-1}<0. \\
	\end{cases}
\end{align}
Here, the work increment at step $t$ can still only take the values $\delta W_{t-1}$ or $-\delta W_{t-1}$. However, while these two values were respectively associated with the fixed probabilities $p_r$ and $1-p_r$ in the no feedback case, here, the corresponding probabilities depend on the sign of $\delta W_{t-1}$. This introduces a bias in the stochastic work increment towards positive values when $1> p_l \geq0$. 

\section{Solution of the run-and-tumble propagators}
\label{SM-Sec:Results_RnTMapping}
\begin{figure}[!htbp]
\centering
\includegraphics[width=15cm]{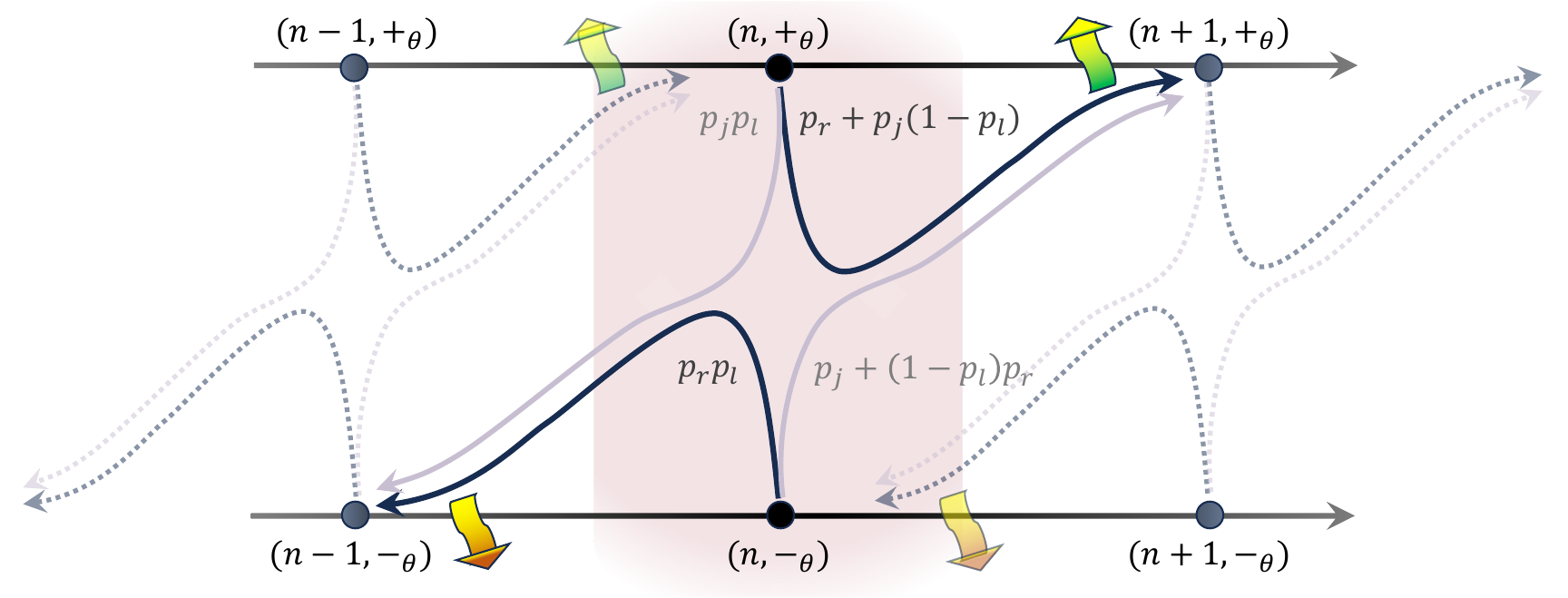}
\caption{Run-and-tumble dynamics of the stochastic work extracted and qubit state at the end of the cycles. The black arrows are the most likely evolutions given $p_j<0.5$. The more information processing is used, i.e., the smaller $p_l\in[0,1]$ is, the more likely it is to extract work and therefore to end the cycle in the state $\ket{+_\theta}$. The lighter arrows correspond to the tumbling events, which occur with probability $p_jp_l$ from $\ket{+_\theta}$ to $\ket{-_\theta}$ and with probability $p_j + (1-p_l)p_r = 1-p_lp_r = 1 -p_l(1-p_j)$ from $\ket{-_\theta}$ to $\ket{+_\theta}$.}
\label{fig_RnT_SM}
\end{figure}

The evolution of the work $W _t$ may be mapped to a discrete-time run and tumble (RnT) process on a one dimensional lattice. 
Discrete-time random walks on lattices of arbitrary dimensions have recently attracted attention, and a series of exact analytical results on propagators and first-passage statistics are obtained~\cite{giuggioli2020exact,sarvaharman2020closed,das2022discrete,dasDynamicsLatticeRandom2023a_SM,sarvaharman2023particle,marris2024persistence}.
These works, barring Ref.~\cite{marris2024persistence}, mainly focused on Markovian random walks in unbounded or confined lattices~\cite{giuggioli2020exact} in the presence of bias~\cite{sarvaharman2020closed}, stochastic resetting~\cite{das2022discrete}, or spatial heterogeneities~\cite{dasDynamicsLatticeRandom2023a_SM,sarvaharman2023particle}. 
In Ref.~\cite{marris2024persistence}, the authors have studied the correlated or persistent random walk as a second-order Markovian process. Here we derive the exact mapping of the evolution of the work extracted $W _t$ to an analogous RnT process, closely related to this last example.
With a slight abuse of notation, we denote the probability of finding the qubit with state $\ket{+_\theta}$ (respectively, $\ket{-_\theta}$) on site $n_t$, i.e., with $W _t = n_t W_\theta$ of work extracted, at the end of cycle $t$ by $P_t(n,+_\theta)$ (respectively, $P_t(n,-_\theta)$). 
If, at every step, the probability to store the measurement outcome and act upon this information with probability $(1-p_l)$, the work extracted and final qubit state dynamic is given by the set of coupled discrete-time Master equations:
\begin{align}
    P_{t+1}(n,+_\theta) &= (1-p_jp_l) P_t(n-1,+_\theta) + (1-p_rp_l) P_t(n-1,-_\theta) , \label{eq:Master+_feedback_SM}\\
    P_{t+1}(n,-_\theta) &= p_r p_l P_t(n+1,-_\theta) + p_j p_lP_t(n+1,+_\theta) , \label{eq:Master-feedback_SM}
\end{align}
where we have $p_r+p_j=1$. Without using information, i.e., $p_l=1$, the above equations reduce to 
\begin{align}
    P_{t+1}(n,+_\theta) &= p_r P_t(n-1,+_\theta) + p_j P_t(n-1,-_\theta), \label{eq:Master+}\\
    P_{t+1}(n,-_\theta) &= p_r P_t(n+1,-_\theta) + p_j P_t(n+1,+_\theta), \label{eq:Master-}
\end{align}
On the other hand, for $p_l = 0$ the Master equations~\eqref{eq:Master+_feedback_SM} and~\eqref{eq:Master-feedback_SM} suggest that the work extracted, irrespective of its initial state, follows a ballistic motion in the positive direction, with the qubit state always ending up in $\ket{+_\theta}$ at the end of every cycle. In general, $p_l\in[0,1]$, the evolution of the state $\ket{\pm_\theta, n_t}$ is illustrated in Fig.~\ref{fig_RnT_SM} where same coloured arrows have the same probability when $p_l=1$ and favour an increase in the work when $p_l\in[0,1)$. 

Let us now proceed to find the propagator generating function. We detail the derivation steps in the simpler $p_l=1$ case. The results when feedback can be applied are derived in exactly the same way and are given both with the qubit starting from the state $\ket{+_\theta}$ and $\ket{-_\theta}$, although we only need to consider the $\ket{+_\theta}$ in the main text, for more generality.\\

\textbf{No information, $p_l=1$} \\
Defining the generating functions with respect to discrete time $t$ as $\widetilde{P}_{u}(n,\pm_\theta) \equiv \sum_{t=0}^{\infty} u^t P_{t} (n,\pm_\theta)$, the Master equations~\eqref{eq:Master+}--\eqref{eq:Master-} may be written in the $\cal{U}$-domain depending on the initial conditions, $P_0(n,+\theta) = \delta_{n,n_0}$ and $P_0(n,-\theta) = 0$, as 
\begin{align}
\widetilde{P}_u(n,+_\theta|n_0,+_\theta) &= \delta_{n,n_0} + u  p_r  \widetilde{P}_u(n-1,+_\theta|n_0,+_\theta) + u p_j \widetilde{P}_u(n-1,-_\theta|n_0,+_\theta)  , \label{eq:Master+inz}\\
    \widetilde{P}_u(n,-_\theta|n_0,+_\theta) &= u  p_r  \widetilde{P}_u(n+1,-_\theta|n_0,+_\theta) + u p_j \widetilde{P}_u(n+1,+_\theta|n_0,+_\theta) . \label{eq:Master-inz}
\end{align}
In writing Eqs.~\eqref{eq:Master+inz} and~\eqref{eq:Master-inz}, we have explicitly mentioned the initial condition in the arguments of the probability quantities such that $\widetilde{P}_u(n,+_\theta|n_0,+_\theta)$ denotes the generating function with respect to time of the propagator to obtain $nW_\theta$ of work and the qubit in state $\ket{+_\theta}$ having started from a work $n_0W_\theta$ and state $\ket{+_\theta}$.
We now perform a similar $\cal{Z}$-transformation on the work quanta $n$ given by $\widehat{\widetilde{P}}_u(z,\pm_\theta|n_0,+_\theta) = \sum_{n=-\infty}^{\infty} z^n{\widetilde{P}}_u(n,\pm_\theta|n_0,+_\theta)$ so that Eqs.~\eqref{eq:Master+inz} and \eqref{eq:Master-inz} become:
\begin{align}
    \widehat{\widetilde{P}}_u(z,+_\theta |n_0,+_\theta) &= z^{n_0} + z u p_r  \widehat{\widetilde{P}}_u(z,+_\theta |n_0,+_\theta) + z u p_j  \widehat{\widetilde{P}}_u(z,-_\theta |n_0,+_\theta), \label{eq:Master+inzu}\\
    \widehat{\widetilde{P}}_u(z,-_\theta|n_0,+_\theta) &= \frac{u p_r }{z} \widehat{\widetilde{P}}_u(z,-_\theta|n_0,+_\theta) + \frac{u p_j }{z} \widehat{\widetilde{P}}_u(z,+_\theta|n_0,+_\theta) . \label{eq:Master-inzu}  
\end{align}
Equations~\eqref{eq:Master+inzu} and~\eqref{eq:Master-inzu} readily yield the solutions
\begin{align}
\widehat{\widetilde{P}}_u(z,+_\theta| n_0,+_\theta) &= \frac{z^{n_0} ( u p_r  - z )}{ u p_r  z^2 -z \left[1+ u^2(p_r-p_j)\right] + u p_r  }, \label{eq:Master+inzu_sol}\\
    \widehat{\widetilde{P}}_u(z,-_\theta|n_0,+_\theta) &= \frac{ - u p_j  z^{n_0} }{ u p_r  z^2 -z \left[1+ u^2(p_r-p_j)\right] + u p_r  } , \label{eq:Master-inzu_sol} 
\end{align}
where we have used $p_r+p_j=1$.
The probability of having a work of $nW_\theta$ at time $t$ irrespective of its state at the end of the cycle (either $\ket{+_\theta}$ or $\ket{-_\theta}$) is given by $P_t(n|n_0,+_\theta) = P_t(n,+_\theta|n_0,+_\theta) + P_t(n,-_\theta|n_0,+_\theta)$. Consequently, we have $\widetilde{P}_u(n|n_0,+_\theta) = \widetilde{P}_u(n,+_\theta|n_0,+_\theta) + \widetilde{P}_u(n,-_\theta|n_0,+_\theta) $ and
\begin{align}
\widehat{\widetilde{P}}_u(z|n_0,+_\theta) = \widehat{\widetilde{P}}_u(z,+_\theta|n_0,+_\theta) + \widehat{\widetilde{P}}_u(z,-_\theta|n_0,+_\theta) = \frac{  u( p_r  - p_j ) z^{n_0}  - z^{n_0+1}}{ u p_r  z^2 -z \left[1+ u^2 (p_r-p_j)\right] + u p_r }, \label{eq:ptildehat}
\end{align}
\\
\textbf{General case: $p_l\in[0,1]$}
\\
\textbf{Starting from state $\ket{+_\theta}$:}
When the work initially is $n_0 W_\theta$ ($n_0=0$ in the main text) and the qubit starts in $\ket{+_\theta}$, the initial conditions are given by $P_0(n,+_\theta) = \delta_{n,n_0}$ and $P_0(n,-_\theta) = 0$. 
One may then rewrite the Master equations~\eqref{eq:Master+_feedback_SM} and~\eqref{eq:Master-feedback_SM} in the $\mathcal{Z}, \mathcal{U}$ domain as
\begin{align}
    \widehat{\widetilde{P}}_{u}(z,+_\theta|n_0,+_\theta) &= z^{n_0} + z u (1 - p_j p_l)  \widehat{\widetilde{P}}_{u}(z,+_\theta|n_0,+_\theta) + z u (1 - p_r p_l )  \widehat{\widetilde{P}}_{u}(z,-_\theta|n_0,+_\theta), \label{eq:Master+inzu+feedback}\\
    \widehat{\widetilde{P}}_{u}(z,-_\theta|n_0,+_\theta) &= \frac{ u p_r p_l }{z} \widehat{\widetilde{P}}_{u}(z,-_\theta|n_0,+_\theta) + \frac{ u p_j p_l }{z} \widehat{\widetilde{P}}_{u}(z,+_\theta|n_0,+_\theta) . \label{eq:Master-inzu+feedback}  
\end{align}
Solving Eqs.~(\ref{eq:Master+inzu+feedback},~\ref{eq:Master-inzu+feedback}) yields the generating functions 
\begin{align}
    \widehat{\widetilde{P}}_{u}(z,+_\theta|n_0,+_\theta) &= \frac{ p_r p_l u z^{n_0} - z^{n_0+1} }{ u (1-p_j p_l)  z^2 -z \left[1+ (p_r-p_j) p_l u^2  \right] +p_r p_l u}, \label{eq:Master++inzu_sol+feedback}\\
    \widehat{\widetilde{P}}_{u}(z,-_\theta|n_0,+_\theta) &= \frac{ - p_j p_l u z^{n_0} }{ u (1-p_j p_l) z^2 -z \left[1+ (p_r-p_j) p_l u^2  \right] +p_r p_l u } . \label{eq:Master-+inzu_sol+feedback}   
\end{align}
Hence, the generating function of the propagator, with feedback, becomes
\begin{align}
    \widehat{\widetilde{P}}_{u}(z|n_0,+_\theta) &= \widehat{\widetilde{P}}_{u}(z,+_\theta|n_0,+_\theta) + \widehat{\widetilde{P}}_{u}(z,-_\theta|n_0,+_\theta) = \frac{   (p_r-p_j) p_l u z^{n_0} - z^{n_0+1} }{ u (1-p_j p_l) z^2 -z \left[1+ (p_r-p_j) p_l u^2  \right] + p_r p_l u }.
    \label{eq:feedback_gen_fun_total+}
\end{align}

\textbf{Starting from state $\ket{-_\theta}$:} On the other hand, when initial work is $n_0W_\theta$ but the qubit starts in $\ket{-_\theta}$, the initial conditions are modified to $P_0(n,+_\theta) = 0$ and $P_0(n,-_\theta) = \delta_{n,n_0}$. Here, the Master equations~\eqref{eq:Master+_feedback_SM} and~\eqref{eq:Master-feedback_SM} in the $\mathcal{Z}, \mathcal{U}$ domain reduce to
\begin{align}
    \widehat{\widetilde{P}}_{u}(z,+_\theta|n_0,-_\theta) &= z u  (1 - p_j p_l) z u \widehat{\widetilde{P}}_{u}(z,+_\theta|n_0,-_\theta) + z u (1 - p_r p_l) \widehat{\widetilde{P}}_{u}(z,-_\theta|n_0,-_\theta), \label{eq:Master-+inzu+feedback}\\
    \widehat{\widetilde{P}}_{u}(z,-_\theta|n_0,-_\theta) &= z^{n_0}+ \frac{u p_r p_l }{z} \widehat{\widetilde{P}}_{u}(z,-_\theta|n_0,-_\theta)+ \frac{u p_j p_l }{z} \widehat{\widetilde{P}}_{u}(z,+_\theta|n_0,-_\theta) , \label{eq:Master--inzu+feedback}  
\end{align}
which yields the generating functions
\begin{align}
    \widehat{\widetilde{P}}_{u}(z,+_\theta|n_0,-_\theta) &= \frac{- (1 - p_r p_l ) u z^{n_0+2}}{ u (1-p_j p_l) z^2 -z \left[1+ (p_r-p_j) p_l u^2  \right] + p_r p_l u }, \label{eq:Master+-inzu_sol+feedback}\\
    \widehat{\widetilde{P}}_{u}(z,-_\theta|n_0,-_\theta) &= \frac{ (1 - p_j p_l) u z^{n_0+2} - z^{n_0+1}}{ u (1-p_j p_l) z^2 -z \left[1+ (p_r-p_j) p_l u^2  \right] +p_r p_l u } . \label{eq:Master--inzu_sol+feedback}  
\end{align}
In this case, the generating function of the propagator, in the most general case, reads
\begin{align}
    \widehat{\widetilde{P}}_u(z|n_0,-_\theta) &= \widehat{\widetilde{P}}_{u}(z,+_\theta|n_0,-_\theta) + \widehat{\widetilde{P}}_{u}(z,-_\theta|n_0,-_\theta) = \frac{   (p_r-p_j) p_l u z^{n_0+2}  - z^{n_0+1}  }{ u (1-p_j p_l) z^2 -z \left[1+ (p_r-p_j) p_l u^2  \right] +p_r p_l u } .
    \label{eq:feedback_gen_fun_total-}
\end{align}

\textbf{Solution in the position basis:} \\
In general, to find the propagator generating functions in the position basis $n$, one has to perform an inverse $\cal{Z}$-transform on the quantities 
$\widehat{\widetilde{P}}_u(z,\pm_\theta|n_0,+_\theta)$ by  computing the complex integral
\begin{align}
    \widetilde{P}_u(n,\pm_\theta | n_0, +_\theta) = \frac{1}{2 \pi \ii} \oint_C \dd{z} \frac{ \widehat{\widetilde{P}}_u(z,\pm_\theta | n_0,+_\theta) }{z^{n+1}} , \label{eq:uinversion}
\end{align}
where $|z|<1$~\cite{gradshteyn2014table,abateNumericalInversionProbability1992_SM}. Here, the complex integral may be computed using the Cauchy residue theorem. To this end, one first takes the closed contour $C$ as the counter-clockwise unit circular path centered at $z=0$ on the complex $z$-plane, and then adding the residues of the integrand at poles lying inside the contour.
In regard to computing such integrals, we refer the readers to Appendix B of Ref.~\cite{dasDynamicsLatticeRandom2023a_SM} and also remind that a few similar transform pairs are given in the beginning of this supplementary material. 
\\
\textbf{Case: $(p_j,p_l)\in (0,1]^2 \setminus \{(1,1)\}$ }\\
In the most general case, i.e., for all values of $p_j$ and $p_l$ except the ones for which $p_jp_l=1$ or $p_rp_l=0$, one obtains, from Eq.~\eqref{eq:Master++inzu_sol+feedback}, that 
\begin{align}
    \widetilde{P}_{u}(n,+_\theta|n_0,+_\theta)   &= \frac{1}{ u (1 - p_j p_l) } \frac{1}{2 \pi \ii} \oint_C \dd{z} \dfrac{   u p_r  p_l  z^{n_0-n-1 } - z^{n_0 - n} }{  z^2 -z \left[ \dfrac{1+ (p_r-p_j) p_l u^2  }{(1 - p_j p_l ) u}\right] + \dfrac{p_r p_l }{1 - p_j p_l }}  \nonumber \\
    &=\frac{1}{ u (1 - p_j p_l)} \frac{1}{2 \pi \ii} \oint_C \dd{z} \dfrac{ u p_r  p_l  z^{n_0-n-1 } - z^{n_0 - n} }{(z-z^-_u)(z-z^+_u)} , \label{eq:feedback_contour_int} 
\end{align}
where the roots of the denominator in the integral, of which we need to compute the residues, are
\begin{align}
   \label{eq:z+-def_feedback}
   z^{\pm}_u = \frac{1+ u^2 p_l (p_r - p_j)  }{2 u (1- p_j p_l)} \pm \sqrt{\left[\frac{1+ u^2 p_l (p_r - p_j)  }{2 u (1- p_j p_l)}\right]^2- \chi }
\end{align}
with
\begin{align}
    \chi \equiv z^+_u z^-_u  = \frac{p_r p_l }{1-p_j p_l}  = \frac{(1-p_j) p_l }{1-p_j p_l}; \quad 0 \leq \chi \leq 1 .
\end{align}
To compute the integral~\eqref{eq:feedback_contour_int}, one needs to find the residue corresponding to the pole lying inside the unit circular contour $C$ on the complex $\cal{Z}$-plane. We denote by $z_u$ the root inside the unit circle to consider in the inverse $\mathcal{Z}$-
transform: 
\begin{equation}
\label{eq:zs_def_feedback}
z_u \equiv  \begin{cases}
    z^+_u &\text{ if } |z^-_u|>1>|z^+_u|  ,\\
    z^-_u &\text{ if } |z^+_u|>1>|z^-_u|  ,
    \end{cases}
\end{equation}
where, since $|z^+_u z^-_u|\leq 1$, the conditions can be simplified by removing the intermediate $1$, as done in the main text, owing to the fact that one of the root is always inside the unit circle and the other one outside.
Using the Cauchy residue theorem, and applying the same reasoning to Eq.~\eqref{eq:Master+-inzu_sol+feedback}, one then obtains~\cite{dasDynamicsLatticeRandom2023a_SM}
\begin{align}
    \label{eq:P++tilde_u-gen-er}
    \widetilde{P}_{u}(n,+_\theta|n_0,+_\theta) &= \frac{ u p_r  p_l \chi^{\frac{n_0-n-1-|n_0-n-1|}{2}} z_u^{|n_0-n-1|+1}   -  \chi^{\frac{n_0-n-|n_0-n|}{2}} z_u^{|n_0-n|+1}  }{u(1 - p_j p_l)  (z_u^2 - \chi) },  \\
    \label{eq:P+-tilde_u-gen-er}
    \widetilde{P}_{u}(n,+_\theta|n_0,-_\theta) &= \frac{ - u (1-p_r p_l) \chi^{\frac{n_0-n+1-|n_0-n+1|}{2}} z_u^{|n_0-n+1|+1}    }{u(1 - p_j p_l)  (z_u^2 - \chi) }.
\end{align}
Since the work $W_t$ extracted from the quantum engine is always chosen such that $n_0 <n $, which implies $n_0 - n - 1 < 0$ and  $n_0 -n +1 \leq 0 $, Eqs~\eqref{eq:P++tilde_u-gen-er} and~\eqref{eq:P+-tilde_u-gen-er} simplify to 
\begin{align}
    \label{eq:P++tilde_u_SM}
    \widetilde{P}_{u}(n,+_\theta|n_0,+_\theta) &= \frac{ u p_r  p_l \chi^{n_0-n-1} z_u^{n-n_0+2}   -  \chi^{n_0-n} z_u^{n-n_0+1}  }{u(1 - p_j p_l)  (z_u^2 - \chi) }, \\
    \label{eq:P+-tilde_u_SM}
    \widetilde{P}_{u}(n,+_\theta|n_0,-_\theta) &= - \frac{   (1-p_r p_l) \chi^{n_0-n+1} z_u^{n-n_0}    }{(1 - p_j p_l)  (z_u^2 - \chi) } ,
\end{align}
which are the main results of this subsection. 

Interestingly, when $p_j=0$, the qubit state never tumbles from $\ket{+_\theta}$ to $\ket{-_\theta}$, irrespective of the value of $p_l$. When the initial work is $n_0W_\theta$ and the initial qubit state $\ket{+_\theta}$, we get from Eqs.~\eqref{eq:Master++inzu_sol+feedback} and~\eqref{eq:Master-+inzu_sol+feedback} that 
\begin{align}
    \widehat{\widetilde{P}}_u(z,+_\theta|n_0,+_\theta) &= \frac{z^{n_0} ( u - z )}{ u z^2 -z \left(1+u^2\right) + u } = \frac{z^{n_0}}{1-uz} \label{eq:0feedbackpj0} \\
     \widehat{\widetilde{P}}_u(z,-_\theta|n_0,+_\theta) &= 0 .
\end{align}
The above equations show that in this case the qubit state always is $\ket{+_\theta}$ at the end of the cycles, and 
when inverted back in $n$ and $t$, we have $P_t(n,-_\theta|n_0,+_\theta)=0$ and $P_t(n,+_\theta|n_0,+_\theta)= \delta_{n_0+t,n}$, which clearly suggests that the work extracted steadily increases with a unit speed.
Notice that Eqs.~\eqref{eq:P++tilde_u_SM} and \eqref{eq:P+-tilde_u_SM} are only valid when $p_jp_l\neq1$ and $p_rp_l\neq0$ so that $\chi\neq0$ and $z_u^\pm$ does not diverge. The missing cases, therefore, have to be treated differently as done in the following. 

\textbf{Case: $p_r p_l=0$}\\
When $p_rp_l=0$, the constant term with respect to $z$ cancels in the denominator of Eq.~\eqref{eq:Master++inzu_sol+feedback} and \eqref{eq:Master-+inzu_sol+feedback} cancel, and therefore this case must be treated separately. 
When $p_l=0$ information is used deterministically to rectify the qubit state after each measurement, Eqs.~\eqref{eq:feedback_gen_fun_total+} and~\eqref{eq:feedback_gen_fun_total-} yield $\widehat{\widetilde{P}}_u(z|n_0,\pm_\theta) = z^{n_0}/(1-u z)$, inverting which back in $n$ and $t$ gives $ P(n,t|n_0,\pm_\theta) = \delta_{n_0+t, n}$, i.e., the evolution of the work extracted is ballistic, irrespective of its initial value and of the qubit initial state. Equivalently, the qubit state is always $\ket{+_\theta}$ at the end of each cycle. 
More generally, when $p_r p_l=0$ we have $\chi=0$ and we have to go back to Eq.~\eqref{eq:uinversion}, which becomes: 
\begin{align}
    \widetilde{P}_{u}(n,+_\theta|n_0,+_\theta) 
    &=\frac{1}{ u (1 - p_j p_l)} \frac{1}{2 \pi \ii} \oint_C \dd{z} \dfrac{ u p_r  p_l  z^{n_0-n-1 } - z^{n_0 - n} }{z-z_u'}, \label{eq:feedback_contour_int_prpl0} 
\end{align}
where $z_u' = \frac{1-p_lu^2}{u(1-p_l)}$.


\textbf{Case: $p_j=1$ and $p_l=1$.}\\
When $p_jp_l=1$, the term proportional to $z^2$ in the denominator of Eq.~\eqref{eq:Master++inzu_sol+feedback} and \eqref{eq:Master-+inzu_sol+feedback} cancel, and therefore this case must be treated separately. 
In this simple case, the qubit state tumbles at every time step, since no feedback rectifies for the systematic measurement induced tumbles. It means that the state of the qubit changes (either from $\ket{+_\theta}$ to $\ket{-_\theta}$ or from $\ket{-_\theta}$ to $\ket{+_\theta}$) at the end of every cycle and Eqs.~\eqref{eq:Master+inzu_sol} and~\eqref{eq:Master-inzu_sol} simplify to 
\begin{align}
    \widehat{\widetilde{P}}_u(z,+_\theta|n_0,+_\theta) &= \frac{z^{n_0} }{1- u^2 } = \frac{z^{n_0}}{2} \qty(\frac{1}{1-u}+ \frac{1}{1+u}) , \label{eq:Master+inzu_sol_nobf_pj1}\\
    \widehat{\widetilde{P}}_u(z,-_\theta|n_0,+_\theta) &= \frac{  u z^{n_0-1} }{  1- u^2 } = \frac{z^{n_0-1}}{2} \qty(\frac{1}{1-u} -  \frac{1}{1+u}) , \label{eq:Master-inzu_sol_nobf_pj1} 
\end{align}
which may be easily inverted back in $n$ and $t$ to yield 
\begin{align}
    P_t(n,+_\theta|n_0,+_\theta) &= \frac{1}{2} \qty[1+(-1)^t ] \delta_{n_0,n}  \, , \label{eq:Master+inzu_sol_nobf_pj1_nt}\\
    P_t(n,-_\theta|n_0,+_\theta) &= \frac{1}{2} \qty[1-(-1)^t ]   \delta_{n_0-1,n}  \, . \label{eq:Master-inzu_sol_nobf_pj1_nt} 
\end{align}
Equations~\eqref{eq:Master+inzu_sol_nobf_pj1_nt} and~\eqref{eq:Master-inzu_sol_nobf_pj1_nt} clearly suggest that the joint work extracted and qubit state oscillates between the work-state pair $(n_0,\ket{+_\theta}) \longleftrightarrow (n_0-1,\ket{-_\theta})$.

\section{Moments of the extracted work}
\label{SM-Sec:averageMSD}
\subsection{Without information processing}
\textbf{Average Work extracted} \\
Given that the average work increments are:  
\begin{align}
	\langle  \delta W_1 \rangle &= (p_r-p_j)W_\theta \nonumber \\
	\langle \delta W_2 \rangle &= p_r	\langle \delta W_2 \rangle -p_j\langle \delta W_2 \rangle =  (p_r-p_j)^2 W_\theta\nonumber \\
	\langle \delta W_3 \rangle &   = (p_r-p_j)^3 W_\theta \nonumber \\
	\langle \delta W_t \rangle &= (p_r-p_j)^{t}W_\theta,
\end{align}
the average work extracted up to the end of cycle $t\geq 1$: $W_t  =  \sum_{k=1}^{t}  \delta W_k$ reads: 
\begin{align}
    \label{Eq:cum_work_av}
	\langle W_t  \rangle &=\sum_{i=1}^{t} \langle \delta W_i \rangle 
	=W_\theta \left[\frac{1-(1-2p_j)^{t+1}}{1-(1-2p_j)} -1 \right]
	= W_\theta \left[\frac{1-(1-2p_j)^{t+1}}{2p_j} -1 \right]
\end{align}
using the fact that $p_r-p_j = 1-2p_j <1$, i.e., $p_j<1/2$. In the long-time limit $t \to \infty$, this quantity tends to the asymptotic value $W_{\infty}=W/2p_j$. 
Given that,
\begin{equation*}
	\langle W_t  \rangle^2  
	= W_\theta^2 \left[\frac{1-(1-2p_j)^{t+1}}{2p_j} -1 \right]^2,
\end{equation*}
to derive the variance of the work,
one way is to compute: 
\begin{align}
\langle (W_t )^2\rangle = \sum_{i=1}^t \sum_{j=1}^t \langle \delta W_i \delta W_j\rangle
\end{align}
Since the increments $\delta W_i$ can only be equal to $W_\theta$ or $-W_\theta$, we have that, for all $i\in [0,t], \langle\delta  W_i^2 \rangle =W^2_\theta$. Moreover, 
\begin{align}
    \delta W_i \delta  W_{i+1} = \begin{cases}
		\delta W_{i}^2  & \text{with probability} \quad p_r \\
		-\delta W_{i}^2    & \text{with probability} \quad p_j
	\end{cases} \Rightarrow 
    \langle \delta W_i \delta W_{i+1}\rangle = \langle \delta W_i^2 \rangle (1-2p_j) = W^2_\theta (1-2p_j).
\end{align}
The same reasoning leads to: 
\begin{align}
    \delta W_i \delta W_{i+2} &= \begin{cases}
		\delta W_{i} \delta W_{i+1}  & \text{with probability} \quad p_r \\
		-\delta W_{i} \delta W_{i+1}    & \text{with probability} \quad p_j
	\end{cases} \nonumber \\
        &= \begin{cases}
		\delta W_{i} \delta W_{i}  & \text{with probability} \quad p_r^2 \\
            -\delta W_{i} \delta W_{i}  & \text{with probability} \quad p_r p_j \\
		\delta W_{i} \delta W_{i}    & \text{with probability} \quad p_j^2 \\
            -\delta W_{i} \delta W_{i}    & \text{with probability} \quad p_j p_r
	\end{cases} \nonumber \\
        &= \begin{cases}
		\delta W_{i}^2   & \text{with probability} \quad p_r^2 + p_j^2 \\
		-\delta W_{i}^2   & \text{with probability} \quad 2 p_j p_r
	\end{cases} \nonumber \\
    \Rightarrow 
    \langle \delta W_i \delta W_{i+2}\rangle &= \langle \delta W_i^2 \rangle ( p_r^2 + p_j^2 -2p_j p_r) = W^2_\theta (p_r-p_j)^2 = W^2_\theta (1-2p_j)^2.
\end{align}
and thus for any $k\in[0,t-i]$, $\langle \delta W_i \delta W_{i+k}\rangle = W^2_\theta (1-2p_j)^k$. Importantly, this equality does not apply with a negative $k$, or applies but with $|k|$ as the power.  
As a result, since $\langle \delta W_i \delta W_j\rangle  = \langle \delta W_j \delta W_i\rangle $, we have:
\begin{align}
    \label{Eq:variance}
	\langle (W_t )^2\rangle = \sum_{i=1}^t \sum_{j=1}^t \langle \delta W_i \delta W_j\rangle 
	&= \sum_{i=1}^t \langle \delta W_i^2\rangle + 2\sum_{i=1}^{t-1} \sum_{k=1}^{t-i} \langle \delta W_i \delta W_{i+k}\rangle \nonumber \\
	&= tW^2_\theta + 2 W^2_\theta\sum_{i=1}^{t-1} \sum_{k=1}^{t-i}  (1-2p_j)^k \nonumber \\
	&= tW^2_\theta + 2 W^2_\theta\sum_{i=1}^{t-1} \left[ \frac{1-(1-2p_j)^{t-i+1}}{2p_j} -1\right] \nonumber \\
	&= -tW^2_\theta + 2W^2_\theta+  \frac{(t-1)W^2_\theta}{p_j}- \frac{W^2_\theta}{p_j}(1-2p_j)^{t+1}\sum_{i=1}^{t-1} \left(\frac{1}{1-2p_j}\right)^{i} 
	\nonumber \\
	&= -tW^2_\theta +  2W^2_\theta +\frac{(t-1)W^2_\theta}{p_j}- \frac{W^2_\theta}{p_j}(1-2p_j)^{t+1}\left[ \frac{1-(1/(1-2p_j))^{t}}{1-(1/(1-2p_j))}- 1\right] \nonumber \\
	&= -(t-2)W^2_\theta +  \frac{(t-1)W^2_\theta}{p_j}+ \frac{W^2_\theta}{p_j}(1-2p_j)^{t+1} - \frac{W^2_\theta}{p_j}(1-2p_j)^{t+2}\left[ \frac{1-(1/(1-2p_j))^{t}}{1-2p_j -1} \right] \nonumber \\
	&= -(t-2)W^2_\theta +  \frac{(t-1)W^2_\theta}{p_j}+ \frac{W^2_\theta}{p_j}(1-2p_j)^{t+1} +\frac{W^2_\theta}{2p_j^2}\left[(1-2p_j)^{t+2}-(1-2p_j)^2 \right] \nonumber \\
 &= W_\theta^2 \left[ \frac{2p_j-tp_j+t-1}{p_j} + \frac{2p_j(1-2p_j)^{t+1}+ (1-2p_j)^{t+2}-(1-2p_j)^2}{2p_j^2} \right] \nonumber \\
 &= W_\theta^2 \left[ \frac{2p_j-tp_j+t-1}{p_j} + \frac{2p_j(1-2p_j)^{t+1}+ (1-2p_j)^{t+1}(1-2p_j)-1-4p_j^2 +4p_j}{2p_j^2} \right] \nonumber \\
 &= W_\theta^2 \left[ \frac{2p_j-tp_j+t-1+2-2p_j}{p_j} + \frac{ (1-2p_j)^{t+1}-1}{2p_j^2} \right] \nonumber \\
 &= W_\theta^2 \left[ \frac{t(1-p_j)+1}{p_j} + \frac{ (1-2p_j)^{t+1}-1}{2p_j^2} \right]. 
\end{align}
We can obtain the variance: 
\begin{align}
	\text{Var}[W_t]&= \langle (W_t )^2 \rangle- \langle W_t  \rangle^2 \nonumber \\
        &= W_\theta^2 \left[ \frac{t(1-p_j)+1}{p_j} + \frac{ (1-2p_j)^{t+1}-1}{2p_j^2} \right] 
        - 
        W_\theta^2 \left[\frac{1-(1-2p_j)^{t+1}}{2p_j} -1 \right]^2 \nonumber \\
         &= W_\theta^2 \left[ \frac{t(1-p_j)+1}{p_j} + \frac{ (1-2p_j)^{t+1}-1}{2p_j^2}  
        - 1-
        \left[\frac{1-(1-2p_j)^{t+1}}{2p_j}\right]^2 +2 \frac{1-(1-2p_j)^{t+1}}{2p_j} \right] \nonumber
\end{align}

As there exists an exact one-to-one correspondence between the dynamics of the work extracted, taken together with the qubit state at the end of each cycle, and that of a RnT process, it is also possible to obtain the average quantities using the generating function of the RnT propagator. 
Let us start with the average work extracted.
Using the generating function given in Eq.~\eqref{eq:ptildehat} and the relation $ \lim_{z\to 1} \pdv{z} \widehat{\widetilde{P}}_u(z|n_0,+) =\sum_{n=-\infty}^{\infty} n \widetilde{P}_u(n|n_0,+) $, one may write
\begin{align}
  \sum_{n=-\infty}^{\infty} n \widetilde{P}_u(n|n_0,+)  =   \frac{u (p_r-p_j) - u n_0 (p_r-p_j)+n_0}{p_r u^2-2 p_r u-p_j u^2+1} = \frac{1-2p_j+2 p_j n_0}{2 p_j (1-u)} - \frac{1-2 p_j}{2 p_j [1-u(1-2 p_j)]} ,
\end{align}
which, when inverted back in time $t$, gives the time-dependent average work extracted from the engine,
\begin{align}
    \label{eq:avgn_zero_feedback}
    \expval{n_t} \equiv \sum_{n=-\infty}^{\infty} n P_t(n|n_0,+) = n_0 - 1 + \frac{1 -(1-2p_j)^{t+1}}{2 p_j}.
\end{align}
Note that for $n_0=0$, we indeed retrieve $\expval{W_t} = W_\theta \expval{n_t} $, see  Eq.~\eqref{Eq:cum_work_av}. 
In a similar fashion, one may also write $ \lim_{z\to 1} \pdv[2]{z} \widehat{\widetilde{P}}_u(z|n_0,+) =\sum_{n=-\infty}^{\infty} n(n-1) \widetilde{P}_u(n|n_0,+) $, which yields
\begin{align}
    \sum_{n=-\infty}^{\infty} n^2 \widetilde{P}_u(n|n_0,+) &=  \sum_{n=-\infty}^{\infty} n \widetilde{P}_u(n|n_0,+) + \lim_{z\to 1} \pdv[2]{z} \widehat{\widetilde{P}}_u(z|n_0,+) \nonumber \\
    &=  \frac{1-p_j}{p_j (1-u)^2} - \frac{1 -2 p_j^2 (1+ n_0^2) - 2 p_j n_0(1-2p_j) }{2 p_j^2 (1-u)}+\frac{(1-2p_j) (1-2p_jn_0)}{2 p_j^2 [1-u(1-2p_j)]}  .
\end{align}
Inverting back the above equation in $t$, we obtain the time-dependent second moment of the work extracted:
\begin{align}
    \label{eq:avgn2_zero_feedback}
    \expval{n_t^2} \equiv \sum_{n=-\infty}^{\infty} n^2 P_t(n|n_0,+) = (n_0-1)^2+ \frac{1}{p_j} (1-p_j) (t+1) + \frac{ (1-2 p_jn_0) [(1-2p_j)^{t+1} -1 ]}{2 p_j^2 },
\end{align}
which for $n_0=0$ is again in agreement with Eq.~\eqref{Eq:variance} as one gets $ \expval{ 
 (W_t)^2} = W^2_\theta \expval{n_t^2}$ in this case.

The time-dependent variance in the work extracted is given by
\begin{align}
    {\rm{Var}}[n_t] = \expval{n_t^2} -\expval{n_t}^2 
    = \frac{4 (1-p_j) \left[p_jt+(1-2 p_j)^{t+1}\right] -(1-2 p_j)^{2 (t+1)} +4 p_j (2-p_j) -3}{4 p_j^2} .
\end{align}
The Mean Square displacement (MSD) of the work extracted obtained using Eqs.~\eqref{eq:avgn_zero_feedback} and~\eqref{eq:avgn2_zero_feedback} is given by
\begin{align}
    \label{eq:msd_zero_feedback}
    \expval{(n_t-n_0)^2} = 1+ \frac{ (t+1)(1-p_j)}{p_j} + \frac{ \left[ (1-2 p_j)^{t+1} -1 \right]}{2 p_j^2} .
\end{align}
In the long-time limit $t\to \infty$, the leading-order behaviour of the moments are given by
\begin{align}
    \expval{n} = n_0 -1 + \frac{1}{2p_j}, \quad  \expval{n^2} = {\rm{Var}}[n]= \frac{t(1-p_j)}{p_j}   .
\end{align}
The above relations show that although the average of the position approaches a constant value, its variance keeps on increasing and becomes linearly proportional to time.

\subsection{With information processing}
\begin{figure}[!htbp]
\subfloat[$p_l=1$ ; $p_j=0.2$\label{color_Wext_pl1}]{\includegraphics[width=0.48\columnwidth]{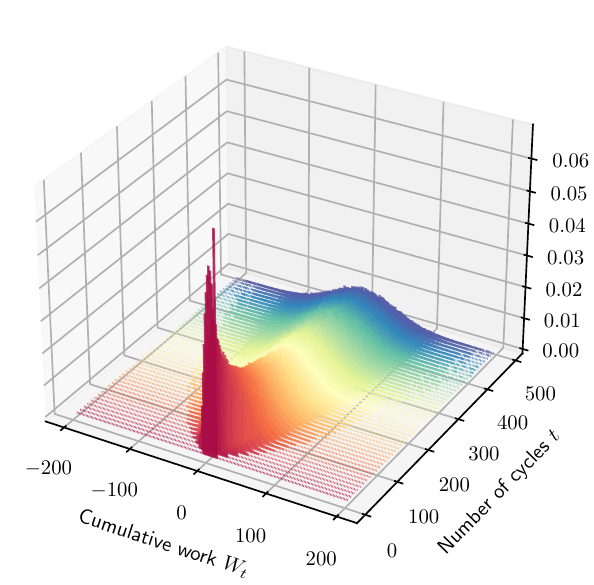}
}
\hspace*{\fill}
\subfloat[$p_l=0.9$ ; $p_j=0.2$\label{color_Wext_pl09}]{\includegraphics[width=0.48\columnwidth]{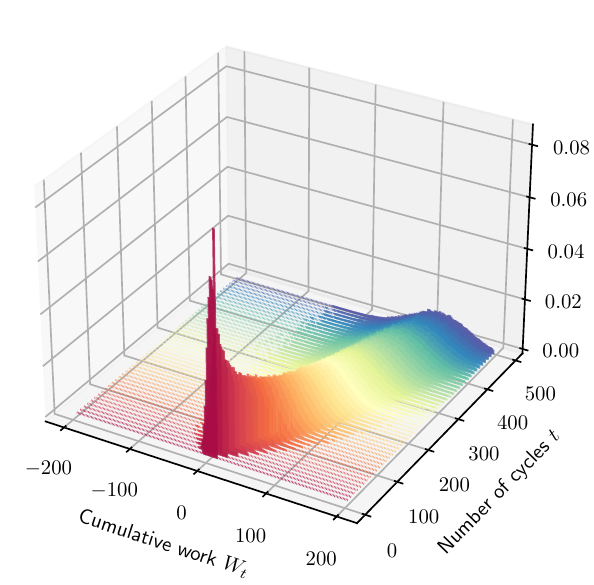}
}
\caption{Snapshots of the work distribution as a function of the number of cycles $t$ obtained from numerical simulations using 50000 trajectories, (a) without information processing, $p_l=1$ or (b) with small information processing, $p_l=0.9$. The colour serves as a guide to the eye for the time coordinate.}
	\label{fig_cum_work_distrib}
\end{figure}

We now proceed to obtain the average position and the mean-squared displacement (MSD) of the work extracted in the presence of feedback. Without losing any generality, here we restrict ourselves to the case in which we initially have $n_0W_\theta$ of work extracted and the qubit in state $\ket{+_\theta}$. The average work extracted in the $\mathcal{U}$-domain may be obtained using the generating function given in Eq.~\eqref{eq:feedback_gen_fun_total+} as
\begin{align}
    \widetilde{\expval{n}}_u &\equiv \sum_{n=-\infty}^{\infty} n \widetilde{P}_u (n|n_0,+)  = \lim_{z \to 1} \pdv{\widehat{\widetilde{P}}_{u}(z|n_0,+)}{z} \nonumber \\
    &= \frac{ (1-p_l)}{ [1-(p_r-p_j)p_l] (1-u)^2}  - \frac{2 p_j (p_r-p_j)p_l^2 }{ [1-(p_r-p_j)p_l]^2 [1 - u(p_r-p_j) p_l ] }\nonumber  \\ 
    &+    \frac{ (n_0-1) [ (p_r-p_j)^2  p_l^2 + p_r (2 (1-p_l) - 1 ) ] +p_j \left(3 n_0 - 2 n_0 (1-p_l) - 1\right)}{ [1-(p_r-p_j)p_l]^2 (1-u)}  ,
\end{align}
which when inverted back in time (using the ${\mathcal{U}}$-transformation relation pairs given in Eq.~\eqref{eq:U_trans_rels}) and simplified by using $p_r=1-p_j$ gives the time-dependent value of the average position
\begin{align}
    \label{eq:avgn_feedback_SM}
    \expval{n} _t 
    &=  n_0 -1 +\frac{ (1-p_l) (t+1)}{ 1-p_l + 2p_j p_l } + \frac{ 2p_j p_l [1- (1-2 p_j)^{t+1} p_l^{t+1}  ]}{(1-p_l + 2 p_jp_l)^2 } .
\end{align}
Note that for zero feedback, i.e., with $p_l = 1$, Eq.~\eqref{eq:avgn_feedback_SM} reduces to Eq.~\eqref{eq:avgn_zero_feedback}. On the other hand, for $p_l=0$, we are in the ballistic regime, i.e., there exists only a single, deterministic trajectory, for which the work extracted grows linearly with time with a unit speed, see Eqs.~\eqref{eq:Master+_feedback_SM} and~\eqref{eq:Master-feedback_SM}. 
In this case, time-dependent average position is given by $\expval{n} _t =  n_0 + t$, as expected for a ballistic evolution. 
Equation~\eqref{eq:avgn_feedback_SM} appears to perfectly correspond to the average obtained by simulating many stochastic trajectories as shown in Fig.~\ref{fig_cum_work_distrib}.

To obtain the second moment of the work extracted with feedback, one may again use the generating function given in Eq.~\eqref{eq:feedback_gen_fun_total+}  along with $  \lim_{z\to 1} \pdv[2]{z} \widehat{\widetilde{P}}_u(z|n_0,+) =\sum_{n=-\infty}^{\infty} n(n-1) \widetilde{P}_u(n|n_0,+) $ and write
\begin{align}
     \widetilde{\expval{n^2}}_u &\equiv \sum_{n=-\infty}^{\infty} n^2 \widetilde{P}_u(n|n_0,+)
    =  \lim_{z\to 1} \qty[  \pdv{z} \widehat{\widetilde{P}}_{u}(z|n_0,+) +  \pdv[2]{z} \widehat{\widetilde{P}}_{u}(z|n_0,+) ] \nonumber \\
    &= \frac{(n_0-1)^2}{1-u} + \frac{2 n_0(1-p_l)}{(1-p_l+2p_jp_l)(1-u)^2} +\frac{2 (1-p_l)^2}{ (1-p_l + 2 p_jp_l)^2 (1-u)^3} +\frac{4 p_j (1-p_l) (1-2 p_j)  p_l^2}{(1-p_l + 2 p_jp_l)^3 [1-(1-2 p_j) p_l u]^2} \nonumber \\
    & -\frac{ (3 -3 p_l + 2 p_j p_l) [ 1  
    -2 p_l + (1 - 2 p_j)^2 p^2_l ]  }{ (1-p_l + 2 p_jp_l)^3 (1-u)^2} \nonumber \\
    & +\frac{4 p_j p_l [ n_0 (1 - p_l + 2 p_j p_l)^2 + 2 p_l (p_j + p_l - 2 p_j p_l) - 2]}{ (1-p_l + 2 p_jp_l)^4 } \qty[\frac{1}{1-u} - \frac{(1-2p_j) p_l}{1-(1-2p_j)p_l u}].
\end{align}
Using the ${\mathcal{U}}$-transformation relation pairs mentioned in Eq.~\eqref{eq:U_trans_rels}, the above equation may be inverted back in time to obtain the time-dependent second moment
\begin{align}
    \label{eq:avgn2_feedback}
    \expval{n_t^2} 
    &= (n_0-1)^2  + \frac{ 2n_0 (1-p_l) (t+1)}{ (1-p_l + 2 p_jp_l) }
    +\frac{ (1-p_l)^2 (t+1) (t+2)}{ (1-p_l + 2 p_jp_l)^2 } +\frac{4 p_j (1-p_l) (1-2 p_j)^{t+1}  p_l^{t+2} (t+1)}{(1-p_l + 2 p_jp_l)^3 }  \nonumber \\
    & - \frac{(t+1)  (3 -3 p_l + 2 p_j p_l) [ 1 - 2 p_l + (1 - 2 p_j)^2 p_l^2 ] }{ (1-p_l + 2 p_jp_l)^3 }  \nonumber \\
    & +\frac{4 p_j p_l  [ n_0 (1 - p_l + 2 p_j p_l)^2 + 2 p_l (p_j + p_l - 2 p_j p_l) -2 ] [1- (1-2p_j)^{t+1} p_l^{t+1}] }{ (1-p_l + 2 p_jp_l)^4 } .
\end{align}
In the ballistic case, i.e. $p_l = 0$, we get 
$\expval{n_t^2} = (n_0 +t)^2$, as expected.

\begin{figure}[!htbp]
\centering
\includegraphics[width=10cm]{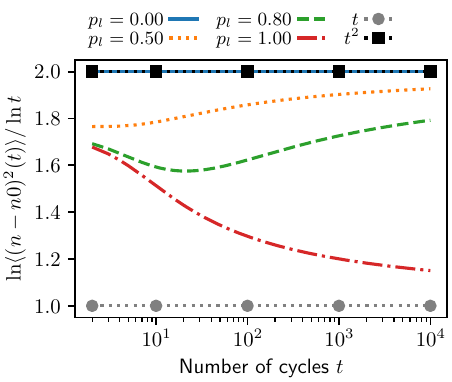}
\caption{Mean Square displacement in logarithmic scale, divided by $\ln t$, with respect to time for $p_j=0.2$ . When $p_l<1$, the behaviour always converges to a ballistic one in the long run whilst for $p_l=1$, it converges to a diffusive one. At intermediate times, the work evolution is superdiffusive.}
\label{fig_MSD_SM}
\end{figure}

The MSD for the work extracted in the presence of feedback may be obtained using Eqs.~\eqref{eq:avgn_feedback_SM} and~\eqref{eq:avgn2_feedback}. To this end, we get
\begin{align}
    \label{eq:MSD_n0}
    \expval{(n_t-n_0)^2}  
    &= 1+  \frac{1}{{(1-p_l + 2 p_jp_l)^4}} \Big[(1-p_l)^4(t^2-1) \nonumber \\
    &+ 4 p_j(t+1) p_l \Big\{  (1-p_l)^2 ((1-p_l)(t-3)+3) + p_j p_l (1-p_l)\left((1-p_l) (t-10)+8  \right) - 4 p_j^3 p_l^3 - 4 p_j^2 (2-3p_l) p_l^2 \Big\}  \nonumber \\
    &-4 p_j p_l \Big\{ 2 \left((2 p_j -1)(1-p_l)^2-(3 p_j-2) (1-p_l)+p_j\right) \left(1-(1-2 p_j)^{t+1} p_l^{t+1}\right) \nonumber \\
    &-(1-p_l) (t+1) (1-p_l + 2 p_jp_l) (1-2 p_j)^{t+1} p_l^{t+1}\Big\} \Big]  .
\end{align}
When $t\gg 1$, if $p_l\in(0,1)$ and $p_j\in(0,0.5)$, the leading factor is $\frac{(1-p_l)^4}{(1-p_l+2p_jp_l)}t^2$ and the MSD tends to a ballistic one. 
For $p_l = 0$, the above equation reduces to $\expval{(n_t-n_0)^2} = t^2$, suggesting a ballistic evolution for the work extracted. On the other hand, in the absence of feedback $(p_l=1)$, we have 
\begin{align}
    \label{eq:MSD_n0_pl1}
    \expval{(n_t-n_0)^2}  
    = 1+  \frac{(1-2 p_j)^{t+1}-1}{2p_j^2} +  (t+1) \frac{(1-  p_j)}{p_j}  \underset{t\rightarrow \infty}{\rightarrow} (t+1) \frac{(1-  p_j)}{p_j},
\end{align}
i.e., a diffusive evolution as found in Eq.~\eqref{eq:msd_zero_feedback}. 
When $p_l<1$, the leading factor when $t\gg1$ scales as $t^2$. In general the evolution is characterized as diffusive, superdiffusive, or ballistic depending on the power $\alpha$ according to which it evolves with time. 
Finding $\alpha$ mean solving $\expval{(n_t-n_0)^2} = t^{\alpha(t)}$ where the power can be time-dependent. Therefore, plotting $\ln(\expval{(n_t-n_0)^2})/\ln(t)$ gives us access to $\alpha(t)$ which will go to $2$ in the ballistic case, $1$ in the diffusive case and in between for the superdiffusive cases, see Fig.~\ref{fig_MSD_SM}.   

\section{First-passage statistics}
\label{SM-sec:FPS}
We now focus on the first-passage time $\mathcal{T}$ when the work extracted reaches a target value $n_TW_\theta$. We provide the general definition and apply it to our particular case, without, \ref{FPT_noi}, and with, \ref{FPT_i}, information processing. Then we briefly discuss the waiting-time distribution to change the final qubit state, \ref{sub_sec:WTD}.  

Note that for $n_{T}>n_0$ (respectively, $n_T<n_0$) the state of the qubit at the end of the cycle when the work reached $n_TW_\theta$ has to be necessarily $\ket{+_\theta}$ (respectively, $\ket{-_\theta}$). 
We may then say that the first-passage time probability $F_{\cal{T}}(n_T|n_0)$, i.e., the probability for the work extracted to reach $n_TW_\theta$ for the first time while starting from site $n_0W_\theta$, is given by
\begin{align}
F_{\cal{T}}(n_T |n_0) = \begin{cases}     F_{{\cal{T}}}(n_T,+|n_0) , \quad \quad n_T > n_0 , \\[0.5ex]
     F_{{\cal{T}}}(n_T,-|n_0) , \quad \quad n_T < n_0.
\end{cases}
\end{align}
In our case, considering $n_T>n_0$ the first-passage time is defined as ${\cal{T}} \equiv \inf \qty{ t\geq 0 | n_t \geq n_T } $.
Following Ref.~\cite{balakrishnanRenewalEquationPersistent1988_SM}, the generating function for the first-passage time probability, defined by $\widetilde{F}_u(n_T|n_0) \equiv   \sum_{{\cal{T}} = 0} ^{\infty} u^{\cal{T}} F_{\cal{T}}(n_T|n_0)$, may be written as
\begin{align}
    \widetilde{F}_u(n_T|n_0) &=\frac{\widetilde{P}_{u}(n',+| n_0,+)+ \widetilde{P}_{u}(n',+| n_0,-)}{ \widetilde{P}_{u}(n',+| n_T,+) } , \quad n_0 < n_T < n',
\end{align}
which in our case reduces to 
\begin{align}
    \label{eq:fpt_gen_our_case}
    \widetilde{F}_u(n_T|n_0) &=\frac{\widetilde{P}_{u}(n',+| n_0,+)}{ \widetilde{P}_{u}(n',+| n_T,+) }, \quad n_0 < n_T < n' ,
\end{align}
owing to the fact that the qubit starts from state $\ket{+_\theta}$.

\subsection{Without information processing}
\label{FPT_noi}
In the absence of feedback ($p_l=1$), we obtain from Eqs.~\eqref{eq:fpt_gen_our_case} and~\eqref{eq:P++tilde_u_SM} that 
\begin{align}
    \label{eq:fpt-gen-0feedback}
    \widetilde{F}_u(n_T|n_0) =  \frac{ u p_r  z_u^{|n_0-n'-1|+1} - z_u^{|n_0-n'|+1} }{u p_r  z_u^{|n_T-n'-1|+1} - z_u^{|n_T-n'|+1}} =  z_u^{n_T-n_0} , 
\end{align}
where we have used $n_0<n_T<n'$ for the last equality and where the quantity $z_{u}$ defined in Eq.~\eqref{eq:zs_def_feedback}. 

Note that for $p_j = 0$ or $1$, the  trajectories in the absence of feedback are deterministic and all trajectories reach the threshold $n_T$ at the same, finite, stopping time.  Moreover, for $p_j=0$, if the qubit starts in $\ket{+_\theta}$, this state does not change during the course of evolution and the work extracted increases by $W_\theta$ in each cycle.

Except for these two extreme jump probabilities, $p_j=0$ or $1$, the mean first-passage time $\expval{\cal{T}} \equiv \lim_{u\to 1} \partial \widetilde{F}_u(n_T|n_0) / \partial u$ for the work extracted to reach the target $n_TW_\theta$ diverges. From simulation, we indeed find that the first-passage probability density decays as $\sim {\cal{T}}^{-3/2}$ as  with free RnT processes in the continuum limit since this scaling is typical in unbiased random walks and Brownian diffusion.

\subsection{With information processing}
\label{FPT_i}
In the presence of feedback ($p_l \neq 1$), the generating function for the first-passage time probability is obtained from  Eqs.~\eqref{eq:fpt_gen_our_case} and~\eqref{eq:P++tilde_u} as
\begin{align}
    \widetilde{F}_u(n_T|n_0) &= \frac{ u p_r  p_l \chi^{n_0-n'-1} z_u^{n'-n_0+2}   -  \chi^{n_0-n'} z_u^{n'-n_0+1}  }{  u p_r  p_l \chi^{n_T-n'-1} z_u^{n'-n_T+2}   -  \chi^{n_T-n'} z_u^{n'-n_T+1} } = \chi^{n_0 - n_T} z_u^{n_T - n_0} ,
    \label{eq:FPT_generating_feedback_SM}
\end{align}
where the quantity $z_{s}$ is defined in Eqs.~\eqref{eq:z+-def_feedback} and~\eqref{eq:zs_def_feedback}. 
We remark that the generating function $\widetilde{F}_u(n_T|n_0)$ both in the absence and the presence of feedback is independent of $n'$, as one expects~\cite{balakrishnanRenewalEquationPersistent1988_SM}. 
From Eq.~\eqref{eq:z+-def_feedback}, we have $\lim_{u\to 1} z_{+} = 1$, $\lim_{u\to 1} z_{-} = \chi$, and therefore $\lim_{u\to 1} \widetilde{F}_u(n_T|n_0)= \chi^{n_0-n_T} \chi^{n_T-n_0} =1$, which shows the normalization of the first-passage probability $F_{\cal{T}}(n_T|n_0)$.
\\
As a result, the mean first-passage time to reach $n_T$ starting from $n_0$  is given by 
\begin{align}
   \label{eq:MFPT_feedback_SM}
   \expval{\mathcal{T}} \equiv \lim_{u\to 1} \pdv{\widetilde{F}_u(n_T|n_0)}{u} = \frac{ (n_T-n_0) (1-p_l+2 p_jp_l) }{1-p_l} .
\end{align}
From Eq.~\eqref{eq:MFPT_feedback_SM} it follows that the mean first-passage time diverges in the absence of feedback ($p_l=1$) and reduces to $(n_T-n_0)$ in the ballistic limit of $p_l=0$.
Here, the second moment of the first-passage time is given by
\begin{align}
    \expval{\mathcal{T}^2} &\equiv  \lim_{u\to 1} \qty[  \pdv{\widetilde{F}_u(n_T|n_0)}{u} + \pdv[2]{\widetilde{F}_u(n_T|n_0)}{u} ] \nonumber \\
    &= (n_T-n_0)^2 + \frac{ 4 p_j (n_T-n_0)p_l (1-p_l+p_jp_l) [(n_T-n_0-1) (1-p_l) - 2 p_j p_l +2 ] }{(1-p_l)^3} ,
\end{align}
and correspondingly the signal-to-noise ratio (SNR) is given by 
\begin{align}
\label{eq:feedback-snr_SM}
    {\rm{SNR}}(\mathcal{T}) \equiv  \frac{\expval{\mathcal{T}}^2}{\expval{\mathcal{T}^2} - \expval{\mathcal{T}}^2 } = \frac{ (1-p_l) (n_T-n_0)  [1-p_l+2 p_j p_l]^2  }{ 4 p_j p_l (1-p_l+p_jp_l) [1+p_l-2 p_j p_l]   },
\end{align}
which for a given $p_l$ varies non-monotonously as a function of $p_j$.

\begin{figure}[!htbp]
\centering
\includegraphics[scale=1.15]{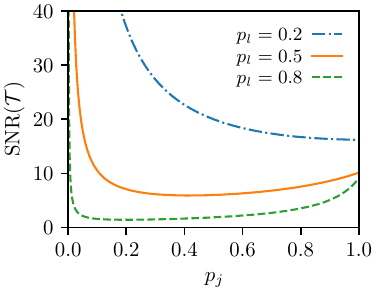}
\includegraphics[scale=1.15]{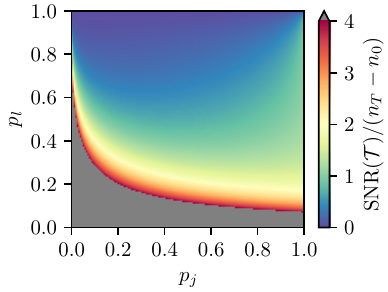}
\caption{Left: The signal-to-noise ratio obtained from Eq.~\eqref{eq:feedback-snr_SM} for $n_0=1$ and $n_T=10$. Right: Scaled signal-to-noise ratio as a function of $p_j$ and $p_l$. The color gray area highlights all values greater than 4. }
\label{fig:snr}
\end{figure}

\subsection{Waiting-time distribution}
\label{sub_sec:WTD}
The waiting time distribution corresponds, for our engine, to the distribution of the number of cycles for which the qubit remains in a given state, i.e., the distribution of the times spent between jumps. 
Although the state of the qubit varies within each cycle, we here consider only the final states of each cycle.\\
\textbf{Without feedback}\\
From the state $\ket{+_\theta}$, the probability to remain in this state for $n$ cycles is simply $p_r^n$. The same goes from the state $\ket{-_\theta}$ and hence, the waiting time distribution follows a power law of the type $(1-p_j)^n \approx e^{-p_j n}$. \\
\textbf{With feedback}\\
From a state $\ket{+_\theta}$, the probability to remain in this state for $n$ cycles is $(p_r+p_j(1-p_l))^n$. Starting from the state $\ket{-_\theta}$ it is $p_r^n(1-pi)^n$. 
Hence, the waiting time distribution is not exactly a power law, but since $p_r+p_j(1-p_l) \geq p_rp_l$, and even more so the greater $(1-p_l)\in[0,1]$ is, the dominant factor is $(p_r+p_j(1-p_l))^n= (1-p_jp_l)^n\approx e^{-(1-p_l)p_ln}$.


\section{Energetics}
\label{SM-sec:Energetics}
In this section, we derive the average power extracted in the steady state and find the optimal value of the laziness probability $p_l$ to maximize it. 
First, the average work increment extracted in the steady state is derived~(Sec~\ref{subsec:avdelta_W}), then the work cost of information processing is obtained~(Sec~\ref{subsec:infoCost}), and then both results are combined to compute the power~(Sec~\ref{subsec:power}) and maximize it over $p_l$~(Sec~\ref{subsec:optpl}). 

\subsection{Average work increment}
\label{subsec:avdelta_W}
From Eq.~\eqref{eq:avgn_feedback_SM}, we have that the mean work extracted until the end of cycle $t$ reads:
\begin{align}
    \label{eq:avgW_feedback}
    \expval{W_t}  
    &= W_\theta \left(n_0 -1 +\frac{ (1-p_l) (t+1)}{ 1-p_l + 2p_j p_l } + \frac{ 2p_j p_l [1- (1-2 p_j)^{t+1} p_l^{t+1}  ]}{(1-p_l + 2 p_jp_l)^2 }\right).
\end{align}

Since the stochastic work extracted during the cycle $t$ is $\delta W_t = W_t -W_{t-1}$, we have, at the average level: 
\begin{align}
    \label{eq:avg_deltaW_feedback}
    \expval{\delta W_t}  
    &= W_\theta \left(\frac{(1-p_l)}{ 1-p_l + 2p_j p_l } + \frac{ 2p_j p_l [- (1-2 p_j)^{t+1} p_l^{t+1} + (1-2 p_j)^{t} p_l^{t}]}{(1-p_l + 2 p_jp_l)^2 }\right)
    \nonumber \\
    &=  W_\theta \left(\frac{(1-p_l)}{ 1-p_l + 2p_j p_l } + \frac{ 2p_j p_l(1-2 p_j)^{t} p_l^{t}  [1- (1-2 p_j) p_l]}{(1-p_l + 2 p_jp_l)^2 }\right)\nonumber \\
    &=  W_\theta \left(\frac{(1-p_l)}{ 1-p_l + 2p_j p_l } + \frac{ 2p_j p_l(1-2 p_j)^{t} p_l^{t} }{(1-p_l + 2 p_jp_l)}\right).
\end{align}
If $p_l\neq 1$, in the steady state limit of $t\rightarrow \infty$, 
\begin{equation}
\expval{\delta W_{\rm ss}} = W_\theta \frac{(1-p_l)}{ 1-p_l + 2p_j p_l }.
\end{equation}

\subsection{Information processing cost}
\label{subsec:infoCost}
\begin{figure}[!htbp]
\centering
\includegraphics[width=15cm]{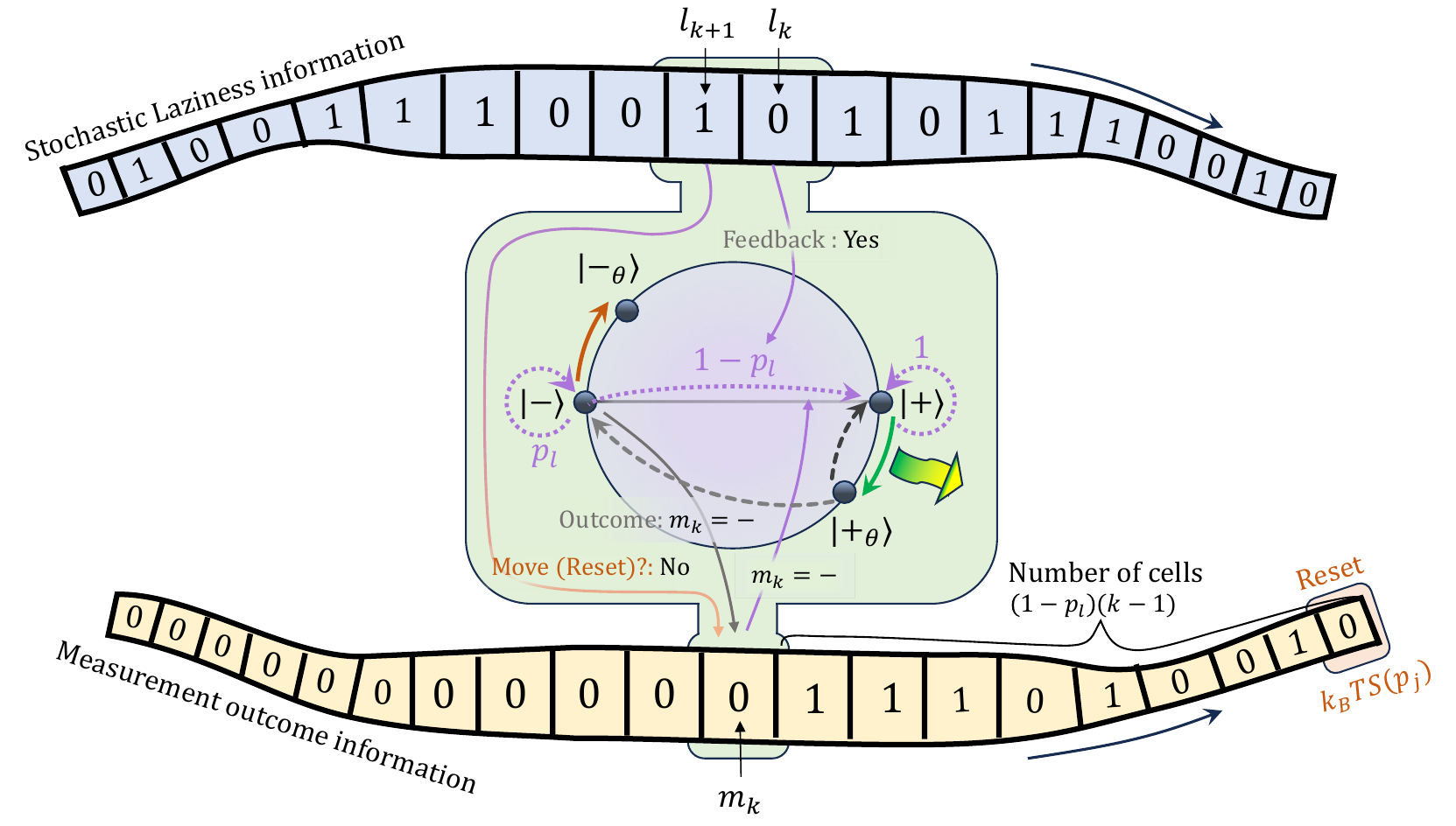}
\caption{LQME with stochastic feedback illustrated with the laziness tape (top) from which the stochastic value of the laziness is taken and the measurement outcome tape (bottom) which stores information on the qubit state in order to potentially use it during the feedback step. The outcome only needs erasure is the next cycle require proper information on the qubit state. 
}
\label{fig:infoCost}
\end{figure}
\begin{figure}[!htbp]
\centering
\includegraphics[width=12cm]{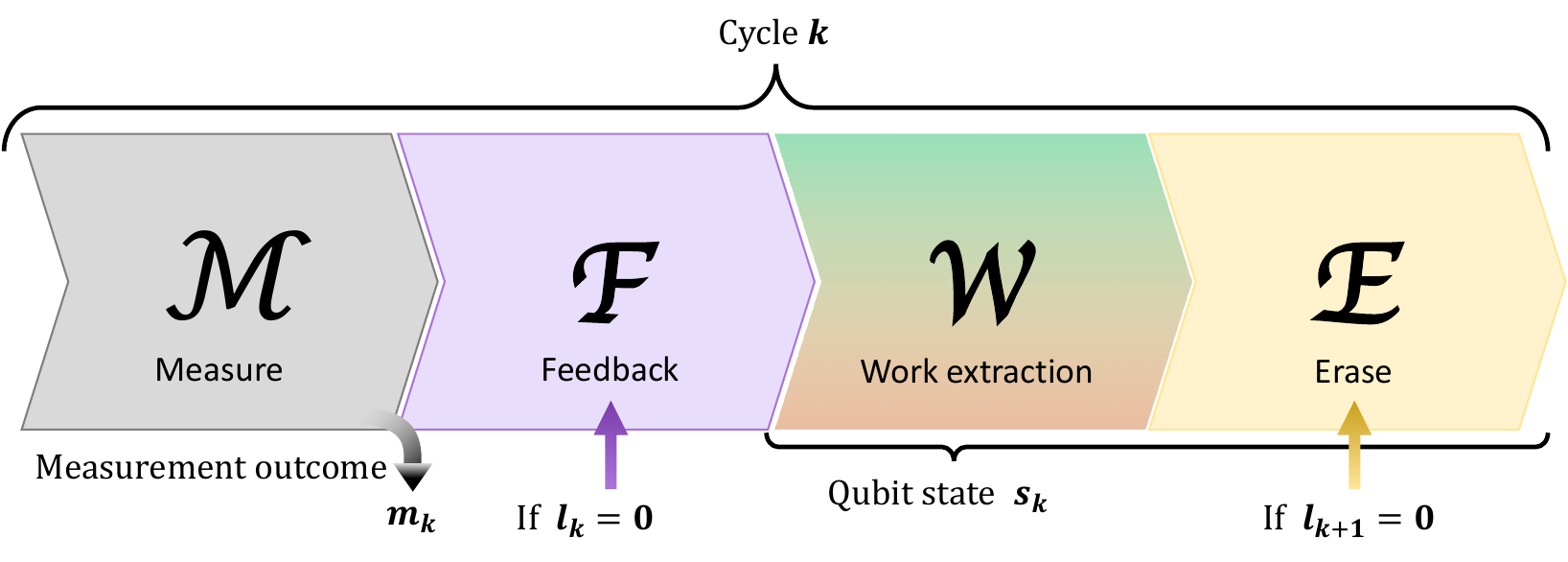}
\caption{A generic cycle $k$ of the engine depending on the laziness stochastic parameters $\ell_k$ and $\ell_{k+1}$. The measurement outcome $m_k$ and final qubit state $s_k$ also appear.}
\label{fig_cycle_protocol_SM}
\end{figure}

In order not to discard any thermodynamic resource, as can be the case when the external agent has access to pure states for information processing, which can be viewed as having access to a zero temperature bath, we must take into account the work cost of resetting all memory cell to their initial value. 
As illustrated in Fig.~\ref{fig:infoCost}, the external agent, or demon, needs two information tapes in order to operate correctly the engine. One stores the measurement (Markovian) outcome results $\{m_1,m_2,...,m_t\}$ and the other one stores the random (i.i.d.) laziness $\ell_k\in\{0,1\}$ where $k\in[0,t]$ for trajectories of $t$-cycles. The tape $\{\ell_1,\ell_2,...,\ell_t\}$ is such that the ratio between the number of $1$ and $0$ converges to $p_l$ when $t\rightarrow\infty$. When $\ell_k=1$, the $k$th cycle should not use the measurement outcome information, i.e., the protocol simply consists of a measurement followed by the work extraction step (no feedback is applied after the measurement), the last erasure step depends on the value of $\ell_{k+1}$. If $\ell_{k+1}=1$, the next cycle will also not need the measurement outcome information and hence there is no use in providing a memory cell in a well-defined pure state to encode this outcome, i.e., there is no need to erase the memory cell containing the outcome of the measurement in the cycle $k$. If, $\ell_{k+1}=0$ however, we need to extract all the information from the measurement in the $k+1$-cycle and hence we should either use a new memory cell in the measurement outcome tape or reset the one containing the measurement outcome of the cycle $k$ (See Fig.~\ref{fig_cycle_protocol_SM} for the cycle step visualization). Since this measurement tape must be identical before and after a trajectory has been implemented, in our case, doing the erasures on the fly when needed or at the end of all the cycles is equivalent if one do not use a need memory cell when it is not needed. 

The tape containing the information about when to use the measurement outcome information or not, $\{\ell_1,\ell_2, \dots,\ell_t\}$, does not require work to be processed because it carries no information on the system and is not changed by the protocol. This tape is the same before and after the experiment, and therefore does not constitute a resource. In other words, in order to know when to use information or not, the external agent just needs to have an appropriate dice giving $0$ with probability $1-p_l$ and $1$ with probability $p_l$.  

The tape containing the measurement outcomes, however, needs erasing. 
If we knew nothing about the protocol, the average information cost per cycle would simply be $k_BT\ln(2)$ as there are only two possible measurement outcomes, hence a bit of information per measurement, and given that we have access to a bath of temperature $T$. 
To reduce this cost, the external agent can use his knowledge of the protocol and of the specific values of $p_l$ and $p_j$. 

First, it should be emphasized that the measurement outcomes $m_k\in\{+,-\}$ are distinct from the states at the end of the cycles given by the $s_k\in\{+_\theta,-_\theta\}$. In the most general case, $0\leq p_l\leq 1$, they are however related by the following relations: 
\begin{align}
\label{eq:skmk}
	\begin{cases}
		m_k = - \begin{casesnewbig}{$p_l$}{$1-p_l$}
			s_k = -_\theta  
			\begin{casesnew}{$1-p_j$}{$p_j$}
				m_{k+1} = -  
					\begin{casesnew}{$p_l$}{$1-p_l$}
						s_{k+1} = -_\theta  \\[50pt]
						s_{k+1} = +_\theta
					\end{casesnew} \\[50pt]
				m_{k+1} = + \rightarrow  s_{k+1} = +
				\\
				\\
				\\
			\end{casesnew}  \\[80pt]
		\\
			\C{s}_k = +_\theta 
			\begin{casesnew}{$p_j$}{$1-p_j$}
				m_{k+1} = -  
					\begin{casesnew}{$p_l$}{$1-p_l$}
						s_{k+1} = -_\theta  \\[40pt]
						s_{k+1} = +_\theta
					\end{casesnew} \\[40pt]
				m_{k+1} = + \rightarrow  s_{k+1} = +
				\\
				\\
				\\
			\end{casesnew} 
		\end{casesnewbig} 
		\\
		m_k = \C{+} 
	\end{cases}
\end{align}
\begin{tikzpicture}[overlay,remember picture,>=latex,shorten >=1pt,shorten <=1pt,very thin]
	\draw[->] (+) to[out=0, in=-140] (s);
\end{tikzpicture}

Where the probabilities are on the arrows and are equal to $1$ if no value is given.
Therefore, the outcomes themselves are such that, 
\vspace{-2pt}
\begin{align}
    \label{eq:mk}
	\hspace{-20pt} m_{k+1}=\begin{cases}
		+ \text{ with probability } 
		\begin{cases}
			1- p_j  &\text{ if }m_{k}=+, 
			\\(1- p_j)(1-p_l)+ p_jp_l &\text{ if } m_{k}=-, 
		\end{cases}
		\\
		- \text{ with probability }
		\begin{cases}
			p_j &\text{ if }m_{k}=+, 
			\\p_l(1- p_j) + p_j(1- p_l) &\text{ if }m_{k}=-, 
		\end{cases}
	\end{cases}
\end{align}
i.e., always are a Markovian process.
Importantly, given that the operator knows if information has been processed in the previous cycle or not, the conditional probability to obtain a specific value for $m_{k+1}$, given $m_k$, reads: 
\begin{align}
\label{eq:p_mk+1}
\begin{matrix*}[l]
&P(m_{k+1}=+\hspace{2pt}|\ell_k=1, m_k=+) = 1-p_j , \quad \quad & 
 P(m_{k+1}=+\hspace{2pt}|\ell_k=0, m_k=+) = 1-p_j ,  \\  
&P(m_{k+1}=+\hspace{2pt}|\ell_k=1, m_k=-) = p_j   , \quad \quad & 
 P(m_{k+1}=+\hspace{2pt}|\ell_k=0, m_k=-) = 1-p_j ,  \\
&P(m_{k+1}=-\hspace{2pt}|\ell_k=1, m_k=+) = p_j   , \quad \quad & 
 P(m_{k+1}=-\hspace{2pt}|\ell_k=0, m_k=+) = p_j   ,  \\
&P(m_{k+1}=-\hspace{2pt}|\ell_k=1, m_k=-) = 1-p_j , \quad \quad & 
 P(m_{k+1}=-\hspace{2pt}|\ell_k=0, m_k=-) = p_j  ,  
\end{matrix*}
\end{align}
as $\ell_k =1$, resp. $\ell_k =0$, amounts in taking $p_l=1$, resp. $p_l=0$, in Eqs.~\eqref{eq:skmk} and \eqref{eq:mk}.

Although the information processing tape does not lead to a work cost, it can be used to reduce the one of processing the measurement outcome tape. Knowing the $\{\ell_1,\cdots,\ell_2\}$ not only affects the measurement outcome probabilities, as shown in Eq.~\ref{eq:p_mk+1}, but it also allows reusing some measurement outcome memory cells, thereby diminishing their total number. Indeed, the memory cell containing the outcome of the cycle $k$ can be used again for the next cycle $k+1$ if $l_{k+1}=1$, i.e. if the cell is not used for the feedback step of the cycle $k+1$. These two effects are taken into account separately. First, we derive the entropy rate per cycle of the measurement outcomes, adapted to an external user using the stochastic value of the laziness~\cite{coverElementsInformationTheory1991}:
\begin{align}
\label{eq:entropyrate}
\dot{S}= \lim_{t\rightarrow\infty} \frac{S(M|L)}{t} &= -\lim_{t\rightarrow\infty} \sum_{\ell_1,\cdots,\ell_t} \sum_{m_1,\cdots,m_t}  \frac{P(m_1,\cdots,m_t,\ell_1,\cdots,\ell_t)\ln(P(m_1,\cdots,m_t|\ell_1,\cdots,\ell_t))}{n} \\
&= -\lim_{t\rightarrow\infty} \sum_{\ell_1,\cdots,\ell_t} P(\ell_1,\cdots,\ell_t) \sum_{m_1,\cdots,m_t}  \frac{P(m_1,\cdots,m_t|\ell_1,\cdots,\ell_t)\ln(P(m_1,\cdots,m_t|\ell_1,\cdots,\ell_t))}{n}.\nonumber 
\end{align}
where $\ell_k \in \{0,1\}$ and $m_k \in \{+,-\}$ for all $k$, $M$ and $L$ are vectors of random variables in the set of all possibles $\{m_1,\cdots, m_t\}$ and $\{\ell_1,\cdots, \ell_t\}$, and $S(X|Y)= - \sum_{x, y} P(x,y) \ln(P(x|y))$ is the conditional entropy or $X$ given $Y$.
Using the Markovianity of the measurement outcomes, we have:
\begin{align}
\label{eq:S(m|l)}
 - \hspace{-5pt}\sum_{m_1, \cdots , m_t}\hspace{-7pt} & P(m_1, ..., m_t|\ell_1,...,\ell_t) \hspace{7pt}\ln(P(m_1, ..., m_t| \ell_1,...,\ell_t) ) \nonumber \\
 	=&- \hspace{-10pt}\sum_{m_1, \cdots, m_t} \hspace{-5pt}P(m_1) P( m_2|m_1,\ell_1)  .. P (m_t|m_{t-1},\ell_{n-1}) \hspace{5pt} \ln(P(m_1) P( m_2|m_1,\ell_1) .. P (m_t|m_{t-1},\ell_{n-1}) ) \nonumber \\ 
 	=&- \sum_{m_1} P(m_1) \ln(P(m_1)) -   \hspace{-10pt}\sum_{m_1, \cdots , m_t} \hspace{-3pt}  P(m_1) P( m_2|m_1,\ell_1)  ... P (m_t|m_{t-1},\ell_{n-1}) \sum_{k=1}^{t-1}  \ln( P( m_{k+1}|m_k,\ell_k))\nonumber \\ 
 	 =& \hspace{5pt} S(p_j)-  \sum_{k=1}^{t-1}  \hspace{5pt} \sum_{m_k, m_{k+1}} \hspace{-5pt} P(m_k) P( m_{k+1}|m_k,\ell_k) \ln( P( m_{k+1}|m_k,\ell_k))
\end{align}
with $S(p_j)$ the binary Shannon entropy with probability $p_j$ (in nats).
Hence, since the $\ell_k$ are i.i.d., i.e., $P(\ell_1,\cdots,\ell_t)= P(\ell_1)P(\ell_2)...P(\ell_t)$, using Eqs.~\eqref{eq:p_mk+1} and \eqref{eq:S(m|l)}: 
\begin{align}
\label{eq:entropyRate_deriv}
- \sum_{\ell_1,\cdots,\ell_t} & P(\ell_1,\cdots,\ell_t) \sum_{m_1,\cdots,m_t}  \frac{P(m_1,\cdots,m_t|\ell_1,\cdots,\ell_t)\ln(P(m_1,\cdots,m_t|\ell_1,\cdots,\ell_t))}{n} \nonumber \\
=&  \sum_{\ell_1,\cdots, \ell_t}  \hspace{-5pt} P(\ell_1,\ell_2,..\ell_t)\frac{S(p_j)-  \sum_{k=1}^{t-1}  \hspace{5pt} \sum_{m_k, m_{k+1}}  P(m_k) P( m_{k+1}|m_k,\ell_k) \ln( P( m_{k+1}|m_k,\ell_k))}{n} \nonumber  \\
=&  \frac{1}{t} \left[S(p_j) -  \sum_{k=1}^{n-1}\sum_{\ell_k}    \hspace{5pt} \sum_{m_k, m_{k+1}}  P(\ell_k) P(m_k) P( m_{k+1}|m_k,\ell_k) \ln( P( m_{k+1}|m_k,\ell_k)) \right] \nonumber \\
=&  \frac{1}{t} \left[ S(p_j) - (1-p_l)(1-p_j) \ln( 1-p_j) \sum_{k=1}^{t-1} P(m_k=+)   \quad - (1-p_l)(1-p_j) \ln(1-p_j) \sum_{k=1}^{t-1} P(m_k=-)   \right. \nonumber \\
& \quad \quad- (1-p_l)p_j \ln(p_j)\sum_{k=1}^{t-1} P(m_k=-)  \quad - (1-p_l)p_j \ln(p_j)\sum_{k=1}^{t-1} P(m_k=+)   \nonumber \\
& \quad \quad- p_l(1-p_j) \ln(1-p_j)\sum_{k=1}^{t-1} P(m_k=+)  \quad - p_l p_j \ln(p_j)\sum_{k=1}^{t-1} P(m_k=-)   \nonumber \\
& \quad \quad \left. -\hspace{2pt} p_l (1-p_j)\ln(1-p_j) \sum_{k=1}^{t-1} P(m_k=-) \quad  - p_lp_j \ln(p_j) \sum_{k=1}^{t-1} P(m_k=+)  \right]  \nonumber \\
=&  \frac{1}{t} \left[ S(p_j) - (t-1)(1-p_l)(1-p_j)\ln( 1-p_j) - (t-1)(1-p_l)p_j\ln(p_j) \right. \nonumber \\
& \quad - \left. (t-1)p_l(1-p_j)\ln(1-p_j) - (t-1)p_lp_j\ln(p_j) \right] \nonumber \\
=& S(p_j)
\end{align}
The entropy rate per cycle therefore is $\dot{S}= S(p_j)$. 
However, due to our protocol, not all this information content needs to be erased, indeed, when $\ell_{k+1}=1$, no erasure needs to be done in cycle $k$. The quantity to consider for the energetic cost of erasure, in place of Eq.~\eqref{eq:S(m|l)} therefore, is 
\begin{align}
\delta_{\ell_2, 0} S(p_j)-  \sum_{k=1}^{t-1}  \hspace{5pt}  \sum_{m_k, m_{k+1}} \hspace{-5pt} \delta_{\ell_{k+1},0} P(m_k) P( m_{k+1}|m_k,\ell_k) \ln( P( m_{k+1}|m_k,\ell_k))
\end{align}
which averaged over all $\ell_1,\cdots, \ell_t$ and divided by $t$ leads to the entropy rate per cycle of $(1-p_l)S(p_j)$ instead of Eq.~\eqref{eq:entropyRate_deriv}, given that $P(\ell_{k+1}=0)= 1-p_l$ for all $k\in\{1,n-1\}$. Equivalently, we can say that $n$-consecutive cycles only require an average of $(1-p_l)n$ binary memory cells for storing the measurement outcomes and hence have an entropy rate per cycle of $(1-p_l)\dot{S}$. 
At a result, the average information processing cost per cycle reads: 
\begin{align}
\langle W_\text{info} \rangle = (1-p_l)k_B T S(p_j).
\end{align}

\subsection{Power extracted}
\label{subsec:power}
From the last two subsections, we can derive the net average work extracted at cycle $t$, which reads 
$ \langle \delta W_t \rangle -\langle W_\text{info} \rangle $. 
In order to obtain the power, one has to divide $ \langle \delta W_t \rangle -\langle W_\text{info} \rangle $ by the duration of a cycle $\tau$. We assume that the measurement duration, erasure duration and feedback duration are negligible compared to the work extraction duration $\tau$. Importantly, we have the following relations: 
\begin{align}
    \theta  = \Omega \tau \, ; \quad 
    W_\theta = \frac{\hbar\omega}{2}\sin(\theta) \, ; \quad
    p_j = \sin^2(\theta/2)
\end{align}
such that: 
\begin{align}
 \langle \dot{W}_t\rangle &= \frac{\langle \delta W_t \rangle -\langle W_\text{info} \rangle }{\tau} \nonumber \\
 &= \frac{\Omega}{\theta}\left[\frac{\hbar\omega}{2}\sin(\theta)\left(\frac{(1-p_l)}{ 1-p_l + 2p_j p_l } + \frac{ 2p_j p_l(1-2 p_j)^{t} p_l^{t} }{(1-p_l + 2 p_jp_l)}\right) - k_BT(1-p_l)S(p_j)\right]. 
\end{align}
When $p_l<1$, there is a bias towards work extraction and therefore the extracted power can reach a non-zero steady state value: 
\begin{align}
\label{eq:Wdotss}
 \langle \dot{W}\rangle_{ss} &= \frac{\langle \delta W \rangle_{ss} -\langle W_\text{info} \rangle }{\tau} \nonumber \\
 &= \frac{\Omega}{\theta}\left[\frac{\hbar\omega}{2}\sin(\theta)\frac{(1-p_l)}{ 1-p_l + 2p_j p_l }  - k_BT(1-p_l)S(p_j)\right] \nonumber \\
 &= \frac{\Omega (1-p_l)}{\theta}
 \left[\frac{\hbar\omega}{2}\frac{\sin(\theta)}{ 1-p_l + 2\sin^2(\theta/2) p_l }  - k_B TS(\sin^2(\theta/2))\right].
\end{align}

\subsection{Optimal $p_l$}
\label{subsec:optpl}
Taking fixed values for the coupling strength to the field $\Omega$, the temperature $T$ and bare qubit energy level spacing $\hbar \omega$, one can find the optimal laziness probability $p_l$ given a value for the angle $\theta$, or equivalently the cycle duration $\tau$ or jump probability $p_j$. 
\begin{figure}[!htbp]
\centering
\includegraphics[width=15cm]{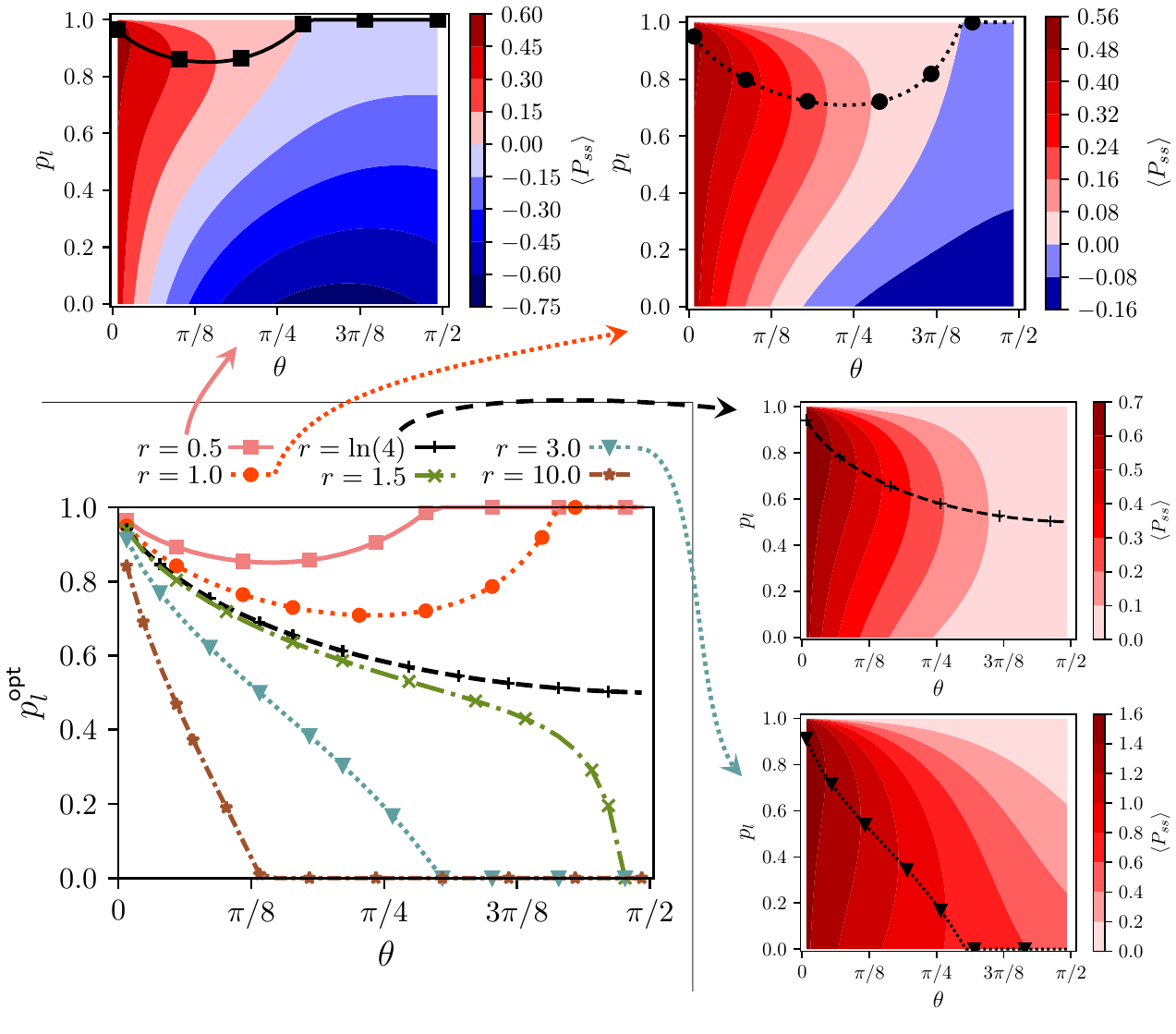}
\vspace{-10pt}
\caption{Colour plots: Analytical average power extracted in the steady state, obtained from Eq.~\eqref{eq:Wdotss}, with respect to $\theta$ and $p_l$ for different values of the ratio $r$. The black lines indicate the optimal $p_l$, obtained from Eq.~\eqref{eq:popt}, as summarized in the bottom left panel (also in the main text).}
\label{fig:SS Power}
\end{figure}

The derivative of $\langle \dot{W}\rangle_{ss}$ with respect to $p_l$ reads: 
\begin{align}
 \frac{\partial\langle \dot{W}\rangle_{ss}}{\partial p_l}
 &= \frac{\Omega \sin(\theta)}{\theta}
 \frac{\hbar\omega}{2}
 \frac{\partial}{\partial p_l}\left(\frac{(1-p_l)}{ 1-p_l + 2\sin^2(\theta/2) p_l }\right)  + \frac{\Omega }{\theta}k_BTS(\sin^2(\theta/2)) \nonumber \\
 &= \frac{\Omega \sin(\theta)}{\theta}
 \frac{\hbar\omega}{2}
 \left(\frac{-(1-p_l + 2\sin^2(\theta/2) p_l )-(1-p_l)(-1 + 2\sin^2(\theta/2))}{ (1-p_l + 2\sin^2(\theta/2) p_l)^2}\right)  + \frac{\Omega }{\theta}k_BTS(\sin^2(\theta/2)) \nonumber \\
 &= \frac{\Omega \sin(\theta)}{\theta}
 \frac{\hbar\omega}{2}
 \left(\frac{-1+p_l - 2\sin^2(\theta/2) p_l +1- 2\sin^2(\theta/2)-p_l+p_l 2\sin^2(\theta/2))}{ (1-p_l + 2\sin^2(\theta/2) p_l)^2}\right)  + \frac{\Omega }{\theta}k_B TS(\sin^2(\theta/2)) \nonumber \\
 &= \frac{\Omega \sin(\theta)}{\theta}
 \frac{\hbar\omega}{2}
 \left(\frac{ - 2\sin^2(\theta/2)}{ (1-p_l + 2\sin^2(\theta/2) p_l)^2}\right)  + \frac{\Omega }{\theta}k_BTS(\sin^2(\theta/2)) \nonumber \\
 &= \frac{\Omega }{\theta}\left[k_BTS(\sin^2(\theta/2))-\frac{ \hbar\omega\sin(\theta)\sin^2(\theta/2)}{ (1-p_l + 2\sin^2(\theta/2) p_l)^2}  \right], 
\end{align}
where $S(\sin^2(\theta/2))= - \sin^2(\theta/2)\ln(\sin^2(\theta/2))- \cos^2(\theta/2)\ln(\cos^2(\theta/2))$ is the binary Shannon entropy with probability $p_j$ (in nats). 
This derivative vanishes when:
\begin{align}
\label{eq:popt}
 \frac{\partial\langle \dot{W}\rangle_{ss}}{\partial p_l} =0 
 \Leftrightarrow & \quad k_BTS(\sin^2(\theta/2))=\frac{ \hbar\omega\sin(\theta)\sin^2(\theta/2)}{ (1-p_l + 2\sin^2(\theta/2) p_l)^2}  \nonumber \\
 \Leftrightarrow & \quad p_l( 2\sin^2(\theta/2) -1) = \sqrt{\frac{\hbar\omega}{k_BT} \frac{\sin(\theta)\sin^2(\theta/2)}{S(\sin^2(\theta/2))}} -1 
 \Leftrightarrow  \quad p_l = p_l^{\rm{opt}} = \frac{1-\sqrt{\frac{\hbar\omega}{k_BT} \frac{\sin(\theta)\sin^2(\theta/2)}{S(\sin^2(\theta/2))}} }{\cos(\theta)}  
\end{align}
We now analyse how this optimal laziness $p_l^{\rm{opt}}$ depends on the ratio $r= \frac{\hbar\omega}{k_B T}$. 
\\
\textbf{When $r= \frac{\hbar\omega}{k_BT}\gg 1$}, the best it to take $p_l=0$, i.e., to use information as much as possible because the erasure cost is negligible compared to the energy gained from the measurement at each cycle. 
\\
\textbf{When $r= \frac{\hbar\omega}{k_BT}\ll 1$}, i.e., in the classical limit, the best it to take $p_l=1$, i.e., to use information as little as possible because the erasure cost is much higher than the energy gained from the measurement at each cycle.
\\
Depending on the value of $r$, when $\theta\rightarrow \pi/2$, the optimal value $p_l^{\rm{opt}}$ diverges to $+\infty$ (hence one should take $p_l=1$) or $-\infty$ (where one should take $p_l=0$), except for one critical value of the ratio: $r_c$ for which: 
\begin{align}
1= r_c \frac{\sin(\theta)\sin^2(\theta/2)}{S(\sin^2(\theta/2))} = r_c \frac{1/2}{\ln(2)}\Leftrightarrow r_c= 2\ln(2)=\ln(4).
\end{align}

\section{Influence of the protocol step order}
\label{SM-sec:Protocol_step_order}
\subsection{Protocol: Measure $\rightarrow$ Feedback $\rightarrow$ Extract $\rightarrow$ Erase (main text)}

When the protocol consists in : (i) Measuring, (ii) applying feedback, (iii) extracting work and (iv) erasing the memory, the Master equation reads: 
\begin{align}
	P_{t+1}(n,+_\theta) &= p_r P_t(n-1,+_\theta) + p_j P_t(n-1,-_\theta), \label{eq:Master+_RTP}\\
	P_{t+1}(n,-_\theta) &= p_r P_t(n+1,-_\theta) + p_j P_t(n+1,+_\theta) , \label{eq:Master-_RTP}.
\end{align}
Notice that we always start and end a cycle in the state $\ket{+_\theta}$ or $\ket{-_\theta}$. 
The advantage of this step ordering is that, since we do the measurement before the work extraction, the first cycle is not different from the other in that the work extracted is stochastic. 
In order for the information processing cost to be $(1-p_l) k_BT S(p_j)$ we need to erase the memory encoding the measurement outcomes only when the next cycle will need to use its measurement outcomes results for the feedback. 
This means that we need to know whether the next cycle will use the measurement information or not in the next cycle to implement this erasure step. This is related to the fact that when $p_l\in(0,1)$, the cycles are dependents on each other. 
Notice that to know when to do the feedback or not, we can simply start a list of "0"s and "1"s such that one of them appears $p_l$ times the total number of digits, over many trajectories. Since this list is unaffected by the experiment, it does not need to be erased and hence has no work cost. 

\subsection{Protocol: Extract $\rightarrow$ Measure $\rightarrow$ Feedback $\rightarrow$ Erase (as in Ref.~\cite{elouardExtractingWorkQuantum2017_SM})}
When the protocol consists in : (i)  extracting work, (ii) Measuring, (iii) applying feedback and (iv) erasing the memory, the Master equation reads: 
\begin{align}
	P_{t+1}(n,+) &= p_r P_t(n-1,+) + p_j P_t(n+1,-), \label{eq:Master+_Elouard}\\
	P_{t+1}(n,-) &= p_r P_t(n+1,-) + p_j P_t(n-1,+) , \label{eq:Master-_Elouard}
\end{align}
Notice that we always start and end a cycle in the state $\ket{+}$ or $\ket{-}$. 

This step order is the one considered in ``Extracting work from quantum measurement in Maxwell’s demon engines"~\cite{elouardExtractingWorkQuantum2017_SM}. Although this former work mainly deals with the perfect and systematic feedback case, $p_l=0$, it briefly comments on the so-called open loop scenario, $p_l=1$. Specifically, it says, ``Reciprocally, it is possible to extract some positive work without feedback during some finite time before the engine switches off", in agreement with our results. 
In their SM, section II, they give analytical results for the average work increments and work, in the deterministic case $p_l=1$ for $\theta \rightarrow 0$, i.e., of a vanishing work extraction step duration. 
Calling the quantities derived in~\cite{elouardExtractingWorkQuantum2017_SM} with the subscript ${\rm EHHA}$ for the name of the authors, the average work read
\begin{align}
\langle \delta W_t \rangle_{\rm EHHA}  &= \frac{\hbar \omega_0}{2} \theta \cos^{t}(\theta)  \\
\langle W_t  \rangle_{\rm EHHA} &= \sum_{k=0}^t \langle \delta W_k \rangle = \frac{\hbar \omega_0}{2} \theta \frac{1-\cos^{t+1}(\theta)}{1-\cos(\theta)}.  
\end{align}
These quantities are consistent with ours, given that, due to the different ordering of the cycle steps, their first cycle deterministically leads to a work extraction of $W_\theta$, $\langle \delta W_{t=0} \rangle_{\rm EHHA} = \hbar \omega_0 \theta/2 \underset{\theta \rightarrow 0} {\approx} W_\theta$, whilst we start at the next cycle, $t=1$. Our work extracted, $W_t$, hence results in summing $\delta W_k$ from $k=1$ to $k=t$, leading to: 
\begin{align}
\langle \delta  W_t \rangle  &= (p_r-p_j)^t W_\theta \\
\langle W_t  \rangle &= \sum_{k=1}^t \langle \delta W_k \rangle = W_\theta \left[\frac{1-(p_r-p_j)^{t+1}}{2p_j} -1 \right]. 
\end{align}
We retrieve the above equation by adding the deterministic contribution to our extracted work, i.e., $\langle W_t  \rangle + W_\theta \underset{\theta \rightarrow 0}{ =} \langle W_t  \rangle_{\rm EHHA}$ given that $W_\theta= \hbar \omega_0 \sin(\theta)/2 \underset{\theta\ll1}{\rightarrow} \hbar \omega_0 \theta/2 $, that 
$p_j = \cos^2(\theta/2)$ and therefore that $(p_r-p_j) = 1-2\cos^2(\theta/2) = \cos\theta$. 

\section{Other generalizations}
\label{SM-sec:Generalization}
\subsection{Meaning of $p_l$}
We introduced $(1-p_l)$ as the probability to process the measurement outcome information during the feedback step. We considered it as a tunable parameter that one can adjust to lower the memory resetting cost, in order to take advantage of the bias given by the initial preparation of the working substance. It can also be a parameter used to model experiments for which the information processing cannot be implemented in each cycle or even at all. Going towards the limit $p_l\rightarrow 1$ also increasingly means that we are using the measurement for its backaction on the measured system state rather than for its information extraction consequence. 

Note that $p_l$ could also be interpreted as an error in the feedback process: wrongly applying the feedback associated to the outcome $+$ when the outcome actually is $-$. The error could also come from the measurement itself, which can mistakenly give the outcome $+$, although the state was $-$. In these two cases, however, all but our energetic analysis would hold,  since the information erasure cost could not be avoided if the occurrence of these effects is not known in advance. 

\subsection{Application to other engines}
Our results are not restricted to the sole single qubit QME considered in the main text. Indeed, they apply equally well even to two-qubits QME, such as the one introduced in ``A two-qubit engine fueled by entangling operations and local measurements"~\cite{bresqueTwoQubitEngineFueled2021_SM}. In this work, only one of the two qubits is measured and there are also two possible measurement outcomes, one leading to work extracting when the correct feedback step is applied and one leading to no work extracting upon applying the appropriate feedback process. The feedback process, in this unlucky case, simply consist in doing nothing. When the feedback is not suited to the post measurement state, i.e., when we try to extract work from the less favourable state, this leads to exactly the same amount of work lost than the work gained when the good feedback is applied to the ``good" state. 

More generally, our results only require an engine for which there are 
\begin{itemize}
    \item Two possible measurement outcomes.
    \item No heat baths, except for the resetting step. 
    \item The same amount of work gained when the correct feedback loop is applied to the ``lucky" state than that which is lost when the same feedback is applied to the ``unlucky" state.
\end{itemize}


\begin{thebibliography}{34}%
\makeatletter
\providecommand \@ifxundefined [1]{%
 \@ifx{#1\undefined}
}%
\providecommand \@ifnum [1]{%
 \ifnum #1\expandafter \@firstoftwo
 \else \expandafter \@secondoftwo
 \fi
}%
\providecommand \@ifx [1]{%
 \ifx #1\expandafter \@firstoftwo
 \else \expandafter \@secondoftwo
 \fi
}%
\providecommand \natexlab [1]{#1}%
\providecommand \enquote  [1]{``#1''}%
\providecommand \bibnamefont  [1]{#1}%
\providecommand \bibfnamefont [1]{#1}%
\providecommand \citenamefont [1]{#1}%
\providecommand \href@noop [0]{\@secondoftwo}%
\providecommand \href [0]{\begingroup \@sanitize@url \@href}%
\providecommand \@href[1]{\@@startlink{#1}\@@href}%
\providecommand \@@href[1]{\endgroup#1\@@endlink}%
\providecommand \@sanitize@url [0]{\catcode `\\12\catcode `\$12\catcode `\&12\catcode `\#12\catcode `\^12\catcode `\_12\catcode `\%12\relax}%
\providecommand \@@startlink[1]{}%
\providecommand \@@endlink[0]{}%
\providecommand \url  [0]{\begingroup\@sanitize@url \@url }%
\providecommand \@url [1]{\endgroup\@href {#1}{\urlprefix }}%
\providecommand \urlprefix  [0]{URL }%
\providecommand \Eprint [0]{\href }%
\providecommand \doibase [0]{https://doi.org/}%
\providecommand \selectlanguage [0]{\@gobble}%
\providecommand \bibinfo  [0]{\@secondoftwo}%
\providecommand \bibfield  [0]{\@secondoftwo}%
\providecommand \translation [1]{[#1]}%
\providecommand \BibitemOpen [0]{}%
\providecommand \bibitemStop [0]{}%
\providecommand \bibitemNoStop [0]{.\EOS\space}%
\providecommand \EOS [0]{\spacefactor3000\relax}%
\providecommand \BibitemShut  [1]{\csname bibitem#1\endcsname}%
\let\auto@bib@innerbib\@empty
\bibitem [{\citenamefont {Leff}\ and\ \citenamefont {Rex}(2002)}]{leff2002maxwell}%
  \BibitemOpen
  \bibfield  {author} {\bibinfo {author} {\bibfnamefont {H.}~\bibnamefont {Leff}}\ and\ \bibinfo {author} {\bibfnamefont {A.~F.}\ \bibnamefont {Rex}},\ }\href {https://doi.org/10.1201/9781420033991} {\emph {\bibinfo {title} {Maxwell's Demon 2 Entropy, Classical and Quantum Information, Computing}}}\ (\bibinfo  {publisher} {CRC Press},\ \bibinfo {year} {2002})\BibitemShut {NoStop}%
\bibitem [{\citenamefont {Parrondo}(2001)}]{parrondoSzilardEngineRevisited2001}%
  \BibitemOpen
  \bibfield  {author} {\bibinfo {author} {\bibfnamefont {J.~M.~R.}\ \bibnamefont {Parrondo}},\ }\bibfield  {title} {\bibinfo {title} {The {{Szilard}} engine revisited: {{Entropy}}, macroscopic randomness, and symmetry breaking phase transitions},\ }\href {https://doi.org/10.1063/1.1388006} {\bibfield  {journal} {\bibinfo  {journal} {Chaos}\ }\textbf {\bibinfo {volume} {11}},\ \bibinfo {pages} {725} (\bibinfo {year} {2001})}\BibitemShut {NoStop}%
\bibitem [{\citenamefont {B{\'e}rut}\ \emph {et~al.}(2012)\citenamefont {B{\'e}rut}, \citenamefont {Arakelyan}, \citenamefont {Petrosyan}, \citenamefont {Ciliberto}, \citenamefont {Dillenschneider},\ and\ \citenamefont {Lutz}}]{berutExperimentalVerificationLandauer2012}%
  \BibitemOpen
  \bibfield  {author} {\bibinfo {author} {\bibfnamefont {A.}~\bibnamefont {B{\'e}rut}}, \bibinfo {author} {\bibfnamefont {A.}~\bibnamefont {Arakelyan}}, \bibinfo {author} {\bibfnamefont {A.}~\bibnamefont {Petrosyan}}, \bibinfo {author} {\bibfnamefont {S.}~\bibnamefont {Ciliberto}}, \bibinfo {author} {\bibfnamefont {R.}~\bibnamefont {Dillenschneider}},\ and\ \bibinfo {author} {\bibfnamefont {E.}~\bibnamefont {Lutz}},\ }\bibfield  {title} {\bibinfo {title} {Experimental verification of {{Landauer}}'s principle linking information and thermodynamics},\ }\href {https://doi.org/10.1038/nature10872} {\bibfield  {journal} {\bibinfo  {journal} {Nature}\ }\textbf {\bibinfo {volume} {483}},\ \bibinfo {pages} {187} (\bibinfo {year} {2012})}\BibitemShut {NoStop}%
\bibitem [{\citenamefont {Parrondo}\ \emph {et~al.}(2015)\citenamefont {Parrondo}, \citenamefont {Horowitz},\ and\ \citenamefont {Sagawa}}]{parrondoThermodynamicsInformation2015}%
  \BibitemOpen
  \bibfield  {author} {\bibinfo {author} {\bibfnamefont {J.~M.~R.}\ \bibnamefont {Parrondo}}, \bibinfo {author} {\bibfnamefont {J.~M.}\ \bibnamefont {Horowitz}},\ and\ \bibinfo {author} {\bibfnamefont {T.}~\bibnamefont {Sagawa}},\ }\bibfield  {title} {\bibinfo {title} {Thermodynamics of information},\ }\href {https://doi.org/10.1038/nphys3230} {\bibfield  {journal} {\bibinfo  {journal} {Nature Physics}\ }\textbf {\bibinfo {volume} {11}},\ \bibinfo {pages} {131} (\bibinfo {year} {2015})}\BibitemShut {NoStop}%
\bibitem [{\citenamefont {Strasberg}\ \emph {et~al.}(2017)\citenamefont {Strasberg}, \citenamefont {Schaller}, \citenamefont {Brandes},\ and\ \citenamefont {Esposito}}]{strasberg2017quantum}%
  \BibitemOpen
  \bibfield  {author} {\bibinfo {author} {\bibfnamefont {P.}~\bibnamefont {Strasberg}}, \bibinfo {author} {\bibfnamefont {G.}~\bibnamefont {Schaller}}, \bibinfo {author} {\bibfnamefont {T.}~\bibnamefont {Brandes}},\ and\ \bibinfo {author} {\bibfnamefont {M.}~\bibnamefont {Esposito}},\ }\bibfield  {title} {\bibinfo {title} {Quantum and information thermodynamics: A unifying framework based on repeated interactions},\ }\href {https://doi.org/10.1103/PhysRevX.7.021003} {\bibfield  {journal} {\bibinfo  {journal} {Physical Review X}\ }\textbf {\bibinfo {volume} {7}},\ \bibinfo {pages} {021003} (\bibinfo {year} {2017})}\BibitemShut {NoStop}%
\bibitem [{\citenamefont {Pekola}\ and\ \citenamefont {Khaymovich}(2019)}]{pekola2019thermodynamics}%
  \BibitemOpen
  \bibfield  {author} {\bibinfo {author} {\bibfnamefont {J.~P.}\ \bibnamefont {Pekola}}\ and\ \bibinfo {author} {\bibfnamefont {I.~M.}\ \bibnamefont {Khaymovich}},\ }\bibfield  {title} {\bibinfo {title} {Thermodynamics in single-electron circuits and superconducting qubits},\ }\href {https://doi.org/10.1146/annurev-conmatphys-033117-054120} {\bibfield  {journal} {\bibinfo  {journal} {Annual Review of Condensed Matter Physics}\ }\textbf {\bibinfo {volume} {10}},\ \bibinfo {pages} {193} (\bibinfo {year} {2019})}\BibitemShut {NoStop}%
\bibitem [{\citenamefont {Bechhoefer}(2021)}]{bechhoefer2021control}%
  \BibitemOpen
  \bibfield  {author} {\bibinfo {author} {\bibfnamefont {J.}~\bibnamefont {Bechhoefer}},\ }\href@noop {} {\emph {\bibinfo {title} {Control theory for physicists}}}\ (\bibinfo  {publisher} {Cambridge University Press},\ \bibinfo {year} {2021})\BibitemShut {NoStop}%
\bibitem [{\citenamefont {Garcia-Millan}\ \emph {et~al.}(2024)\citenamefont {Garcia-Millan}, \citenamefont {Sch{\"u}ttler}, \citenamefont {Cates},\ and\ \citenamefont {Loos}}]{garcia2024optimal}%
  \BibitemOpen
  \bibfield  {author} {\bibinfo {author} {\bibfnamefont {R.}~\bibnamefont {Garcia-Millan}}, \bibinfo {author} {\bibfnamefont {J.}~\bibnamefont {Sch{\"u}ttler}}, \bibinfo {author} {\bibfnamefont {M.~E.}\ \bibnamefont {Cates}},\ and\ \bibinfo {author} {\bibfnamefont {S.~A.}\ \bibnamefont {Loos}},\ }\bibfield  {title} {\bibinfo {title} {Optimal closed-loop control of active particles and a minimal information engine},\ }\href@noop {} {\bibfield  {journal} {\bibinfo  {journal} {arXiv preprint arXiv:2407.18542}\ } (\bibinfo {year} {2024})}\BibitemShut {NoStop}%
\bibitem [{\citenamefont {Ding}\ \emph {et~al.}(2018)\citenamefont {Ding}, \citenamefont {Yi}, \citenamefont {Kim},\ and\ \citenamefont {Talkner}}]{dingMeasurementdrivenSingleTemperature2018}%
  \BibitemOpen
  \bibfield  {author} {\bibinfo {author} {\bibfnamefont {X.}~\bibnamefont {Ding}}, \bibinfo {author} {\bibfnamefont {J.}~\bibnamefont {Yi}}, \bibinfo {author} {\bibfnamefont {Y.~W.}\ \bibnamefont {Kim}},\ and\ \bibinfo {author} {\bibfnamefont {P.}~\bibnamefont {Talkner}},\ }\bibfield  {title} {\bibinfo {title} {Measurement-driven single temperature engine},\ }\bibfield  {journal} {\bibinfo  {journal} {Physical Review E}\ }\textbf {\bibinfo {volume} {98}},\ \href {https://doi.org/10.1103/PhysRevE.98.042122} {10.1103/PhysRevE.98.042122} (\bibinfo {year} {2018})\BibitemShut {NoStop}%
\bibitem [{\citenamefont {Yi}\ \emph {et~al.}(2017)\citenamefont {Yi}, \citenamefont {Talkner},\ and\ \citenamefont {Kim}}]{yiSingletemperatureQuantumEngine2017}%
  \BibitemOpen
  \bibfield  {author} {\bibinfo {author} {\bibfnamefont {J.}~\bibnamefont {Yi}}, \bibinfo {author} {\bibfnamefont {P.}~\bibnamefont {Talkner}},\ and\ \bibinfo {author} {\bibfnamefont {Y.~W.}\ \bibnamefont {Kim}},\ }\bibfield  {title} {\bibinfo {title} {Single-temperature quantum engine without feedback control},\ }\bibfield  {journal} {\bibinfo  {journal} {Physical Review E}\ }\textbf {\bibinfo {volume} {96}},\ \href {https://doi.org/10.1103/PhysRevE.96.022108} {10.1103/PhysRevE.96.022108} (\bibinfo {year} {2017})\BibitemShut {NoStop}%
\bibitem [{\citenamefont {Buffoni}\ \emph {et~al.}(2019)\citenamefont {Buffoni}, \citenamefont {Solfanelli}, \citenamefont {Verrucchi}, \citenamefont {Cuccoli},\ and\ \citenamefont {Campisi}}]{buffoniQuantumMeasurementCooling2019}%
  \BibitemOpen
  \bibfield  {author} {\bibinfo {author} {\bibfnamefont {L.}~\bibnamefont {Buffoni}}, \bibinfo {author} {\bibfnamefont {A.}~\bibnamefont {Solfanelli}}, \bibinfo {author} {\bibfnamefont {P.}~\bibnamefont {Verrucchi}}, \bibinfo {author} {\bibfnamefont {A.}~\bibnamefont {Cuccoli}},\ and\ \bibinfo {author} {\bibfnamefont {M.}~\bibnamefont {Campisi}},\ }\bibfield  {title} {\bibinfo {title} {Quantum {{Measurement Cooling}}},\ }\href@noop {} {\bibfield  {journal} {\bibinfo  {journal} {Physical Review Letters}\ }\textbf {\bibinfo {volume} {122}},\ \bibinfo {pages} {5} (\bibinfo {year} {2019})}\BibitemShut {NoStop}%
\bibitem [{\citenamefont {Bresque}\ \emph {et~al.}(2021)\citenamefont {Bresque}, \citenamefont {Camati}, \citenamefont {Rogers}, \citenamefont {Murch}, \citenamefont {Jordan},\ and\ \citenamefont {Auff{\`e}ves}}]{bresqueTwoQubitEngineFueled2021}%
  \BibitemOpen
  \bibfield  {author} {\bibinfo {author} {\bibfnamefont {L.}~\bibnamefont {Bresque}}, \bibinfo {author} {\bibfnamefont {P.~A.}\ \bibnamefont {Camati}}, \bibinfo {author} {\bibfnamefont {S.}~\bibnamefont {Rogers}}, \bibinfo {author} {\bibfnamefont {K.}~\bibnamefont {Murch}}, \bibinfo {author} {\bibfnamefont {A.~N.}\ \bibnamefont {Jordan}},\ and\ \bibinfo {author} {\bibfnamefont {A.}~\bibnamefont {Auff{\`e}ves}},\ }\bibfield  {title} {\bibinfo {title} {Two-{{Qubit Engine Fueled}} by {{Entanglement}} and {{Local Measurements}}},\ }\href {https://doi.org/10.1103/PhysRevLett.126.120605} {\bibfield  {journal} {\bibinfo  {journal} {Phys. Rev. Lett.}\ }\textbf {\bibinfo {volume} {126}},\ \bibinfo {pages} {120605} (\bibinfo {year} {2021})}\BibitemShut {NoStop}%
\bibitem [{\citenamefont {Jussiau}\ \emph {et~al.}(2023)\citenamefont {Jussiau}, \citenamefont {Bresque}, \citenamefont {Auff{\`e}ves}, \citenamefont {Murch},\ and\ \citenamefont {Jordan}}]{jussiauManybodyQuantumVacuum2023}%
  \BibitemOpen
  \bibfield  {author} {\bibinfo {author} {\bibfnamefont {{\'E}.}~\bibnamefont {Jussiau}}, \bibinfo {author} {\bibfnamefont {L.}~\bibnamefont {Bresque}}, \bibinfo {author} {\bibfnamefont {A.}~\bibnamefont {Auff{\`e}ves}}, \bibinfo {author} {\bibfnamefont {K.~W.}\ \bibnamefont {Murch}},\ and\ \bibinfo {author} {\bibfnamefont {A.~N.}\ \bibnamefont {Jordan}},\ }\bibfield  {title} {\bibinfo {title} {Many-body quantum vacuum fluctuation engines},\ }\href {https://doi.org/10.1103/PhysRevResearch.5.033122} {\bibfield  {journal} {\bibinfo  {journal} {Phys. Rev. Research}\ }\textbf {\bibinfo {volume} {5}},\ \bibinfo {pages} {033122} (\bibinfo {year} {2023})}\BibitemShut {NoStop}%
\bibitem [{\citenamefont {Elouard}\ \emph {et~al.}(2017{\natexlab{a}})\citenamefont {Elouard}, \citenamefont {{Herrera-Mart{\'i}}}, \citenamefont {Huard},\ and\ \citenamefont {Auff{\`e}ves}}]{elouardExtractingWorkQuantum2017}%
  \BibitemOpen
  \bibfield  {author} {\bibinfo {author} {\bibfnamefont {C.}~\bibnamefont {Elouard}}, \bibinfo {author} {\bibfnamefont {D.}~\bibnamefont {{Herrera-Mart{\'i}}}}, \bibinfo {author} {\bibfnamefont {B.}~\bibnamefont {Huard}},\ and\ \bibinfo {author} {\bibfnamefont {A.}~\bibnamefont {Auff{\`e}ves}},\ }\bibfield  {title} {\bibinfo {title} {Extracting {{Work}} from {{Quantum Measurement}} in {{Maxwell}}'s {{Demon Engines}}},\ }\bibfield  {journal} {\bibinfo  {journal} {Physical Review Letters}\ }\textbf {\bibinfo {volume} {118}},\ \href {https://doi.org/10.1103/PhysRevLett.118.260603} {10.1103/PhysRevLett.118.260603} (\bibinfo {year} {2017}{\natexlab{a}})\BibitemShut {NoStop}%
\bibitem [{\citenamefont {Brandner}\ \emph {et~al.}(2015)\citenamefont {Brandner}, \citenamefont {Bauer}, \citenamefont {Schmid},\ and\ \citenamefont {Seifert}}]{brandnerCoherenceenhancedEfficiencyFeedbackdriven2015}%
  \BibitemOpen
  \bibfield  {author} {\bibinfo {author} {\bibfnamefont {K.}~\bibnamefont {Brandner}}, \bibinfo {author} {\bibfnamefont {M.}~\bibnamefont {Bauer}}, \bibinfo {author} {\bibfnamefont {M.~T.}\ \bibnamefont {Schmid}},\ and\ \bibinfo {author} {\bibfnamefont {U.}~\bibnamefont {Seifert}},\ }\bibfield  {title} {\bibinfo {title} {Coherence-enhanced efficiency of feedback-driven quantum engines},\ }\href {https://doi.org/10.1088/1367-2630/17/6/065006} {\bibfield  {journal} {\bibinfo  {journal} {New J. Phys.}\ }\textbf {\bibinfo {volume} {17}},\ \bibinfo {pages} {065006} (\bibinfo {year} {2015})}\BibitemShut {NoStop}%
\bibitem [{\citenamefont {Elouard}\ \emph {et~al.}(2017{\natexlab{b}})\citenamefont {Elouard}, \citenamefont {{Herrera-Mart{\'i}}}, \citenamefont {Clusel},\ and\ \citenamefont {Auff{\`e}ves}}]{elouardRoleQuantumMeasurement2017}%
  \BibitemOpen
  \bibfield  {author} {\bibinfo {author} {\bibfnamefont {C.}~\bibnamefont {Elouard}}, \bibinfo {author} {\bibfnamefont {D.~A.}\ \bibnamefont {{Herrera-Mart{\'i}}}}, \bibinfo {author} {\bibfnamefont {M.}~\bibnamefont {Clusel}},\ and\ \bibinfo {author} {\bibfnamefont {A.}~\bibnamefont {Auff{\`e}ves}},\ }\bibfield  {title} {\bibinfo {title} {The role of quantum measurement in stochastic thermodynamics},\ }\href {https://doi.org/10.1038/s41534-017-0008-4} {\bibfield  {journal} {\bibinfo  {journal} {npj Quantum Information}\ }\textbf {\bibinfo {volume} {3}},\ \bibinfo {pages} {9} (\bibinfo {year} {2017}{\natexlab{b}})}\BibitemShut {NoStop}%
\bibitem [{\citenamefont {Sagawa}\ and\ \citenamefont {Ueda}(2009)}]{sagawaMinimalEnergyCost2009}%
  \BibitemOpen
  \bibfield  {author} {\bibinfo {author} {\bibfnamefont {T.}~\bibnamefont {Sagawa}}\ and\ \bibinfo {author} {\bibfnamefont {M.}~\bibnamefont {Ueda}},\ }\bibfield  {title} {\bibinfo {title} {Minimal {{Energy Cost}} for {{Thermodynamic Information Processing}}: {{Measurement}} and {{Information Erasure}}},\ }\bibfield  {journal} {\bibinfo  {journal} {Physical Review Letters}\ }\textbf {\bibinfo {volume} {102}},\ \href {https://doi.org/10.1103/PhysRevLett.102.250602} {10.1103/PhysRevLett.102.250602} (\bibinfo {year} {2009})\BibitemShut {NoStop}%
\bibitem [{\citenamefont {Mamede}\ \emph {et~al.}(2024)\citenamefont {Mamede}, \citenamefont {Singh}, \citenamefont {Pal}, \citenamefont {Fiore},\ and\ \citenamefont {Proesmans}}]{Mamede_2024}%
  \BibitemOpen
  \bibfield  {author} {\bibinfo {author} {\bibfnamefont {I.~N.}\ \bibnamefont {Mamede}}, \bibinfo {author} {\bibfnamefont {P.}~\bibnamefont {Singh}}, \bibinfo {author} {\bibfnamefont {A.}~\bibnamefont {Pal}}, \bibinfo {author} {\bibfnamefont {C.~E.}\ \bibnamefont {Fiore}},\ and\ \bibinfo {author} {\bibfnamefont {K.}~\bibnamefont {Proesmans}},\ }\bibfield  {title} {\bibinfo {title} {Work statistics at first-passage times},\ }\href {https://doi.org/10.1088/1367-2630/ad313d} {\bibfield  {journal} {\bibinfo  {journal} {New Journal of Physics}\ }\textbf {\bibinfo {volume} {26}},\ \bibinfo {pages} {033034} (\bibinfo {year} {2024})}\BibitemShut {NoStop}%
\bibitem [{\citenamefont {Berg}(2004)}]{bergColiMotion2004}%
  \BibitemOpen
  \bibfield  {author} {\bibinfo {author} {\bibfnamefont {H.~C.}\ \bibnamefont {Berg}},\ }\href@noop {} {\emph {\bibinfo {title} {E. Coli in Motion}}},\ Biological and Medical Physics, Biomedical Engineering\ (\bibinfo  {publisher} {Springer [u.a.]},\ \bibinfo {address} {New York Berlin Heidelberg},\ \bibinfo {year} {2004})\BibitemShut {NoStop}%
\bibitem [{\citenamefont {Tailleur}\ and\ \citenamefont {Cates}(2008)}]{tailleurStatisticalMechanicsInteracting2008}%
  \BibitemOpen
  \bibfield  {author} {\bibinfo {author} {\bibfnamefont {J.}~\bibnamefont {Tailleur}}\ and\ \bibinfo {author} {\bibfnamefont {M.~E.}\ \bibnamefont {Cates}},\ }\bibfield  {title} {\bibinfo {title} {Statistical {{Mechanics}} of {{Interacting Run-and-Tumble Bacteria}}},\ }\href {https://doi.org/10.1103/PhysRevLett.100.218103} {\bibfield  {journal} {\bibinfo  {journal} {Phys. Rev. Lett.}\ }\textbf {\bibinfo {volume} {100}},\ \bibinfo {pages} {218103} (\bibinfo {year} {2008})},\ \Eprint {https://arxiv.org/abs/0803.1069} {arXiv:0803.1069 [cond-mat, q-bio]} \BibitemShut {NoStop}%
\bibitem [{\citenamefont {Angelani}\ \emph {et~al.}(2014)\citenamefont {Angelani}, \citenamefont {Di~Leonardo},\ and\ \citenamefont {Paoluzzi}}]{angelaniFirstpassageTimeRuntumble2014}%
  \BibitemOpen
  \bibfield  {author} {\bibinfo {author} {\bibfnamefont {L.}~\bibnamefont {Angelani}}, \bibinfo {author} {\bibfnamefont {R.}~\bibnamefont {Di~Leonardo}},\ and\ \bibinfo {author} {\bibfnamefont {M.}~\bibnamefont {Paoluzzi}},\ }\bibfield  {title} {\bibinfo {title} {First-passage time of run-and-tumble particles},\ }\href {https://doi.org/10.1140/epje/i2014-14059-4} {\bibfield  {journal} {\bibinfo  {journal} {Eur. Phys. J. E}\ }\textbf {\bibinfo {volume} {37}},\ \bibinfo {pages} {59} (\bibinfo {year} {2014})}\BibitemShut {NoStop}%
\bibitem [{\citenamefont {Shreshtha}\ and\ \citenamefont {Harris}(2019)}]{shreshthaThermodynamicUncertaintyRuntumble2019}%
  \BibitemOpen
  \bibfield  {author} {\bibinfo {author} {\bibfnamefont {M.}~\bibnamefont {Shreshtha}}\ and\ \bibinfo {author} {\bibfnamefont {R.~J.}\ \bibnamefont {Harris}},\ }\bibfield  {title} {\bibinfo {title} {Thermodynamic uncertainty for run-and-tumble type processes},\ }\href {https://doi.org/10.1209/0295-5075/126/40007} {\bibfield  {journal} {\bibinfo  {journal} {EPL}\ }\textbf {\bibinfo {volume} {126}},\ \bibinfo {pages} {40007} (\bibinfo {year} {2019})},\ \Eprint {https://arxiv.org/abs/1903.01972} {arXiv:1903.01972 [cond-mat]} \BibitemShut {NoStop}%
\bibitem [{\citenamefont {Tucci}\ \emph {et~al.}(2022)\citenamefont {Tucci}, \citenamefont {Gambassi}, \citenamefont {Majumdar},\ and\ \citenamefont {Schehr}}]{tucciFirstpassageTimeRuntumble2022}%
  \BibitemOpen
  \bibfield  {author} {\bibinfo {author} {\bibfnamefont {G.}~\bibnamefont {Tucci}}, \bibinfo {author} {\bibfnamefont {A.}~\bibnamefont {Gambassi}}, \bibinfo {author} {\bibfnamefont {S.~N.}\ \bibnamefont {Majumdar}},\ and\ \bibinfo {author} {\bibfnamefont {G.}~\bibnamefont {Schehr}},\ }\bibfield  {title} {\bibinfo {title} {First-passage time of run-and-tumble particles with noninstantaneous resetting},\ }\href {https://doi.org/10.1103/PhysRevE.106.044127} {\bibfield  {journal} {\bibinfo  {journal} {Phys. Rev. E}\ }\textbf {\bibinfo {volume} {106}},\ \bibinfo {pages} {044127} (\bibinfo {year} {2022})}\BibitemShut {NoStop}%
\bibitem [{\citenamefont {Das}\ and\ \citenamefont {Giuggioli}(2023)}]{dasDynamicsLatticeRandom2023a}%
  \BibitemOpen
  \bibfield  {author} {\bibinfo {author} {\bibfnamefont {D.}~\bibnamefont {Das}}\ and\ \bibinfo {author} {\bibfnamefont {L.}~\bibnamefont {Giuggioli}},\ }\bibfield  {title} {\bibinfo {title} {Dynamics of lattice random walk within regions composed of different media and interfaces},\ }\href {https://doi.org/10.1088/1742-5468/aca8f9} {\bibfield  {journal} {\bibinfo  {journal} {J. Stat. Mech.}\ }\textbf {\bibinfo {volume} {2023}},\ \bibinfo {pages} {013201} (\bibinfo {year} {2023})}\BibitemShut {NoStop}%
\bibitem [{\citenamefont {Metzler}\ \emph {et~al.}(2014)\citenamefont
  {Metzler}, \citenamefont {Jeon}, \citenamefont {G.~Cherstvy},\ and\
  \citenamefont {Barkai}}]{metzlerAnomalousDiffusionModels2014}%
  \BibitemOpen
  \bibfield  {author} {\bibinfo {author} {\bibfnamefont {R.}~\bibnamefont
  {Metzler}}, \bibinfo {author} {\bibfnamefont {J.-H.}\ \bibnamefont {Jeon}},
  \bibinfo {author} {\bibfnamefont {A.}~\bibnamefont {G.~Cherstvy}},\ and\
  \bibinfo {author} {\bibfnamefont {E.}~\bibnamefont {Barkai}},\ }\bibfield
  {title} {\bibinfo {title} {Anomalous diffusion models and their properties:
  Non-stationarity, non-ergodicity, and ageing at the centenary of single
  particle tracking},\ }\href {https://doi.org/10.1039/C4CP03465A} {\bibfield
  {journal} {\bibinfo  {journal} {Physical Chemistry Chemical Physics}\
  }\textbf {\bibinfo {volume} {16}},\ \bibinfo {pages} {24128} (\bibinfo {year}
  {2014})}\BibitemShut {NoStop}%
\bibitem [{\citenamefont {Metzler}\ and\ \citenamefont
  {Klafter}(2000)}]{metzlerRandomWalksGuide2000}%
  \BibitemOpen
  \bibfield  {author} {\bibinfo {author} {\bibfnamefont {R.}~\bibnamefont
  {Metzler}}\ and\ \bibinfo {author} {\bibfnamefont {J.}~\bibnamefont
  {Klafter}},\ }\bibfield  {title} {\bibinfo {title} {The random walk's guide
  to anomalous diffusion: A fractional dynamics approach},\ }\href
  {https://doi.org/10.1016/S0370-1573(00)00070-3} {\bibfield  {journal}
  {\bibinfo  {journal} {Physics Reports}\ }\textbf {\bibinfo {volume} {339}},\
  \bibinfo {pages} {1} (\bibinfo {year} {2000})}\BibitemShut {NoStop}%
\bibitem [{\citenamefont {Balakrishnan}\ \emph {et~al.}(1988)\citenamefont {Balakrishnan}, \citenamefont {Lakshmibala},\ and\ \citenamefont {Van Den~Broeck}}]{balakrishnanRenewalEquationPersistent1988}%
  \BibitemOpen
  \bibfield  {author} {\bibinfo {author} {\bibfnamefont {V.}~\bibnamefont {Balakrishnan}}, \bibinfo {author} {\bibfnamefont {S.}~\bibnamefont {Lakshmibala}},\ and\ \bibinfo {author} {\bibfnamefont {C.}~\bibnamefont {Van Den~Broeck}},\ }\bibfield  {title} {\bibinfo {title} {The renewal equation for persistent diffusion},\ }\href {https://doi.org/10.1016/0378-4371(88)90101-X} {\bibfield  {journal} {\bibinfo  {journal} {Physica A: Statistical Mechanics and its Applications}\ }\textbf {\bibinfo {volume} {153}},\ \bibinfo {pages} {57} (\bibinfo {year} {1988})}\BibitemShut {NoStop}%
\bibitem [{\citenamefont {Abate}\ and\ \citenamefont {Whitt}(1992)}]{abateNumericalInversionProbability1992}%
  \BibitemOpen
  \bibfield  {author} {\bibinfo {author} {\bibfnamefont {J.}~\bibnamefont {Abate}}\ and\ \bibinfo {author} {\bibfnamefont {W.}~\bibnamefont {Whitt}},\ }\bibfield  {title} {\bibinfo {title} {Numerical inversion of probability generating functions},\ }\href {https://doi.org/10.1016/0167-6377(92)90050-D} {\bibfield  {journal} {\bibinfo  {journal} {Operations Research Letters}\ }\textbf {\bibinfo {volume} {12}},\ \bibinfo {pages} {245} (\bibinfo {year} {1992})}\BibitemShut {NoStop}%
\bibitem [{\citenamefont {Malakar}\ \emph {et~al.}(2018)\citenamefont {Malakar}, \citenamefont {Jemseena}, \citenamefont {Kundu}, \citenamefont {Kumar}, \citenamefont {Sabhapandit}, \citenamefont {Majumdar}, \citenamefont {Redner},\ and\ \citenamefont {Dhar}}]{malakarSteadyStateRelaxation2018}%
  \BibitemOpen
  \bibfield  {author} {\bibinfo {author} {\bibfnamefont {K.}~\bibnamefont {Malakar}}, \bibinfo {author} {\bibfnamefont {V.}~\bibnamefont {Jemseena}}, \bibinfo {author} {\bibfnamefont {A.}~\bibnamefont {Kundu}}, \bibinfo {author} {\bibfnamefont {K.~V.}\ \bibnamefont {Kumar}}, \bibinfo {author} {\bibfnamefont {S.}~\bibnamefont {Sabhapandit}}, \bibinfo {author} {\bibfnamefont {S.~N.}\ \bibnamefont {Majumdar}}, \bibinfo {author} {\bibfnamefont {S.}~\bibnamefont {Redner}},\ and\ \bibinfo {author} {\bibfnamefont {A.}~\bibnamefont {Dhar}},\ }\bibfield  {title} {\bibinfo {title} {Steady state, relaxation and first-passage properties of a run-and-tumble particle in one-dimension},\ }\href {https://doi.org/10.1088/1742-5468/aab84f} {\bibfield  {journal} {\bibinfo  {journal} {J. Stat. Mech.}\ }\textbf {\bibinfo {volume} {2018}},\ \bibinfo {pages} {043215} (\bibinfo {year} {2018})}\BibitemShut {NoStop}%
\bibitem [{\citenamefont {Mori}\ \emph {et~al.}(2020)\citenamefont {Mori}, \citenamefont {Le~Doussal}, \citenamefont {Majumdar},\ and\ \citenamefont {Schehr}}]{moriUniversalPropertiesRuntumble2020}%
  \BibitemOpen
  \bibfield  {author} {\bibinfo {author} {\bibfnamefont {F.}~\bibnamefont {Mori}}, \bibinfo {author} {\bibfnamefont {P.}~\bibnamefont {Le~Doussal}}, \bibinfo {author} {\bibfnamefont {S.~N.}\ \bibnamefont {Majumdar}},\ and\ \bibinfo {author} {\bibfnamefont {G.}~\bibnamefont {Schehr}},\ }\bibfield  {title} {\bibinfo {title} {Universal properties of a run-and-tumble particle in arbitrary dimension},\ }\href {https://doi.org/10.1103/PhysRevE.102.042133} {\bibfield  {journal} {\bibinfo  {journal} {Phys. Rev. E}\ }\textbf {\bibinfo {volume} {102}},\ \bibinfo {pages} {042133} (\bibinfo {year} {2020})}\BibitemShut {NoStop}%
\bibitem [{\citenamefont {Bruyne}\ \emph {et~al.}(2021)\citenamefont {Bruyne}, \citenamefont {Majumdar},\ and\ \citenamefont {Schehr}}]{bruyneSurvivalProbabilityRuntumble2021}%
  \BibitemOpen
  \bibfield  {author} {\bibinfo {author} {\bibfnamefont {B.~D.}\ \bibnamefont {Bruyne}}, \bibinfo {author} {\bibfnamefont {S.~N.}\ \bibnamefont {Majumdar}},\ and\ \bibinfo {author} {\bibfnamefont {G.}~\bibnamefont {Schehr}},\ }\bibfield  {title} {\bibinfo {title} {Survival probability of a run-and-tumble particle in the presence of a drift},\ }\href {https://doi.org/10.1088/1742-5468/abf5d5} {\bibfield  {journal} {\bibinfo  {journal} {J. Stat. Mech.}\ }\textbf {\bibinfo {volume} {2021}},\ \bibinfo {pages} {043211} (\bibinfo {year} {2021})}\BibitemShut {NoStop}%
\bibitem [{\citenamefont {{Garcia-Millan}}\ and\ \citenamefont {Pruessner}(2021)}]{garcia-millanRuntumbleMotionHarmonic2021}%
  \BibitemOpen
  \bibfield  {author} {\bibinfo {author} {\bibfnamefont {R.}~\bibnamefont {{Garcia-Millan}}}\ and\ \bibinfo {author} {\bibfnamefont {G.}~\bibnamefont {Pruessner}},\ }\bibfield  {title} {\bibinfo {title} {Run-and-tumble motion in a harmonic potential: Field theory and entropy production},\ }\href {https://doi.org/10.1088/1742-5468/ac014d} {\bibfield  {journal} {\bibinfo  {journal} {J. Stat. Mech.}\ }\textbf {\bibinfo {volume} {2021}},\ \bibinfo {pages} {063203} (\bibinfo {year} {2021})}\BibitemShut {NoStop}%
\bibitem [{\citenamefont {Gu{\'e}neau}\ \emph {et~al.}(2024)\citenamefont {Gu{\'e}neau}, \citenamefont {Majumdar},\ and\ \citenamefont {Schehr}}]{gueneauRuntumbleParticleOnedimensional2024}%
  \BibitemOpen
  \bibfield  {author} {\bibinfo {author} {\bibfnamefont {M.}~\bibnamefont {Gu{\'e}neau}}, \bibinfo {author} {\bibfnamefont {S.~N.}\ \bibnamefont {Majumdar}},\ and\ \bibinfo {author} {\bibfnamefont {G.}~\bibnamefont {Schehr}},\ }\href {https://doi.org/10.48550/arXiv.2409.16951} {\bibinfo {title} {Run-and-tumble particle in one-dimensional potentials: Mean first-passage time and applications}} (\bibinfo {year} {2024}),\ \Eprint {https://arxiv.org/abs/2409.16951} {arXiv:2409.16951 [cond-mat, physics:math-ph]} \BibitemShut {NoStop}%
\bibitem [{\citenamefont {Maes}\ \emph {et~al.}(2022)\citenamefont {Maes}, \citenamefont {Meerts},\ and\ \citenamefont {Struyve}}]{maes2022diffraction}%
  \BibitemOpen
  \bibfield  {author} {\bibinfo {author} {\bibfnamefont {C.}~\bibnamefont {Maes}}, \bibinfo {author} {\bibfnamefont {K.}~\bibnamefont {Meerts}},\ and\ \bibinfo {author} {\bibfnamefont {W.}~\bibnamefont {Struyve}},\ }\bibfield  {title} {\bibinfo {title} {Diffraction and interference with run-and-tumble particles},\ }\href {https://doi.org/10.1016/j.physa.2022.127323} {\bibfield  {journal} {\bibinfo  {journal} {Physica A: Statistical Mechanics and its Applications}\ }\textbf {\bibinfo {volume} {598}},\ \bibinfo {pages} {127323} (\bibinfo {year} {2022})}\BibitemShut {NoStop}%
\bibitem [{\citenamefont {Krekels}\ \emph {et~al.}(2024)\citenamefont {Krekels}, \citenamefont {Maes}, \citenamefont {Meerts},\ and\ \citenamefont {Struyve}}]{krekels2024zig}%
  \BibitemOpen
  \bibfield  {author} {\bibinfo {author} {\bibfnamefont {S.}~\bibnamefont {Krekels}}, \bibinfo {author} {\bibfnamefont {C.}~\bibnamefont {Maes}}, \bibinfo {author} {\bibfnamefont {K.}~\bibnamefont {Meerts}},\ and\ \bibinfo {author} {\bibfnamefont {W.}~\bibnamefont {Struyve}},\ }\bibfield  {title} {\bibinfo {title} {Zig-zag dynamics in a stern--gerlach spin measurement},\ }\href {https://doi.org/10.1098/rspa.2023.0861} {\bibfield  {journal} {\bibinfo  {journal} {Proceedings of the Royal Society A}\ }\textbf {\bibinfo {volume} {480}},\ \bibinfo {pages} {20230861} (\bibinfo {year} {2024})}\BibitemShut {NoStop}%
\bibitem [{\citenamefont {Kewming}\ \emph {et~al.}(2024)\citenamefont {Kewming}, \citenamefont {Kiely}, \citenamefont {Campbell},\ and\ \citenamefont {Landi}}]{kewmingFirstPassageTimes2024a}%
  \BibitemOpen
  \bibfield  {author} {\bibinfo {author} {\bibfnamefont {M.~J.}\ \bibnamefont {Kewming}}, \bibinfo {author} {\bibfnamefont {A.}~\bibnamefont {Kiely}}, \bibinfo {author} {\bibfnamefont {S.}~\bibnamefont {Campbell}},\ and\ \bibinfo {author} {\bibfnamefont {G.~T.}\ \bibnamefont {Landi}},\ }\bibfield  {title} {\bibinfo {title} {First passage times for continuous quantum measurement currents},\ }\href {https://doi.org/10.1103/PhysRevA.109.L050202} {\bibfield  {journal} {\bibinfo  {journal} {Phys. Rev. A}\ }\textbf {\bibinfo {volume} {109}},\ \bibinfo {pages} {L050202} (\bibinfo {year} {2024})}\BibitemShut {NoStop}%
\end{thebibliography}

\begin{thebibliography}{13}%
\makeatletter
\providecommand \@ifxundefined [1]{%
 \@ifx{#1\undefined}
}%
\providecommand \@ifnum [1]{%
 \ifnum #1\expandafter \@firstoftwo
 \else \expandafter \@secondoftwo
 \fi
}%
\providecommand \@ifx [1]{%
 \ifx #1\expandafter \@firstoftwo
 \else \expandafter \@secondoftwo
 \fi
}%
\providecommand \natexlab [1]{#1}%
\providecommand \enquote  [1]{``#1''}%
\providecommand \bibnamefont  [1]{#1}%
\providecommand \bibfnamefont [1]{#1}%
\providecommand \citenamefont [1]{#1}%
\providecommand \href@noop [0]{\@secondoftwo}%
\providecommand \href [0]{\begingroup \@sanitize@url \@href}%
\providecommand \@href[1]{\@@startlink{#1}\@@href}%
\providecommand \@@href[1]{\endgroup#1\@@endlink}%
\providecommand \@sanitize@url [0]{\catcode `\\12\catcode `\$12\catcode
  `\&12\catcode `\#12\catcode `\^12\catcode `\_12\catcode `\%12\relax}%
\providecommand \@@startlink[1]{}%
\providecommand \@@endlink[0]{}%
\providecommand \url  [0]{\begingroup\@sanitize@url \@url }%
\providecommand \@url [1]{\endgroup\@href {#1}{\urlprefix }}%
\providecommand \urlprefix  [0]{URL }%
\providecommand \Eprint [0]{\href }%
\providecommand \doibase [0]{https://doi.org/}%
\providecommand \selectlanguage [0]{\@gobble}%
\providecommand \bibinfo  [0]{\@secondoftwo}%
\providecommand \bibfield  [0]{\@secondoftwo}%
\providecommand \translation [1]{[#1]}%
\providecommand \BibitemOpen [0]{}%
\providecommand \bibitemStop [0]{}%
\providecommand \bibitemNoStop [0]{.\EOS\space}%
\providecommand \EOS [0]{\spacefactor3000\relax}%
\providecommand \BibitemShut  [1]{\csname bibitem#1\endcsname}%
\let\auto@bib@innerbib\@empty
\bibitem [{\citenamefont {Elouard}\ \emph {et~al.}(2017)\citenamefont
  {Elouard}, \citenamefont {{Herrera-Mart{\'i}}}, \citenamefont {Huard},\ and\
  \citenamefont {Auff{\`e}ves}}]{elouardExtractingWorkQuantum2017_SM}%
  \BibitemOpen
  \bibfield  {author} {\bibinfo {author} {\bibfnamefont {C.}~\bibnamefont
  {Elouard}}, \bibinfo {author} {\bibfnamefont {D.}~\bibnamefont
  {{Herrera-Mart{\'i}}}}, \bibinfo {author} {\bibfnamefont {B.}~\bibnamefont
  {Huard}},\ and\ \bibinfo {author} {\bibfnamefont {A.}~\bibnamefont
  {Auff{\`e}ves}},\ }\bibfield  {title} {\bibinfo {title} {Extracting {{Work}}
  from {{Quantum Measurement}} in {{Maxwell}}'s {{Demon Engines}}},\ }\bibfield
   {journal} {\bibinfo  {journal} {Physical Review Letters}\ }\textbf {\bibinfo
  {volume} {118}},\ \href {https://doi.org/10.1103/PhysRevLett.118.260603}
  {10.1103/PhysRevLett.118.260603} (\bibinfo {year} {2017})\BibitemShut
  {NoStop}%
\bibitem [{\citenamefont {Gradshteyn}\ and\ \citenamefont
  {Ryzhik}(2014)}]{gradshteyn2014table}%
  \BibitemOpen
  \bibfield  {author} {\bibinfo {author} {\bibfnamefont {I.~S.}\ \bibnamefont
  {Gradshteyn}}\ and\ \bibinfo {author} {\bibfnamefont {I.~M.}\ \bibnamefont
  {Ryzhik}},\ }\href@noop {} {\emph {\bibinfo {title} {Table of integrals,
  series, and products}}}\ (\bibinfo  {publisher} {Academic press},\ \bibinfo
  {year} {2014})\BibitemShut {NoStop}%
\bibitem [{\citenamefont {Abate}\ and\ \citenamefont
  {Whitt}(1992)}]{abateNumericalInversionProbability1992_SM}%
  \BibitemOpen
  \bibfield  {author} {\bibinfo {author} {\bibfnamefont {J.}~\bibnamefont
  {Abate}}\ and\ \bibinfo {author} {\bibfnamefont {W.}~\bibnamefont {Whitt}},\
  }\bibfield  {title} {\bibinfo {title} {Numerical inversion of probability
  generating functions},\ }\href {https://doi.org/10.1016/0167-6377(92)90050-D}
  {\bibfield  {journal} {\bibinfo  {journal} {Operations Research Letters}\
  }\textbf {\bibinfo {volume} {12}},\ \bibinfo {pages} {245} (\bibinfo {year}
  {1992})}\BibitemShut {NoStop}%
\bibitem [{\citenamefont {Kammerlander}\ and\ \citenamefont
  {Anders}(2016)}]{kammerlander2016coherence}%
  \BibitemOpen
  \bibfield  {author} {\bibinfo {author} {\bibfnamefont {P.}~\bibnamefont
  {Kammerlander}}\ and\ \bibinfo {author} {\bibfnamefont {J.}~\bibnamefont
  {Anders}},\ }\bibfield  {title} {\bibinfo {title} {Coherence and measurement
  in quantum thermodynamics},\ }\href@noop {} {\bibfield  {journal} {\bibinfo
  {journal} {Scientific reports}\ }\textbf {\bibinfo {volume} {6}},\ \bibinfo
  {pages} {22174} (\bibinfo {year} {2016})}\BibitemShut {NoStop}%
\bibitem [{\citenamefont {Giuggioli}(2020)}]{giuggioli2020exact}%
  \BibitemOpen
  \bibfield  {author} {\bibinfo {author} {\bibfnamefont {L.}~\bibnamefont
  {Giuggioli}},\ }\bibfield  {title} {\bibinfo {title} {Exact spatiotemporal
  dynamics of confined lattice random walks in arbitrary dimensions: a century
  after smoluchowski and p{\'o}lya},\ }\href
  {https://doi.org/10.1103/PhysRevX.10.021045} {\bibfield  {journal} {\bibinfo
  {journal} {Physical Review X}\ }\textbf {\bibinfo {volume} {10}},\ \bibinfo
  {pages} {021045} (\bibinfo {year} {2020})}\BibitemShut {NoStop}%
\bibitem [{\citenamefont {Sarvaharman}\ and\ \citenamefont
  {Giuggioli}(2020)}]{sarvaharman2020closed}%
  \BibitemOpen
  \bibfield  {author} {\bibinfo {author} {\bibfnamefont {S.}~\bibnamefont
  {Sarvaharman}}\ and\ \bibinfo {author} {\bibfnamefont {L.}~\bibnamefont
  {Giuggioli}},\ }\bibfield  {title} {\bibinfo {title} {Closed-form solutions
  to the dynamics of confined biased lattice random walks in arbitrary
  dimensions},\ }\href {https://doi.org/10.1103/PhysRevE.102.062124} {\bibfield
   {journal} {\bibinfo  {journal} {Physical Review E}\ }\textbf {\bibinfo
  {volume} {102}},\ \bibinfo {pages} {062124} (\bibinfo {year}
  {2020})}\BibitemShut {NoStop}%
\bibitem [{\citenamefont {Das}\ and\ \citenamefont
  {Giuggioli}(2022)}]{das2022discrete}%
  \BibitemOpen
  \bibfield  {author} {\bibinfo {author} {\bibfnamefont {D.}~\bibnamefont
  {Das}}\ and\ \bibinfo {author} {\bibfnamefont {L.}~\bibnamefont
  {Giuggioli}},\ }\bibfield  {title} {\bibinfo {title} {Discrete space-time
  resetting model: Application to first-passage and transmission statistics},\
  }\href {https://iopscience.iop.org/article/10.1088/1751-8121/ac9765}
  {\bibfield  {journal} {\bibinfo  {journal} {Journal of Physics A:
  Mathematical and Theoretical}\ }\textbf {\bibinfo {volume} {55}},\ \bibinfo
  {pages} {424004} (\bibinfo {year} {2022})}\BibitemShut {NoStop}%
\bibitem [{\citenamefont {Das}\ and\ \citenamefont
  {Giuggioli}(2023)}]{dasDynamicsLatticeRandom2023a_SM}%
  \BibitemOpen
  \bibfield  {author} {\bibinfo {author} {\bibfnamefont {D.}~\bibnamefont
  {Das}}\ and\ \bibinfo {author} {\bibfnamefont {L.}~\bibnamefont
  {Giuggioli}},\ }\bibfield  {title} {\bibinfo {title} {Dynamics of lattice
  random walk within regions composed of different media and interfaces},\
  }\href {https://doi.org/10.1088/1742-5468/aca8f9} {\bibfield  {journal}
  {\bibinfo  {journal} {J. Stat. Mech.}\ }\textbf {\bibinfo {volume} {2023}},\
  \bibinfo {pages} {013201} (\bibinfo {year} {2023})}\BibitemShut {NoStop}%
\bibitem [{\citenamefont {Sarvaharman}\ and\ \citenamefont
  {Giuggioli}(2023)}]{sarvaharman2023particle}%
  \BibitemOpen
  \bibfield  {author} {\bibinfo {author} {\bibfnamefont {S.}~\bibnamefont
  {Sarvaharman}}\ and\ \bibinfo {author} {\bibfnamefont {L.}~\bibnamefont
  {Giuggioli}},\ }\bibfield  {title} {\bibinfo {title} {Particle-environment
  interactions in arbitrary dimensions: A unifying analytic framework to model
  diffusion with inert spatial heterogeneities},\ }\href
  {https://doi.org/10.1103/PhysRevResearch.5.043281} {\bibfield  {journal}
  {\bibinfo  {journal} {Phys. Rev. Res.}\ }\textbf {\bibinfo {volume} {5}},\
  \bibinfo {pages} {043281} (\bibinfo {year} {2023})}\BibitemShut {NoStop}%
\bibitem [{\citenamefont {Marris}\ and\ \citenamefont
  {Giuggioli}(2024)}]{marris2024persistence}%
  \BibitemOpen
  \bibfield  {author} {\bibinfo {author} {\bibfnamefont {D.}~\bibnamefont
  {Marris}}\ and\ \bibinfo {author} {\bibfnamefont {L.}~\bibnamefont
  {Giuggioli}},\ }\bibfield  {title} {\bibinfo {title} {Persistent and
  anti-persistent motion in bounded and unbounded space: resolution of the
  first-passage problem},\ }\href {https://doi.org/10.1088/1367-2630/ad5d85}
  {\bibfield  {journal} {\bibinfo  {journal} {New Journal of Physics}\ }\textbf
  {\bibinfo {volume} {26}},\ \bibinfo {pages} {073020} (\bibinfo {year}
  {2024})}\BibitemShut {NoStop}%
\bibitem [{\citenamefont {Balakrishnan}\ \emph {et~al.}(1988)\citenamefont
  {Balakrishnan}, \citenamefont {Lakshmibala},\ and\ \citenamefont {Van
  Den~Broeck}}]{balakrishnanRenewalEquationPersistent1988_SM}%
  \BibitemOpen
  \bibfield  {author} {\bibinfo {author} {\bibfnamefont {V.}~\bibnamefont
  {Balakrishnan}}, \bibinfo {author} {\bibfnamefont {S.}~\bibnamefont
  {Lakshmibala}},\ and\ \bibinfo {author} {\bibfnamefont {C.}~\bibnamefont {Van
  Den~Broeck}},\ }\bibfield  {title} {\bibinfo {title} {The renewal equation
  for persistent diffusion},\ }\href
  {https://doi.org/10.1016/0378-4371(88)90101-X} {\bibfield  {journal}
  {\bibinfo  {journal} {Physica A: Statistical Mechanics and its Applications}\
  }\textbf {\bibinfo {volume} {153}},\ \bibinfo {pages} {57} (\bibinfo {year}
  {1988})}\BibitemShut {NoStop}%
\bibitem [{\citenamefont {Cover}\ and\ \citenamefont
  {Thomas}(1991)}]{coverElementsInformationTheory1991}%
  \BibitemOpen
  \bibfield  {author} {\bibinfo {author} {\bibfnamefont {T.~M.}\ \bibnamefont
  {Cover}}\ and\ \bibinfo {author} {\bibfnamefont {J.~A.}\ \bibnamefont
  {Thomas}},\ }\href {https://doi.org/10.1002/0471200611} {\emph {\bibinfo
  {title} {Elements of {{Information Theory}}}}},\ edited by\ \bibinfo {editor}
  {\bibfnamefont {D.~L.}\ \bibnamefont {Schilling}},\ Wiley {{Series}} in
  {{Telecommunications}}\ (\bibinfo  {publisher} {John Wiley \& Sons, Inc.},\
  \bibinfo {address} {New York, USA},\ \bibinfo {year} {1991})\BibitemShut
  {NoStop}%
\bibitem [{\citenamefont {Bresque}\ \emph {et~al.}(2021)\citenamefont
  {Bresque}, \citenamefont {Camati}, \citenamefont {Rogers}, \citenamefont
  {Murch}, \citenamefont {Jordan},\ and\ \citenamefont
  {Auff{\`e}ves}}]{bresqueTwoQubitEngineFueled2021_SM}%
  \BibitemOpen
  \bibfield  {author} {\bibinfo {author} {\bibfnamefont {L.}~\bibnamefont
  {Bresque}}, \bibinfo {author} {\bibfnamefont {P.~A.}\ \bibnamefont {Camati}},
  \bibinfo {author} {\bibfnamefont {S.}~\bibnamefont {Rogers}}, \bibinfo
  {author} {\bibfnamefont {K.}~\bibnamefont {Murch}}, \bibinfo {author}
  {\bibfnamefont {A.~N.}\ \bibnamefont {Jordan}},\ and\ \bibinfo {author}
  {\bibfnamefont {A.}~\bibnamefont {Auff{\`e}ves}},\ }\bibfield  {title}
  {\bibinfo {title} {Two-{{Qubit Engine Fueled}} by {{Entanglement}} and
  {{Local Measurements}}},\ }\href
  {https://doi.org/10.1103/PhysRevLett.126.120605} {\bibfield  {journal}
  {\bibinfo  {journal} {Phys. Rev. Lett.}\ }\textbf {\bibinfo {volume} {126}},\
  \bibinfo {pages} {120605} (\bibinfo {year} {2021})}\BibitemShut {NoStop}%
\end{thebibliography}
\end{document}